\documentclass[journal=jacsat,manuscript=communication]{achemso}
\usepackage{graphicx}
\usepackage{color}
\usepackage{titlecaps}
\usepackage{caption}
\usepackage{amssymb}
\usepackage{amsmath}
\usepackage{gensymb}
\usepackage{booktabs}
\usepackage{multicol}
\usepackage[table]{xcolor}
\usepackage{etoolbox}
\makeatletter
\def\acs@contact@details{
E-mail: \acs@email@list
}
\makeatother

\title{Fine-tuning Microporosity of Crystalline Vanadomolybdate Frameworks for Selective Adsorptive Separation of Kr from Xe}
\setkeys{acs}{
  abbreviations = false,
  articletitle  = true,
  biblabel      = brackets,
  biochem       = false,
  doi           = false,
  etalmode      = truncate,
  keywords      = false,
  maxauthors    = 95,
  super         = true
}
\author{Suchona Akter}
\affiliation{Division of Energy, Matter and Systems, School of Science and Engineering, University of Missouri $-$ Kansas City, Kansas City 64110, MO United States}
\author{Yong Li}
\affiliation{Division of Energy, Matter and Systems, School of Science and Engineering, University of Missouri $-$ Kansas City, Kansas City 64110, MO United States}
\author{Minbum Kim}
\affiliation{Physical and Computational Science Directorate, Pacific Northwest National Laboratory, Richland, WA 99352, USA}
\author{Md Omar Faruque}
\affiliation{Division of Energy, Matter and Systems, School of Science and Engineering, University of Missouri $-$ Kansas City, Kansas City 64110, MO United States}
\author{Zhonghua Peng}
\email{pengz@umkc.edu}
\affiliation{Division of Energy, Matter and Systems, School of Science and Engineering, University of Missouri $-$ Kansas City, Kansas City 64110, MO United States}
\author{Praveen K. Thallapally}
\email{praveen.thallapally@pnnl.gov}
\affiliation{Physical and Computational Science Directorate, Pacific Northwest National Laboratory, Richland, WA 99352, USA}
\author{Mohammad R. Momeni}
\affiliation{Division of Energy, Matter and Systems, School of Science and Engineering, University of Missouri $-$ Kansas City, Kansas City 64110, MO United States}
\email{mmomenitaheri@umkc.edu}
\date{}

\begin{document}

\begin{abstract}
Selective adsorptive capture and separation of chemically inert Kr and Xe noble gases with very low ppmv concentrations in air and industrial off-gases constitute an important technological challenge. Here, using a synergistic combination of experiment and theory, the microporous crystalline vanadomolybdates (MoVO$_x$) as highly selective Kr sorbents are studied in detail. By varying the Mo/V ratios, we show for the first time that their one-dimensional pores can be fine-tuned for the size-selective adsorption of Kr over the larger Xe with selectivities reaching $>$100. Using extensive electronic structure calculations and grand canonical Monte-Carlo simulations, the competition between Kr uptake with CO$_2$ and N$_2$ was also investigated. As most materials reported so far are selective toward the larger, more polarizable Xe than Kr, this work constitutes an important step toward robust Kr-selective sorbent materials. This work highlights the potential use of porous crystalline transition metal oxides as energy-efficient and selective noble gas capture sorbents for industrial applications.
\end{abstract}

%%%%%%%%%%%%%%%%%%%%%%%%%%%%%%%%%%%%%%%%%%%%%
% SECTION 1: INTRODUCTION
%%%%%%%%%%%%%%%%%%%%%%%%%%%%%%%%%%%%%%%%%%%%%

\section{1. Introduction}
\textbf{1. Introduction}\newline
Radioactive krypton (Kr) and xenon (Xe) noble gas mixtures produced during the reprocessing of the used nuclear fuel are often responsible for nuclear reactor accidents.\cite{lin2021multiscale} $^{85}$Kr, the most significant radioactive noble gas in the off-gas stream, emits beta (99.6\%) and gamma (0.4\%) radiations, direct exposure to which harms human health and increases cancer risk.\cite{abe2009distillation, none1975krypton} Therefore, the entrapment and separation of these noble gases from off-gas streams are demanded by environmental concerns and are also driven by economic incentives due to the scarcity and high commodity of both Kr and Xe.\cite{sunder1997calculation, liu2010mechanistic, feng2016kr} For example, Xe is used in healthcare, \cite{cullen1951anesthetic, sanders2005xenon} imaging in the biomedical industry, \cite{albert1994biological} and as a satellite propellant in the space industry.\cite{beattie1989xenon} Both Kr and Xe are used in lighting, \cite{yeralan2005advantages} lasers, \cite{bridges1965visible} double glazing for insulation, \cite{manz2008minimizing} and as carrier gases in analytical chemistry.\cite{lipsky1963use} This is due to their desirable properties, including high density, low chemical reactivity and solubility, and very low boiling temperature and conductivity. \cite{banerjee2018xenon} Nevertheless, due to the chemically inert nature of these noble gases, they are difficult to capture and separate. Currently, the only viable option for the separation of Kr and Xe from the spent nuclear fuel is the energy-intensive cryogenic distillation.\cite{zhao2023high}

The Kr/Xe noble gases also possess very low abundances (ppmv level) in air and industrial off-gases, making them even more difficult to capture and separate.\cite{wei2022efficient} Again, the highly energy-intensive cryogenic distillation is the common method for their capture and separation from the atmosphere despite their minuscule presence (1.14 and 0.086 ppmv for Kr and Xe, respectively).\cite{kerry2007industrial}. A more energy-efficient alternative can be physisorption-based separation using microporous materials. Solid adsorbents for capturing noble gases studied so far have been limited to activated carbon\cite{marsh2006activated}, carbon nanotubes (CNTs) \cite{majumdar2018adsorptive, yang2020monte}, zeolites\cite{smit2008molecular}, polymers \cite{hong2020troger}, covalent-organic frameworks (COFs)\cite{waller2015chemistry}, and metal-organic frameworks (MOFs).\cite{banerjee2018xenon, li2009selective} Despite significant advancements in the development of new adsorbents for the capture and separation of noble gases, the need for novel materials beyond those few studied families is immense. As such, significant research endeavors have recently been devoted to the design and discovery of next-generation solid adsorbents for noble gas separation.\cite{osti_1844866} 

Given the size and shape of an adsorbent’s pores and/or the binding affinity of its adsorption sites, the adsorptive separation of gas mixtures can be realized by differentiating the adsorbate molecular sizes, shapes, polarities, polarizabilities, coordination abilities, and conformations.\cite{li2009selective} Accordingly, one can classify separation mechanisms as thermodynamic separation, kinetic separation, conformational separation, and molecular sieving. The selectivity, therefore, depends on several factors, such as the adsorbent-adsorbate interactions, the adsorbent's pore size, and the size/shape of the adsorbate molecules.\cite{gu2002simulation} Not only a high specific surface area but also pore sizes appropriate to the diameter of one of the components in the gas mixture are crucial for enhanced adsorption capacity. In such a condition, some gas molecules are selectively adsorbed in the pores, while others are excluded (i.e., the size sieving mechanism). Consequently, porous materials with tunable pore sizes and high specific surface areas are highly desirable solid sorbents for the selective adsorptive separation of gas mixtures.\cite{sun2016recent, gadipelli2015graphene}

There are currently two key challenges in developing next-generation effective materials for noble gas separation: (i) the demand for structural robustness under different operating conditions, e.g., in radiolytic, corrosive, and chemically reactive environments,\cite{moyer2022innovative} and (ii) the need for microporous materials with tunable pore sizes given the inert nature of noble gases with negligible charge transfer and polarization effects which limits the separation mechanism to size sieving. Adsorbents based on activated carbon and zeolites have been tested for noble gas capture but have shown low capacity, selectivity, and lack of modularity.\cite{thallapally2012facile, bazan2011adsorption, jameson1997competitive} 
From a Kr/Xe separation viewpoint,\cite{sikora2012thermodynamic, simon2015best} it is now well-established that (i) selectivity for Xe is maximized in materials with uniform pores slightly larger than Xe (with a kinetic diameter of $\approx$4.1~\AA) and (ii) selective Kr capture is achieved in materials with uniform pores smaller than Xe but larger than Kr (with a kinetic diameter of $\approx$3.7~\AA). 

\begin{figure}[!t]
    \centering
    \includegraphics[width=0.99\linewidth]{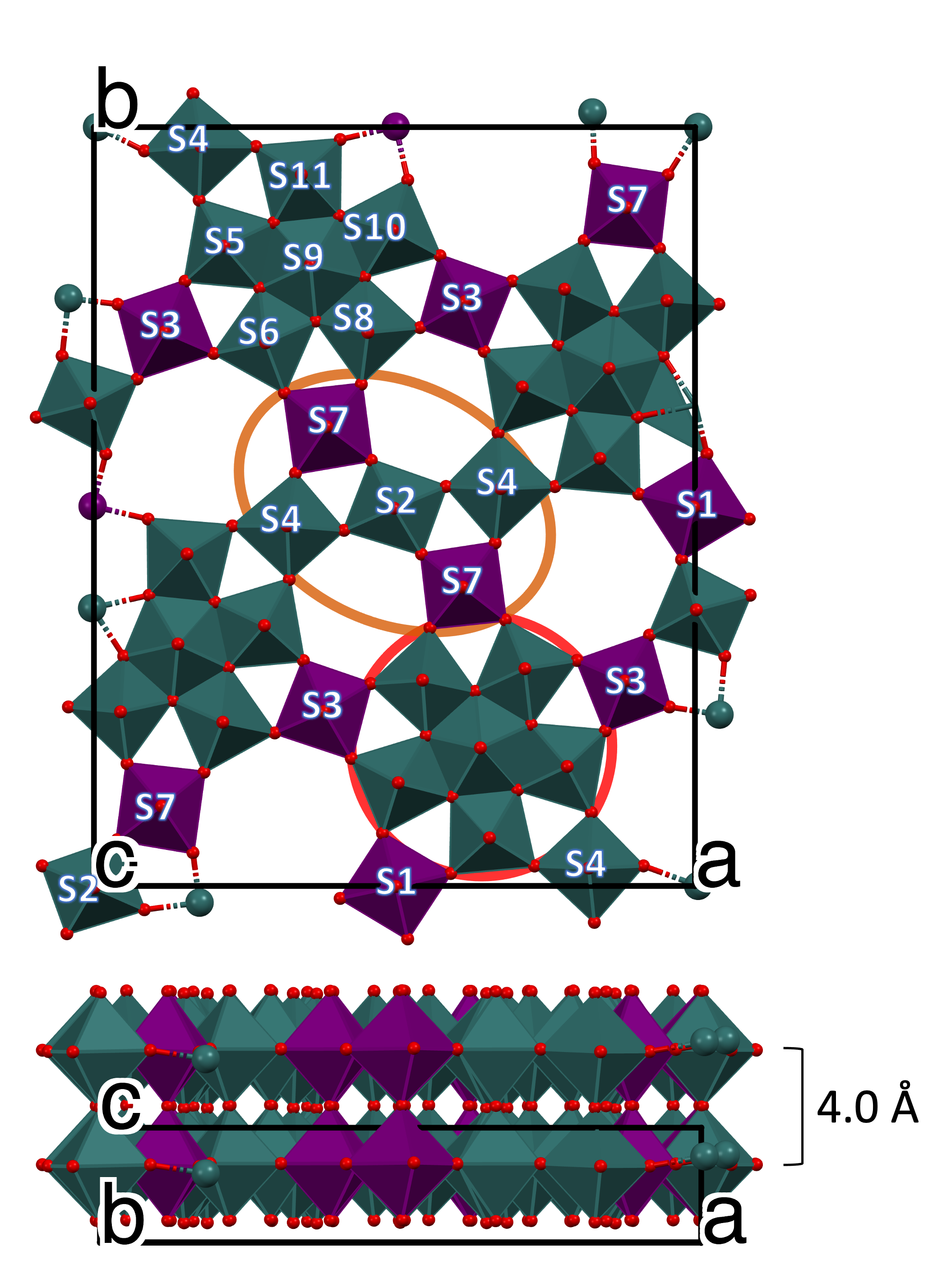}
    \caption{The layered crystal strucutre of the representative orthorhombic Mo$_{30}$V$_{10}$O$_{112}$ vanadomolybdate along the crystallographic ``c" (top) and ``a" (bottom) axes, respectively. The unit cell contains four hexagonal and four heptagonal pores. The red and orange circles show pentagonal and pentameric units, respectively. The interlayer distance is also given in~\AA.}
    \label{unit_cell}
\end{figure}

In this work, the selective adsorptive separation of Kr over Xe is studied using all inorganic crystalline MoVO$_x$ vanadomolybdate adsorbents via a combination of experimental syntheses and characterizations and theoretical calculations (Figure \ref{unit_cell}). Crystalline vanadomolybdates MoVO$_x$ are a unique class of intrinsically porous inorganic materials that have been extensively studied for catalytic applications.\cite{ishikawa2015redox, konya2013orthorhombic} However, they are yet to be explored as sorbents for Kr/Xe noble gas capture and separation applications.
The variability of the fractional occupancy in MoVO$_x$ is a desirable feature of these materials recently utilized to design novel structures with modified Mo/V ratios\cite{pyrz2010atomic} with V centers known to be reduced first.\cite{lopez2005highly} A recent XPS analysis has shown that under steady-state conditions, V in these materials remains predominantly in the oxidation state of +4 while Mo adapts an oxidation state of +6.\cite{andrushkevich1993heterogeneous} In this work, different orthorhombic Mo$_{40-n}$V$_n$O$_{112}$ were designed by systematically increasing V contents, resulting in frameworks ranging from $n=$0 (i.e., Mo$_{40}$O$_{112}$ with no V atoms) to $n=$16 where all the linking sites were replaced with V (i.e., Mo$_{24}$V$_{16}$O$_{112}$; S1, S2, S3, S4, and S7 sites in Figure \ref{unit_cell}). Using the obtained lowest energy configurations, the structural, electronic, and Kr/Xe uptakes and selectivities of these materials were studied in detail. Our results show that the introduced changes significantly affect the stability and selectivity of these materials toward Kr/Xe adsorptive separation. Experimental synthesis and characterizations of the representative Mo$_{30}$V$_{10}$O$_{110}$ material were also performed, and in agreement with the theory, a higher selectivity toward Kr capture than Xe was obtained. This work is organized as follows: in Section 2, experimental syntheses, characterizations, and adsorption measurements are given; simulation details are outlined in Section 3; results and discussions are presented in Section 4, followed by conclusions and future works.

%%%%%%%%%%%%%%%%%%%%%%%%%%%%%%%%%%%%%%%%%%%%%
% SECTION 1: METHODS
%%%%%%%%%%%%%%%%%%%%%%%%%%%%%%%%%%%%%%%%%%%%%

\section{2. Experimental Section}
\textbf{2. Experimental Section}\newline
\subsection{2.1. Synthesis and Characterization of Crystalline Vanadomolybdate Frameworks}
\textbf{2.1. Synthesis and Characterization of Crystalline Vanadomolybdate Frameworks}\newline 
Orthorhombic Mo$_{30}$V$_{10}$O$_{112}$ was prepared following a modified literature procedure.\cite{acie_47_2493} Ammonium molybdate tetrahydrate [(NH$_4$)$_6$Mo$_7$O$_{24}$·4H$_2$O; Sigma-Aldrich] (3.7054 g, 3.00 mmol) was dissolved in deionized water (50.4 mL). Vanadium(IV) oxide sulfate hydrate (VOSO$_4$·xH$_2$O; Sigma-Aldrich) (1.3774 g) was dissolved in deionized water (50.4 mL). The above two solutions were mixed at room temperature and the resulting mixture was stirred for 10 min and then transferred to a 125-mL Teflon liner of a stainless steel autoclave. The reaction mixture was purged with nitrogen for 10 min and then heated at 175 $^{\circ}$C for 48 h. After cooling to room temperature, the reaction mixture was then filtered. The obtained solid was washed with deionized water and then dried at 80 $^{\circ}$C overnight. A gray solid (1.3970 g) crude product was obtained. The crude solid was added to an aqueous solution of oxalic acid (0.4 M, 35 mL), and this mixture was stirred at 60 $^{\circ}$C for 30 min. The mixture was then subjected to vacuum filtration. The solid obtained was washed with deionized water and then dried at 80 $^{\circ}$C overnight. The solid was then moved to a conventional furnace and heated at 10 $^{\circ}$C.min$^{-1}$ to 400 $^{\circ}$C in air. After calcination at 400 $^{\circ}$C for 2 h, a solid product (0.80 g) was obtained.
Crystal structure characterization was carried out by X-ray diffraction measurement on a PANalytical Empyrean diffractometer with Cu target at a tube voltage/current of 45 kV/40 mA, see Supporting Information (SI) Figure S1. The theta-2theta scan range was performed from 5 to 70 degrees in continuous scan mode. The detector used was a 1D detector (PIXcel) with 255 channels with a step size of 0.01313 degrees. A divergence slit of 1/8 degrees was used to limit the length of the sample illuminated with the incident X-ray beam, whereas an anti-scatter slit of 1/4 degrees was used to reduce the amount of background radiation.
The XRD pattern of the final synthesized orthorhombic Mo$_{30}$V$_{10}$O$_{112}$ crystals (SI Figure S1) matches very well with those reported in the literature.\cite{acie_47_2493,konya2013orthorhombic}

\subsection{2.2. Kr/Xe Adsorption Measurements}
\begin{figure}[!h]
    \centering
    \includegraphics[width=0.99\linewidth]{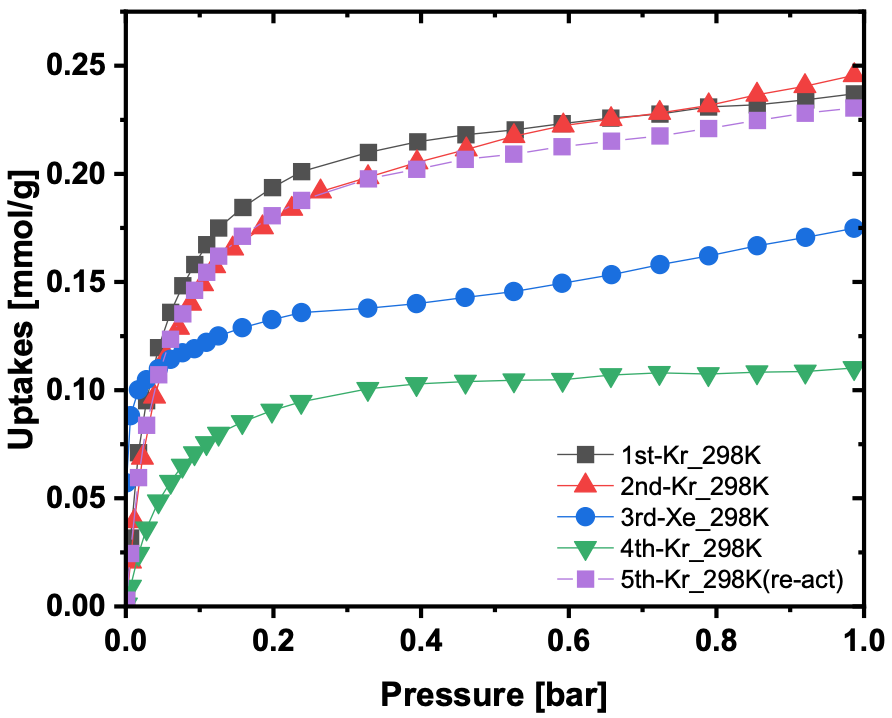}
    \caption{Measured Kr and Xe single-component adsorption isotherms for orthorhombic Mo$_{30}$V$_{10}$O$_{112}$ at 298 K. See the text for details.}
    \label{exp_isotherm}
\end{figure}

\textbf{2.2. Kr/Xe Adsorption Measurements}\newline 
The single-component Kr and Xe adsorption isotherms were measured using a 3Flex (Micromeritics instruments, USA). The sample (approx. 100 mg) was activated at 100 $^{\circ}$C for 16 h followed by 300 $^{\circ}$C for 2 h under vacuum. 
The first adsorption experiment (1$^{st}$-Kr, Figure \ref{exp_isotherm}), conducted on the activated sample for Kr at 298 K, resulted in a Kr adsorption uptake of approximately 0.24 mmol/g at 1 bar.
Subsequently, Kr was desorbed using only vacuum without heating, followed by a second Kr adsorption experiment (2$^{nd}$-Kr, Figure \ref{exp_isotherm}) measured at the same 298 K temperature. In the second Kr adsorption experiment, the sample exhibited a negligible decrease in Kr adsorption at low pressure and nearly identical Kr adsorption uptake at 1 bar. After degassing the sample using a vacuum without heating, a third adsorption experiment (3$^{rd}$-Xe, Figure \ref{exp_isotherm}) was conducted for Xe at 298 K. At 1 bar, the sample exhibited a Xe adsorption uptake of approximately 0.17 mmol/g, which was lower than the Kr adsorption uptake (0.24 mmol/g). After the Xe adsorption experiment, the sample was degassed with a simple vacuum, and a fourth experiment (4$^{th}$-Kr, Figure \ref{exp_isotherm}) was performed for Kr at 298 K. Interestingly, the sample exhibited a significantly decreased Kr uptake (0.1 mmol/g) compared to the previous Kr uptake (1$^{st}$ and 2$^{nd}$ experiments). The decrease in Kr uptake could be attributed to the Xe molecules partially occupying the sample's pores, which were not completely desorbed by applying only vacuum without heating. To address the decreased Kr uptake, the sample was re-activated at 100 $^{\circ}$C for 2 h. Subsequently, the Kr adsorption was measured again at 298 K (5$^{th}$-Kr, Figure \ref{exp_isotherm}), confirming that the Kr uptake was almost similar to the previous Kr uptake. 

\section{3. Theory and Simulation Details}
\textbf{3. Theory and Simulation Details}\newline
\subsection{2.1. Electronic Structure Calculations}
\textbf{3.1. Electronic Structure Calculations}\newline
Comparative adsorptive performance of vanadomolybdate frameworks with different Mo/V ratios has not been reported previously. Here, microporous orthorhombic Mo$_{40-n}$V$_n$O$_{112}$ materials with different V contents are considered with n$=$0, 2, 4, 6, 8, 10, 12, 14, and 16 encompassing Mo$_{40}$O$_{112}$ system with no V atoms, to the V-rich Mo$_{24}$V$_{16}$O$_{112}$ system where all the octahedral linking sites are doped with V (see Figure \ref{unit_cell}). Using our in-house codes, each system was initially created by placing the V atoms on all possible linking sites. The lowest energy configuration was located for all systems by subjecting all isomers to full minimizations using dispersion-corrected density functional theory (DFT) calculations in VASP.\cite{Kresse1993,Kresse1994,Kresse1996,Kresse19962,kresse1996efficient} Periodic boundary conditions were used in all calculations. For n$=$0 and 16 systems, there is only one possible isomer. The lowest energy isomer for n$=$10, i.e., Mo$_{30}$V$_{10}$O$_{112}$, has already been identified and was taken from Ref. \citenum{li2018distribution}.

Interactions between electrons and ions were described by Projector Augmented Wave (PAW) potentials \cite{PAW1, PAW2, blochl1994projector} with the energy cutoff of 500 eV. Gaussian smearing was adopted in all atomic position and cell vector minimizations with a smearing width of 0.05 eV. The convergence criteria were set to 10$^{-5}$ for SCF and $10^{-6}$ for electronic property calculations. The $k$-point mesh in the Monkhorst--Pack \cite{monkhorst1976special} scheme was set to 1$\times$1$\times$6 in the geometry optimizations and twice denser for the electronic property calculations. The precision for all calculations was set to “high”, which increases the plane-wave cutoff energy by 30\%. The spin polarization of the unpaired electrons was considered with the  ISPIN and NUPDOWN parameters in VASP. Different spin states were considered, and the lowest-energy states were identified for all studied systems whenever applicable.

Comprehensive benchmarks were performed for the representative Mo$_{30}$V$_{10}$ system against previously reported\cite{li2018distribution} dispersion-corrected hybrid B3LYP-D2 data.\cite{jcp_98_5648, grimme2010consistent} These include cutoff energy, $k$-point mesh, and different GGA functionals, including Perdew-Burke-Ernzerhof (PBE),\cite{perdew1996generalized} revised PBE (RPBE), PBE for solids (PBEsol),\cite{perdew2008restoring} dispersion-corrected PBE-D2 and PBE-D3,\cite{grimme2010consistent} as well as PBE with Hubbard U correction (PBE-U), with varying U values (see SI Figure S2 and Table S6). The Hubbard-U correction was added for the Mo and V transition metals to compensate for the over-delocalization of $d$-electrons commonly encountered with GGA functionals, as well as for the $p$ orbitals of the O atoms.\cite{moore2024high} The rotationally invariant Hubbard U parameter, as proposed in Ref. \citenum{dudarev1998electron}, reflects the strength of the on-site Coulomb interactions, whereas the parameter J adjusts the exchange interactions. These two parameters are usually combined into a single parameter U$_{eff}$=U–J. Here, different U$_{eff}$ parameters were benchmarked according to the ranges specified in Ref. \citenum{moore2024high} and are tabulated in the SI Figure S2. For these benchmarks, different U$_{eff}$ parameters were tested for reproducing the unit cell and atomic positions as reported in Ref. \citenum{li2018distribution}. Our extensive benchmarks for the representative Mo$_{30}$V$_{10}$ system showed that the PBE+U method combined with Grimme's DFT-D3 dispersion correction and Becke-Johnson damping \cite{johnson2006post} worked well and closely reproduced that of the reference (see SI Figure S2 and Table S6). This methodology was, therefore, adapted for the remainder of the systems.

Formation energy ($\Delta E_f$) is among the essential descriptors for predicting the thermodynamic stability of an unknown system and is defined as the energy required to form a material from its constituent elements. The formation energies of all considered systems were calculated using the following equation:\cite{kirklin2015open}
\begin{equation}\label{eqn:form}
    \Delta E_f = E_{tot} - \sum_i \mu_i x_i
\end{equation}
where $E_{tot}$ is the DFT calculated total energy of the system, with $\mu_i$ being the chemical potential of element $i$ and $x_i$ the quantity of element $i$ in the material.
To probe the effects of different Mo/V contents on the structural properties of the lowest energy systems, a variety of geometric features, including density, porosity, gravimetric and volumetric surface areas (GSA and VSA), largest cavity diameter (LCD), and pore limiting diameter (PLD) were calculated using Zeo++.\cite{willems2012algorithms, martin2012addressing} A He probe with a vdW radius of 1.4~\AA~was used.

\subsection{3.2. Molecular Simulations}
\textbf{3.2. Molecular Simulations}\newline
Single-component adsorption isotherms for Kr, Xe, and the competing CO$_2$ and N$_2$ adsorbates were calculated at 298 K using Grand Canonical ($\mu$VT ensemble) Monte Carlo (GCMC)\cite{sikora2012thermodynamic} simulations as implemented in RASPA.\cite{dubbeldam2016raspa, dubbeldam2018iraspa} The selectivity for Kr capture in 50:50 and 80:20 Kr/Xe mixtures was simulated at 298 K. Equimolar mixtures of Kr with the competing CO$_2$ and N$_2$ adsorbates were also simulated using GCMC simulations in RASPA.
The universal force field (UFF)\cite{casewit1992application, lin2021multiscale} was used for all studied vanadomolybdates except for the O atoms, which was taken from Ref. \citenum{cuadros1996determination}. The rigid framework approximation was used for all studied materials with their atomic positions fixed at their PBE-U-D3(BJ) minimized positions.\cite{ozturk2015hydrogen, wu2012effects} Peng-Robinson equation of state was used to calculate the relationship between the chemical potential and the pressure of the systems (fugacites). Periodic boundary conditions were employed in all simulations to remove surface effects. An atom-atom cutoff distance of 12.0~\AA~was employed for truncating the short-range interactions, while long-range electrostatics were calculated using the Ewald summation technique.\cite{Leach:2001} Simulations were performed in cells with sufficient repeat units such that all side lengths were greater than 24~\AA. Partial atomic charges were calculated using Bader charge analysis with the exception that a charge of zero was assigned to noble gasses given their chemically inert nature.\cite{henkelman2006fast} Among the adsorbates, Kr and Xe were modeled using parameters from Ref. \citenum{ryan2011computational} and N$_2$ and CO$_2$ were modeled using the TraPPE force field.\cite{rai2008application} The 6-12 Lennard-Jones (LJ) potential was employed to describe adsorbent-adsorbate and adsorbate-adsorbate interactions with all cross terms computed using the Lorentz-Berthelot mixing rules. To achieve higher accuracies, the $\epsilon_{Kr-O}$ and $\epsilon_{Xe-O}$ parameters were adjusted to reproduce our experimental adsorption isotherms (see SI Figure S3). 1$\times$10$^6$ Monte Carlo equilibration cycles were found to suffice for all systems, whereas 2$\times$10$^6$ production cycles were used to calculate ensemble averages. Our extensive GCMC benchmarks for all systems are included in SI Figures S4-S20.
Four types of Monte Carlo (MC) moves in equal probabilities, including swap (addition/deletion), reinsertion, random rotation, and translation of the adsorbate molecules, were considered for single-component adsorption isotherms. In addition to these MC moves, the identity change probability (e.g., changing Kr to Xe) was also considered in two-component adsorption isotherms. All simulations were performed at 298 K and pressures of 0.02, 0.04, 0.06, 0.08, 0.10, 0.15, 0.20, 0.25, 0.30, 0.35, 0.40, 0.45, 0.50, 0.55, 0.60, 0.65, 0.70, 0.75, 0.80, 0.85, 0.90 and 1.00 bar. The He void fractions for the structures were calculated using widom insertions.\cite{myers2002adsorption} All potential parameters are listed in SI Tables S2-S5.

Adsorption selectivity is an important parameter for probing the separation capability of a potential adsorbent material. For the two-component mixtures of Kr and Xe, the selectivity $S_{Kr/Xe}$ was calculated using the following equation: 
\begin{equation}\label{eqn:selectivity}
    S_{Kr/Xe} = \frac{x_{Kr} / x_{Xe}}{y_{Kr} / y_{Xe}}
\end{equation}
where $x$ and $y$ are adsorbed gas loadings in the material and mole fraction of adsorbates in bulk, respectively. The same equation was used to calculate the selectivity of Kr and Xe with an equimolar mixture of N$_2$ and CO$_2$.
As a measure of the adsorbate-adsorbent attraction strengths, the fluctuation method \cite{nicholson1982computer} was used to calculate the isosteric heat of adsorption (Q$_{st}$) at 1 bar as
  \begin{equation}\label{eqn:qst}
   Q_{st} = \frac{\langle UN \rangle - \langle U \rangle \langle N \rangle}{\langle N^2 \rangle - \langle N \rangle^2} - RT
\end{equation}
where R, T, U, and N are the universal gas constant, temperature, energy, and number of particles, respectively, with brackets representing ensemble averages.

%%%%%%%%%%%%%%%%%%%%%%%%%%%%%%%%%%%%%%%%%%%%%
% RESULTS
%%%%%%%%%%%%%%%%%%%%%%%%%%%%%%%%%%%%%%%%%%%%%
\section{4. Results and Discussion}
\textbf{4. Results and Discussion}\newline
\subsection{4.1. Structural and Thermodynamic Properties of V-Doped Vanadomolybdates}
\textbf{4.1. Structural and Thermodynamic Properties of V-Doped Vanadomolybdates}\newline 
Vanadomolybdate frameworks are known to crystallize in four different phases: orthorhombic, trigonal, tetragonal, and amorphous, each with different porosities and surface areas.\cite{fjermestad2018acrolein} Out of these, orthorhombic MoVO$_x$ oxides have attracted much attention due to their outstanding catalytic performance, for example, in oxidative activation of short-chain alkanes.\cite{ishikawa2015redox, konya2013orthorhombic} Orthorhombic MoVO$_x$ features slabs comprising 1D hexagonal and heptagonal pores of corner-sharing MO$_6$ octahedra (M = Mo or V) and pentagonal Mo$_6$O$_{21}$ units with MoO$_7$ pentagonal bipyramidal units and five edge-sharing MO$_6$ octahedra (see Figure \ref{unit_cell}).\cite{ishikawa2019multi, ishikawa2018unit} Stacking of the six and seven-membered rings constructs 1D channel structures, with the seven-membered channel forming a micropore with a diameter of around 4~\AA~ that allows small molecules such as methane, ethane, N$_2$, Ar, and CO$_2$ to enter.\cite{sadakane2008molybdenum} This orthorhombic M1 phase of MoVO$_x$ is known to crystallize in the Pba2 space group which closely resembles that of the so-called M1 phase of MoVTeNbO$_x$.\cite{lunkenbein2015direct} It shows a characteristic layered structure stacked in the ``c" direction (Figure \ref{unit_cell}). Per unit cell, this microporous material contains two pentameric units, four pentagonal units, and six octahedral linker sites.\cite{grasselli2014catalytic, li2019reactivity} The space group Pba2 imposes constraints on the M1 MoVO$_x$ structure such that only 11 among the 40 transition metal centers are symmetry inequivalent, labeled as S1, S2, ..., S11 (see Figure \ref{unit_cell}).\cite{li2019reactivity}

As mentioned above, it has already been established that the stability of the M1 phases of MoVO$_x$ and MoVNbTeO depend on the distribution of the V centers.\cite{li2018distribution, arce2020teo, cheng2015silico} We extensively surveyed V site occupancies in MoVO$_x$ as linker sites in these materials are known to host both Mo and V atoms.\cite{willingerdirect} 
Initial experiments suggested that, the pentagonal sites S5, S6, S8, S9, S10, and S11 exhibit the highest occupancy for Mo followed by the linking S2, S4, S3, S7, and S1 sites, with the order S9$\approx$S5$\approx$S6$>$S8$\approx$S10$\approx$S11$>$S2$>$S4$\approx$S3$\approx$S7$>$S1\cite{desanto2003structural, desanto2004structural}. These preliminary results were first deduced based on powder diffraction data, but later on, scanning transmission electron microscopy (STEM) data generated a more accurate occupancy scenario for similar systems.\cite{pyrz2010atomic} For example, it was demonstrated that all octahedra sites connecting the pentagonal units (S1, S2, S3, and S4, Figure \ref{unit_cell}) contain appreciable amounts of V.\cite{li2011improvement, grasselli2014catalytic} The S5, S6, and S11 sites were also found to contain about 5, 12, and 5\% V, respectively, but the sites S8 and S10 were shown to contain only Mo. The V atoms were found to strongly prefer the corner-sharing octahedra rather than the edge-sharing sites. Later experiments reported that the highest V occupancies belong to the S1, S3, and S7 sites,\cite{epicier2017spatial} with S4 being a highly Mo-dominant site with Mo occupancies close to 100\%.\cite{arce2020teo}

Guided by these results, a full survey of all possible V-doped structures was performed for the extended family of the vanadomolybdate MoVO$_x$ materials using dispersion-corrected PBE-U-D3 calculations in VASP. The presence of V atom in the S4 sites was found to always result in relatively high energy isomers, which agrees with the experimental observation of significantly lower V occupancy in S4 sites for the M1 orthorhombic phase of vanadomolybdates as well as for the MoVTeO and MoVTeNbO materials mentioned above.\cite{li2011improvement, epicier2017spatial} 
The lowest energy isomer for Mo$_{30}$V$_{10}$ contains V in S1, S3, and S7 sites with the ordering of S1 $>$ S3 $>$ S7, which also agrees with experiment \cite{arce2020teo}. In the lowest energy isomer of Mo$_{28}$V$_{12}$, S1, S2, S3, and S7 sites are occupied by V. There is no V in any of the four S4 sites in the unit cell.

\begin{figure}[!t]
    \centering
    \includegraphics[width=0.99\linewidth]{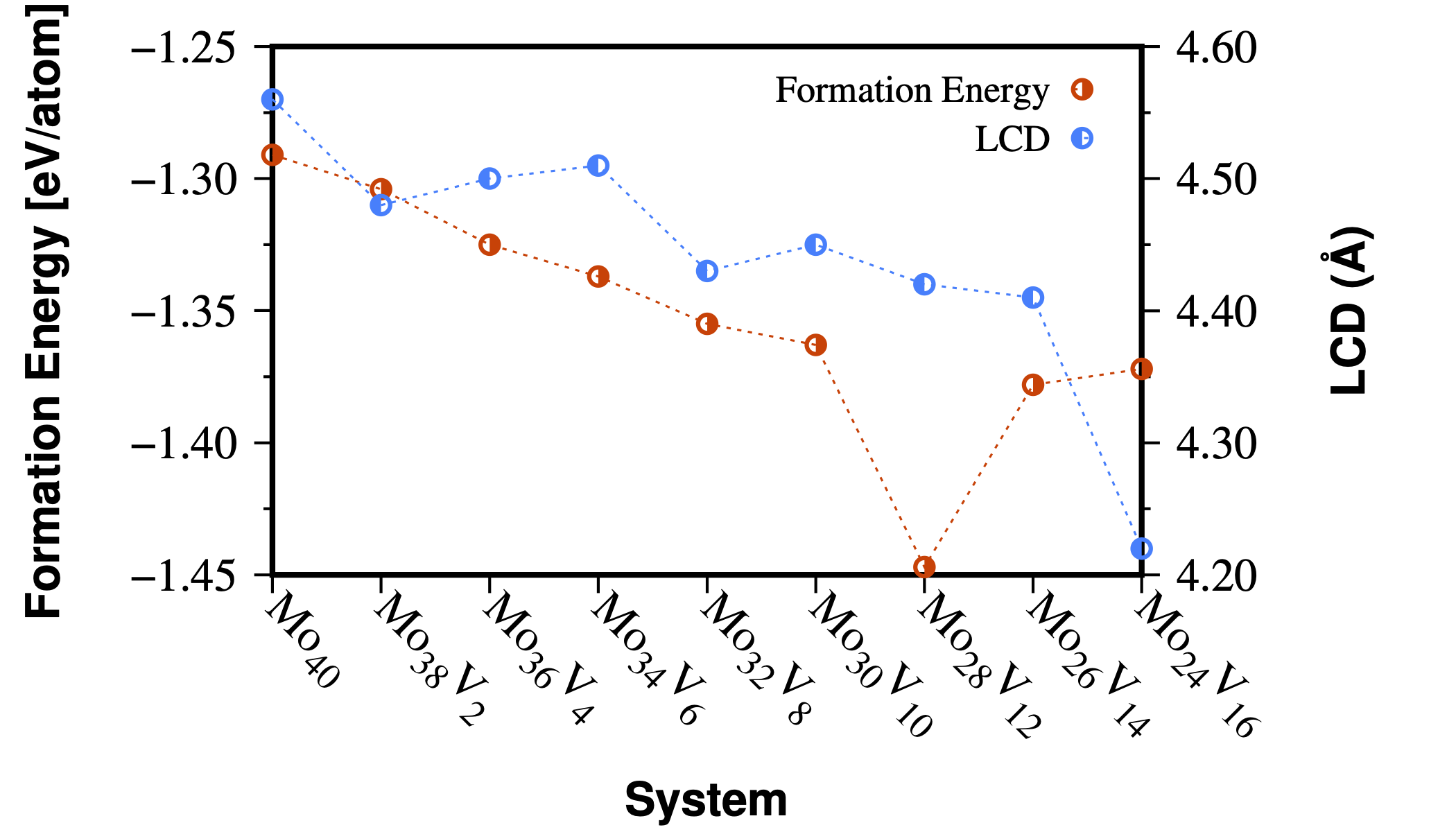}
    \caption{Calculated (a) formation energies (in eV/atom) and (b) LCD (in~\AA) of all systems considered in this work.}
    \label{lcd_form}
\end{figure}
Figure \ref{lcd_form} shows the calculated formation energy increases (i.e., becomes more negative) as the V content increases up to $n=$14, and then plateaus at $n=$16 with the Mo$_{28}$V$_{12}$ system being an outlier with the lowest calculated formation energy of -1.447 eV/atom.
Our computed trend of formation energies here overall agrees very well with Ref. \citenum{cheng2015silico}, where for the similar M1 phase of MoVNbTeO, DFT calculations showed that the structure stabilizes by increasing the concentration of V. Interestingly, as for the Mo$_{28}$V$_{12}$ with the lowest calculated formation energy, we found that its unit cell volume is also the highest among all studied systems. This increased volume is resulted from an elongated ``c" axis in this material compared to the others. We attribute this rather significant unit cell change in Mo$_{28}$V$_{12}$ to the Jahn-Teller (JT) effect,\cite{jahn1937stability} most commonly observed in octahedral transition metal complexes.\cite{raabe2024heteroelements} According to the JT effect, a distortion happens in nonlinear complexes, splitting the degenerate energy levels and lowering the overall symmetry, which results in higher stabilities.\cite{jahn1937stability} As a result, the average VO$_6$ octahedra length is the highest (4.3~\AA) for Mo$_{28}$V$_{12}$, wheres for the other systems it ranges from 3.9~\AA~in Mo$_{26}$V$_{14}$ to 4.1~\AA~in Mo$_{36}$V$_4$.
For all studied systems, one can see a conspicuous trend in their calculated geometric parameters with an increase in the V content. Specifically, the calculated LCD, PLD, GSA, and VSA are the highest for Mo$_{40}$ with no V and the lowest for the V-rich Mo$_{24}$V$_{16}$ system (see SI Table S7). This is due to the smaller ionic radius of V compared to that of Mo (0.54~\AA~for V$^{+5}$ vs. 0.59~\AA~for Mo$^{+6}$), resulting in slightly smaller pore sizes as more Mo atoms are replaced with V. As such, the expectation for these materials is for them to become more selective toward the smaller Kr than Xe as the pore sizes are reduced upon V doping. More specifically, Mo$_{24}$V$_{16}$ with a computed LCD of 4.22~\AA~is expected to effectively exclude the larger Xe atoms with a kinetic diameter of $\approx$4.1~\AA~compared to Mo$_{40}$ with the calculated largest LCD value of 4.56~\AA. Kr/Xe adsorption and selectivity and their possible correlations with these geometric features are outlined in detail in the next section.

\begin{figure*}[!h]
    \centering
    \includegraphics[width=0.99\linewidth]{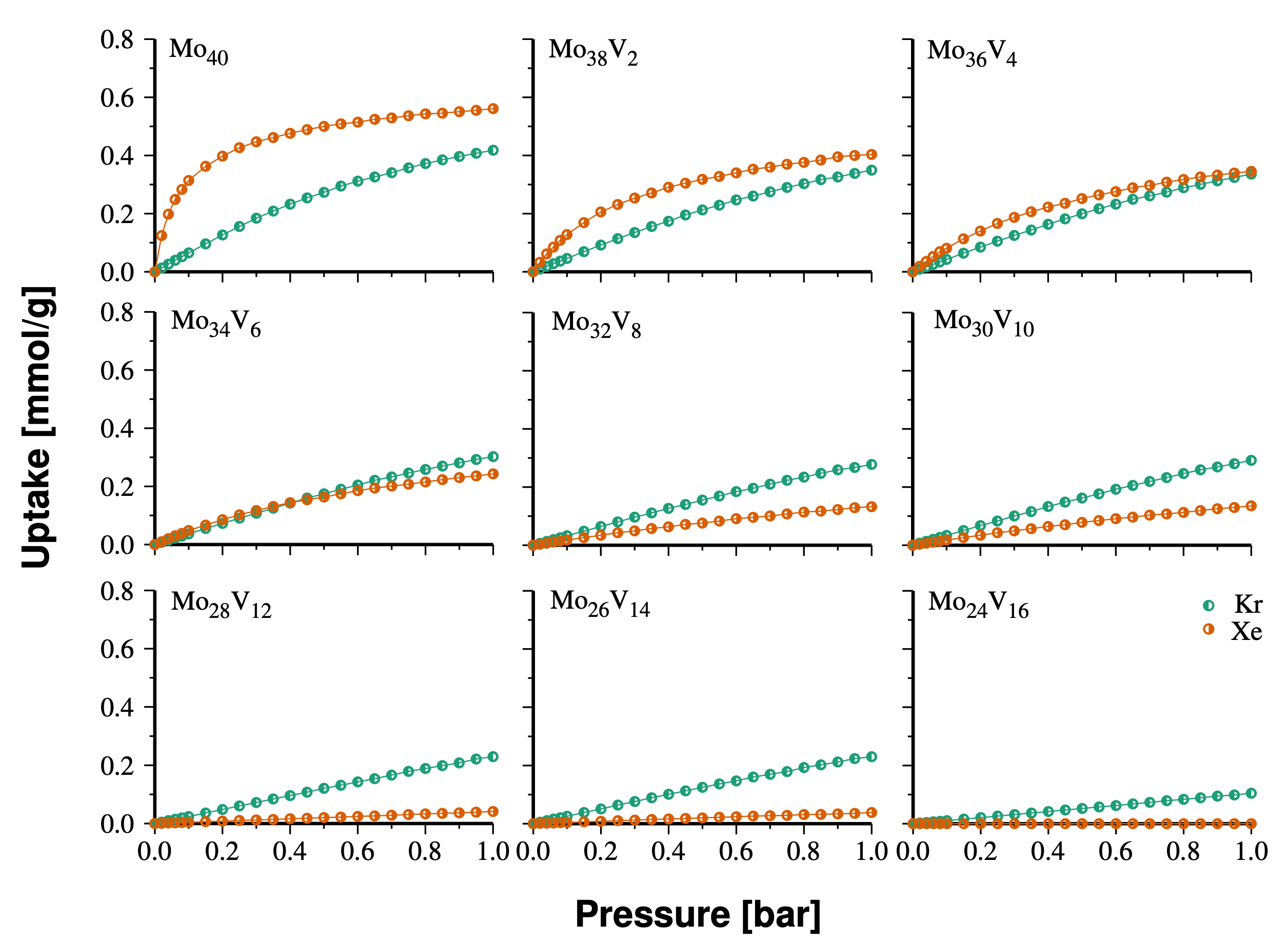}
    \caption{GCMC calculated Kr and Xe single-component adsorption isotherms of all considered systems at 298 K using RASPA.}
    \label{isotherm}
\end{figure*}
\subsection{4.2. Kr/Xe Adsorption in V-Doped Vanadomolybdates}
\textbf{4.2. Kr/Xe Adsorption in V-Doped Vanadomolybdates}\newline
Single-component adsorption isotherms of Kr and Xe were calculated using GCMC simulations at 298 K in RASPA (Figure \ref{isotherm}). Overall, increasing the V content in the studied MoVO$_x$ vanadomolybdates was found to lower the uptake for both Kr and Xe. This is consistent with the computed decreasing trend in the calculated surface area and porosity of these materials as more Mo ions are replaced by the smaller V ones. Interestingly, the Xe uptake, which is initially higher than Kr for $n=$0 and 2, was found to become similar at the higher 1 bar pressure for n$=$4 in the Mo$_{36}$V$_{4}$ system. And, starting from Mo$_{34}$V$_{6}$, the order was calculated to reverse with Kr uptake becoming higher than Xe. The Xe uptake was found to continuously go down as more V atoms are introduced to the material till for Mo$_{24}$V$_{16}$, where the material can virtually no longer adsorb Xe. 

\begin{figure}[!h]
    \centering
    \includegraphics[width=0.99\linewidth]{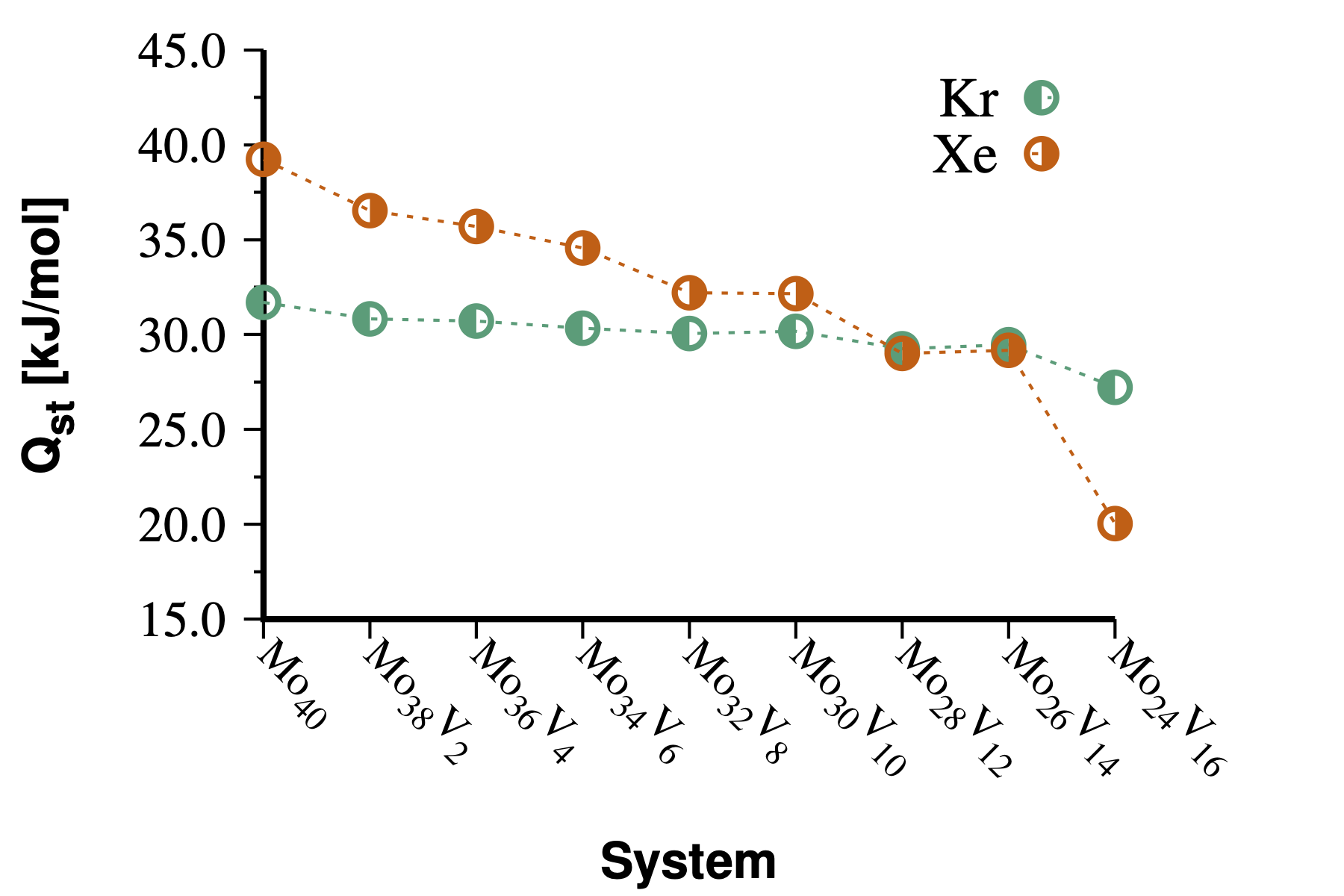}
    \caption{GCMC calculated Q$_{st}$ for Kr and Xe at 298 K and 1 bar for all systems considered in this work.}
    \label{qst}
\end{figure}

The isosteric heat of adsorption (Q$_{st}$) of both Kr and Xe were calculated at 1 bar, the results of which are given in Figure \ref{qst}.
The Q$_{st}$ for Kr is calculated to stay almost constant throughout the considered systems (ranging from 31.7 to 27.2 kJ/mol), whereas for Xe, it starts off from 39.2 kJ/mol for Mo$_{40}$ and continuously goes down and plateaus at Mo$_{26}$V$_{14}$ and then sharply decreases to 20.0 kJ/mol for Mo$_{24}$V$_{16}$ (Figure \ref{qst} and SI Table S8). This calculated trend for Xe can be easily understood based on its larger kinetic diameter of $\approx$4.1~\AA~compared to that of $\approx$3.7~\AA~for Kr. 
As mentioned above, V doping leads to overall lower pore sizes as the ionic radius of the V atoms is smaller than that of Mo. As a result, the heptagonal pores continuously shrink as V content increases with the material ultimately being able to effectively exclude the larger Xe atoms. Our higher calculated Q$_{st}$ value for Xe than Kr in the case of Mo$_{30}$V$_{10}$ (32.2 vs. 30.2 kJ/mol, SI Table S8) agrees qualitatively with our experimental characterizations that applying heat is necessary in order the evacuate the adsorbed Xe from the micropores of this material (see Figure \ref{exp_isotherm}). Since V-doping leads to higher stabilities, as judged by the calculated formation energy trends (Figure \ref{lcd_form}), the V-rich vanadomolybdates are expected to be viable candidates for synthesis and characterization. 
\begin{figure*}[!h]
    \centering
    \includegraphics[width=0.99\linewidth]{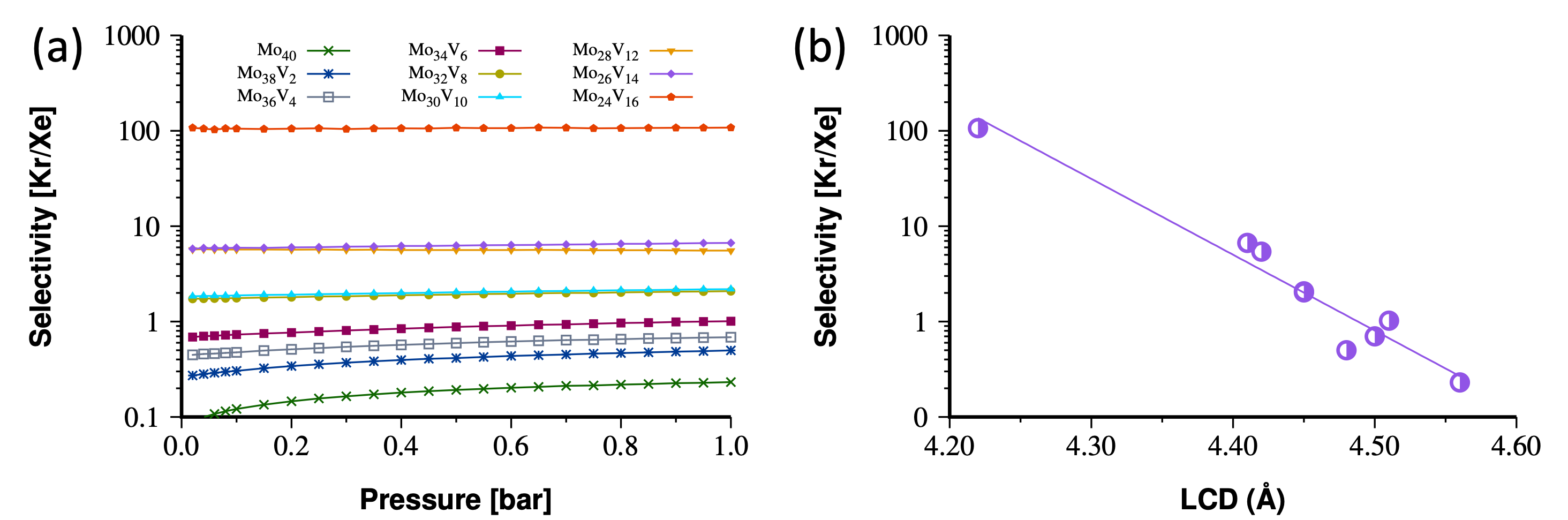}
    \caption{GCMC calculated (a) Kr/Xe selectivity for the 50:50 mixture at 298 K and (b) correlation (R$^2=$0.81) between Kr/Xe selectivity (logarithmic scale) for the 50:50 mixture and LCD. A line is added to guide the eye. The corresponding data for the 80:20 mixture is given in the SI Figure S21.}
    \label{selectivity}
\end{figure*}

As demonstrated in Figure \ref{selectivity}a, Kr selectivity is found to increase with an increase in V doping, with Mo$_{24}$V$_{16}$ being the highest Kr selective material with a calculated selectivity of 106.5 for an equimolar Kr/Xe mixture (SI Table S7). Moreover, Mo$_{26}$V$_{14}$ and Mo$_{28}$V$_{12}$ show around six times more Kr selectivity than Xe, with Mo$_{30}$V$_{10}$ and Mo$_{32}$V$_{8}$ showing a Kr/Xe selectivity of around two. The rest of the systems show more selectivity towards Xe, with the highest Xe selectivity in Mo$_{40}$.
By far, MOFs are the most studied materials for Kr and Xe adsorption separation. However, most reported MOFs are exclusively Xe selective\cite{easun2017structural, wang2018sensing, zhao2018metal}. For example, SBMOF-1 exhibits a high Xe adsorption of 13.2 mmol/kg and an Xe/Kr selectivity of 16.\cite{banerjee2016metal} Z11CBF-1000-2 is another MOF with a 20.6 mmol/kg uptake capacity for Xe and an Xe/Kr selectivity of 19.7.\cite{gong2018metal} JAVTAC and KAXQIL are another two examples with high Xe adsorption as well.\cite{simon2015best} As for Kr selective materials, FMOF-Cu was the first reported material with a high Kr/Xe selectivity of 36, at 203 K and 0.1 bar.\cite{fernandez2012switching} In addition, UTSA-280 and CECYOY are two theoretically predicted materials with Kr uptakes of 1.48 and 0.23 mmol/g, respectively, and Kr/Xe selectivity of 72.1 and 248.3, respectively.\cite{xiong2020new, lin2021multiscale} 
Overall, based on selectivity for Kr/Xe adsorption, we have found that the Mo$_{24}$V$_{16}$ (106.5),  Mo$_{26}$V$_{14}$ (6.7) and Mo$_{28}$V$_{12}$ (5.4) are among the top performing materials for Kr selective adsorptive separation.

Expectedly, the calculated Kr/Xe selectively was found to correlate well with the calculated LCDs (see Figure \ref{selectivity}b). As mentioned previously, doping with more V incrementally decreases the calculated LCDs, subsequently making the Xe adsorption more unfavorable compared to that of the smaller Kr. As discussed above, this is consistent with the molecular sieving mechanism for Kr/Xe adsorptive separation in these materials. Calculated radial distribution functions (RDFs) for the distances between the O atoms of different solid sorbents and Kr and Xe adsorbates are given in the SI Figure S24. The calculated Kr-O distances were found to be uniform and relatively constant, centered around $\approx$3.8~\AA, for all studied materials, whereas a rather large range of 3.6~\AA-4.2~\AA~ was found for Xe-O distances (SI Figure S24).

\begin{figure}[!h]
    \centering
    \includegraphics[width=0.99\linewidth]{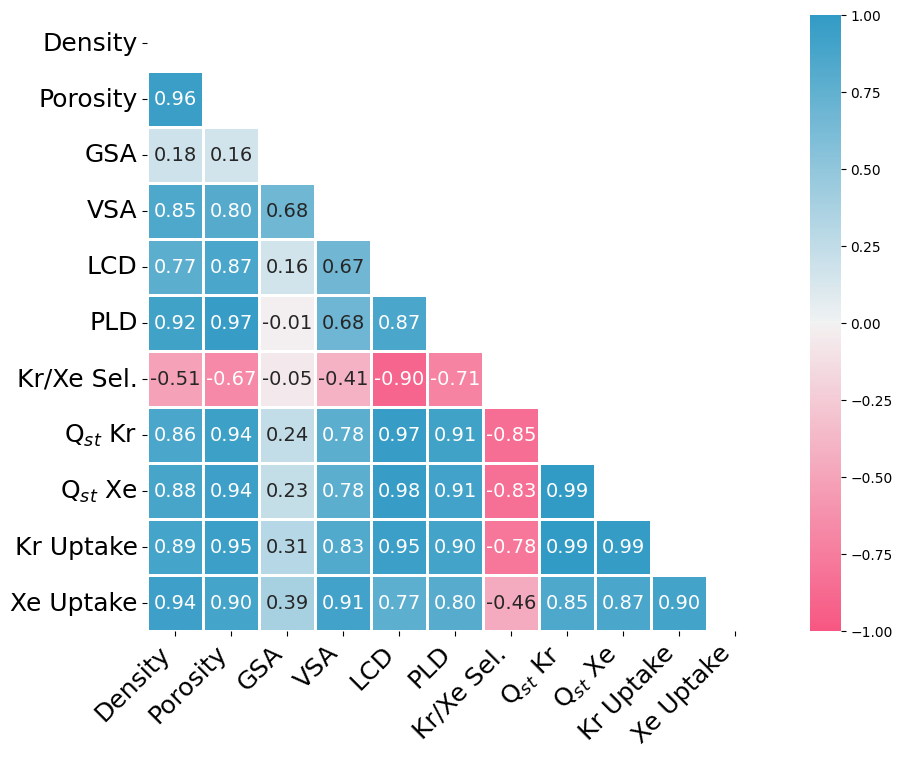}
    \caption{The Pearson correlation coefficient matrix for all systems studied in this work.}
    \label{corr}
\end{figure}

Figure \ref{corr} shows the calculated Pearson correlation matrix across different features of the studied materials.
As discussed, a good correlation exists between calculated selectivity from GCMC simulations and LCD from Zeo++ that can be used for high-throughput screening purposes, bypassing the need for expensive GCMC simulations. They could also be used as useful features in developing machine learning models, which are currently under investigation in our labs.

\subsection{4.3. Competition with CO$_2$ and N$_2$ }
\textbf{4.3. Competition with CO$_2$ and N$_2$ }\newline
CO$_{2}$  and N$_{2}$ molecules are abundant in air and can therefore compete with Kr uptake. Subsequently, pure component adsorption isotherms for the competing CO$_{2}$  and N$_{2}$ molecules were calculated at 298 K for all considered systems (see SI Figure S25). The CO$_2$ uptake was found not to follow the computed trend for LCDs and instead to be similar for all systems except for Mo$_{24}$V$_{16}$ and Mo$_{28}$V$_{12}$, which are about half compared to the rest. Instead for N$_2$, except Mo$_{30}$V$_{10}$, the adsorption was found to follow the trend for calculated LCDs with Mo$_{40}$ and Mo$_{24}$V$_{16}$ showing the highest and lowest uptakes, respectively. To rationalize these trends and outliers, we first note that in contrast to the studied Kr and Xe noble gasses with zero assigned charges, both CO$_2$ and N$_2$ carry charges, and therefore, electrostatic interactions might play a major role. To further investigate this, the calculated Bader charges of all systems were analyzed (see SI Table S5). The calculated average charges of the framework O atoms were found to be the highest for the Mo$_{30}$V$_{10}$ system. On the other hand, Mo$_{28}$V$_{12}$ and Mo$_{24}$V$_{16}$ were found to have the lowest charges for these O atoms. As such, compared to the rest of the studied systems, electrostatic interactions are anticipated to be stronger between N$_{2}$ and Mo$_{30}$V$_{10}$ but weaker for CO$_2$ interactions with Mo$_{28}$V$_{12}$ and Mo$_{24}$V$_{16}$. 

To study the competition between Kr uptake in the presence of the competing CO$_2$ and N$_2$ adsorbates, the selectivity of all studied systems for binary equimolar mixtures of Kr/CO$_2$ and Kr/N$_2$ was calculated. We note that as the main focus of this work is to develop materials with high Kr selectivity, we focus here only on Kr selectivity in the presence of these competing adsorbates. As shown in the SI Figure S25, except Mo$_{30}$V$_{10}$, all considered materials exhibit higher selectivity toward Kr adsorption than N$_2$ (see our rationale for Mo$_{30}$V$_{10}$ above). In agreement with their calculated CO$_{2}$ uptakes, the Mo$_{28}$V$_{12}$ and Mo$_{24}$V$_{16}$ systems were found to be the least CO$_{2}$ selective materials followed by Mo$_{30}$V$_{10}$ compared to the rest of the considered systems. Overall, none of the studied systems was found to be more selective towards Kr than CO$_2$. This implies that CO$_2$ could potentially compete with Kr adsorption and should, therefore, be removed from the mixture before these materials can be used for separating Kr.

As mentioned before, the studied orthorhombic vanadomolybdates contain four heptagonal and four hexagonal pores in their unit cells. Given the kinetic diameters of Kr and Xe and the calculated PLDs (ranging from 3.05~\AA~to 3.43~\AA) and LCDs (ranging from 4.22~\AA~to 4.56~\AA) for the hexagonal and heptagonal pores, respectively, the adsorption was found to only occur in the larger heptagonal channels. To further confirm this, Kr/Xe uptakes were calculated after blocking the heptagonal pores, which led to zero uptakes not only for Kr and Xe but also for the competing CO$_2$ and N$_2$ adsorbates. On the other hand, when the hexagonal pores were blocked, adsorption isotherms similar to those of the unblocked simulations were obtained (see the calculated Kr and Xe density distribution plots in SI Figure S26).

%%%%%%%%%%%%%%%%%%%%%%%%%%%%%%%%%%%%%%%%%%%%%
% SECTION 1: CONCLUSIONS
%%%%%%%%%%%%%%%%%%%%%%%%%%%%%%%%%%%%%%%%%%%%%

\section{5. Conclusions}
\textbf{5. Conclusion}\newline
Extensive electronic structure calculations and molecular simulations combined with targeted synthesis and characterizations are used to explore vanadium doping as an effective tool for fine-tuning the microporosity of an underexplored family of mixed metal oxide vanadomolybdates for selective Kr uptake from Kr/Xe mixtures in the presence of other competing CO$_2$ and N$_2$ adsorbates. Unprecedented Kr selectivities as high as $>$100 were found by effective size exclusion of the larger Xe adsorbates in V-rich materials. Useful correlations were found between Kr/Xe uptakes and selectivity and LCD, porosity, and isosteric heat of adsorption. These correlations are very useful as excellent features for training machine learning models or fast high-throughput screening of these materials. 
One possibility to increase Kr uptake is to explore other phases of vanadomolybdates, such as the trigonal phase, which has already been synthesized and used for catalytic oxidation purposes. This trigonal phase contains a higher heptagonal/hexagonal ratio of 3/2, compared to that of 1/1 for the studied orthorhombic vanadomolybdates in this work. This should lead to higher Kr uptakes in the former, considering similar levels of V-doping. We hope that this study is helpful in providing useful insights into a new class of chemically robust all-inorganic materials for selective Kr adsorptive separation under extreme operating conditions.

\begin{suppinfo}
Benchmark of our electronic structure calculations, force field parameters used in RASPA simulations, convergence tests using different number of cycles for GCMC simulations, and simulated uptakes and selectivities for CO$_2$ and N$_2$.
\end{suppinfo}

\begin{acknowledgement}
This work is partially supported by the U.S. Department of Energy Office of Science, Basic Energy Sciences Program (Grant No. DE-SC0024512). PKT thanks DOE Nuclear Energy – Materials Research and Waste Form Development Campaign. Particularly, PKT thanks Dr. Ken Marsden (INL), Mrs. Amy Welty (INL), and Mrs. Kimberly Gray (DOE Nuclear Energy). ZP and MRM thank UMKC Tier 1 Funding For Excellence (FFE) grant for support. We thank Dr. James B. Murowchick of UMKC for his assistance in powder X-ray measurements. Simulations presented in this work used resources from Bridges-2 at Pittsburgh Supercomputing Center through allocation PHY230099 from the Advanced Cyberinfrastructure Coordination Ecosystem: Services \& Support (ACCESS) program,\cite{access} which is supported by National Science Foundation grants \#2138259, \#2138286, \#2138307, \#2137603, and \#2138296. Technical support and computing resources provided by the HPC center at UMKC are also gratefully acknowledged.
\end{acknowledgement}

\Addlcwords{is with of in the an a iv v as for on and by to at}
\bibliography{bib}

\end{document}

\newpage
\tableofcontents
\newpage

\clearpage
\begin{figure}[!h]
\addcontentsline{toc}{figure}{Figure \ref{xrd}. Comparison of XRD diffractograms of the orthorhombic Mo$_{30}$V$_{10}$O$_{112}$ system with the literature}
    \centering
    \includegraphics[width=0.99\linewidth]{../figs/xrd}
    \caption{Comparison of XRD diffractograms of the orthorhombic Mo$_{30}$V$_{10}$O$_{112}$ system reported in the literature (up)\cite{acie_47_2493,konya2013orthorhombic} and this work (bottom).}
    \label{xrd}
\end{figure}

\clearpage
\begin{table}
\addcontentsline{toc}{table}{Table \ref{mo32isomer}. Calculated relative energies (in kJ/mol) of all Mo$_{32}$V$_{8}$O$_{112}$ isomers}
\centering
  \resizebox{0.99\textwidth}{!}{
\begin{tabular}{c c c c c c} \hline     
Sl & Isomer & Relative energy (kJ/mol) & Sl & Isomer & Relative energy (kJ/mol) \\ \hline
1 & S1abS3abS3cdS7ad & 0.00 & 33 & S1abS2abS3abS7bd & -93.12 \\
2 & S1abS2abS3adS3bc & -16.63 & 34 & S1abS2abS3cdS7bd & -93.24 \\
3 & S1abS3abS3cdS7ac & -23.79 & 35 & S1abS2abS3abS7bc & -98.08 \\
4 & S1abS3abS3cdS7bd & -23.80 & 36 & S1abS2abS3cdS7bc & -98.29 \\
5 & S1abS3abS3cdS7ab & -37.92 & 37 & S2abS3abS3cdS7cd & -103.57 \\
6 & S1abS3abS3cdS7cd & -37.93 & 38 & S2abS3abS3cdS7ab & -103.70 \\
7 & S1abS3abS3cdS7bc & -46.13 & 39 & S1abS2abS3abS7ab & -103.97 \\
8 & S2abS3abS3cdS7bd & -50.68 & 40 & S1abS2abS3cdS7cd & -104.00 \\
9 & S2abS3abS3cdS7ac & -51.21 & 41 & S1abS2abS3bcS7ab & -106.91 \\
10 & S2abS3abS3cdS7bc & -55.10 & 42 & S1abS2abS3bcS7cd & -106.92 \\
11 & S2abS3abS3cdS7ad & -55.42 & 43 & S1abS2abS3acS7ac & -107.75 \\
12 & S3abS3cdS7abS7cd & -61.03 & 44 & S1abS2abS3bdS7bd & -107.90 \\
13 & S1abS2abS3bcS7ad & -83.46 & 45 & S1abS2abS3bdS7ad & -112.75 \\
14 & S1abS2abS3adS7bd & -85.08 & 46 & S1abS2abS3acS7ad & -112.97 \\
15 & S1abS3adS7adS7bc & -85.10 & 47 & S1abS2abS3bdS7bc & -119.91 \\
16 & S1abS2abS3adS7ac & -85.35 & 48 & S1abS2abS3acS7bc & -119.96 \\
17 & S1abS2abS3adS7bc & -85.62 & 49 & S1abS2abS3adS7cd & -140.02 \\
18 & S1abS3bdS7abS7cd & -87.37 & 50 & S1abS2abS3adS7ab & -140.02 \\
19 & S1abS2abS3cdS7ac & -89.18 & 51 & S1abS2abS3bdS7ab & -141.38 \\
20 & S1abS2abS3bcS7bd & -89.43 & 52 & S1abS2abS3acS7cd & -141.42 \\
21 & S1abS2abS3bcS7ac & -89.50 & 53 & S1abS2abS3abS7cd & -144.56 \\
22 & S1abS2abS3bdS7ac & -89.91 & 54 & S1abS2abS3cdS7ab & -144.69 \\
23 & S1abS2abS3acS7bd & -90.07 & 55 & S1abS2abS7adS7bc & -145.07 \\
24 & S1abS2abS3cdS7ad & -90.12 & 56 & S1abS2abS3acS7ab & -152.37 \\
25 & S1abS2abS3abS7ad & -90.13 & 57 & S1abS2abS3bdS7cd & -152.53 \\
26 & S1abS3cdS7adS7bc & -91.59 & 58 & S2abS3adS7adS7bc & -197.90 \\
27 & S1abS3abS7abS7cd & -91.63 & 59 & S2abS3cdS7adS7bc & -206.54 \\
28 & S1abS2abS3bcS7bc & -92.68 & 60 & S2abS3abS7abS7cd & -206.61 \\
29 & S1abS3bcS7adS7bc & -92.86 & 61 & S2abS3bcS7adS7bc & -210.92 \\
30 & S1abS2abS3abS7ac & -93.02 & 62 & S2abS3acS7adS7bc & -212.97 \\
31 & S1abS2abS3adS7ad & -93.04 & 63 & S2abS3bdS7acS7bd & -213.02 \\
32 & S1abS3acS7adS7bc & -93.11 &  &  &  \\ \hline
\end{tabular}}
\caption{Calculated relative energies (in kJ/mol) of all Mo$_{32}$V$_{8}$O$_{112}$ isomers.}
\label{mo32isomer}
\end{table}

\clearpage
\begin{table}
\addcontentsline{toc}{table}{Table \ref{lj1}. Lennard-Jones parameters for POMFs and noble gases}
\centering
\resizebox{0.3\linewidth}{!}{
\begin{tabular}{ccc} \hline 
Atom type & $\sigma$ (\AA) & $\epsilon$ (K) \\ \hline
V & 2.80 & 8.1 \\
Mo & 2.72 & 28.2 \\
O & 3.65 & 113.3 \\
Kr & 3.64 & 166.4 \\
Xe & 4.10 & 221.0 \\ \hline
\end{tabular}}
\caption{Lennard-Jones parameters for the framework atoms and noble gases.}
\label{lj1}
\end{table}

\begin{table}
\addcontentsline{toc}{table}{Table \ref{lj2}. Fitted $\epsilon_{Kr-O}$ and $\epsilon_{Xe-O}$ terms to the experiment}
\centering
\resizebox{0.2\linewidth}{!}{
\begin{tabular}{cc} \hline 
Interactions & $\epsilon$ (K) \\ \hline
Kr-O & 110.0 \\
Xe-O & 190.0 \\ \hline
\end{tabular}}
\caption{Fitted $\epsilon_{Kr-O}$ and $\epsilon_{Xe-O}$ terms to the experiment.}
\label{lj2}
\end{table}

\begin{table}
\addcontentsline{toc}{table}{Table \ref{lj3}. Force field parameters for CO$_2$ and N$_2$}
\centering
\resizebox{0.4\linewidth}{!}{
\begin{tabular}{cccc} \hline 
Atom type & $\sigma$ (\AA) & $\epsilon$ (K) & Charge\\ \hline
C\_co2 & 2.80 & 27.0 & 0.700 \\
O\_co2 & 3.05 & 79.0 & -0.350 \\
N\_n2 & 3.31 & 36.0 & -0.482 \\
N\_com & - & - & 0.964 \\ \hline
\end{tabular}}
\caption{Force field parameters for CO$_2$ and N$_2$.}
\label{lj3}
\end{table}

\begin{table}
\addcontentsline{toc}{table}{Table \ref{bader}. Calculated Bader charges for all systems}
\centering
\resizebox{0.98\linewidth}{!}{
\begin{tabular}{cccccccccccccccccc} \hline
\multicolumn{2}{c} {Mo$_{40}$} & \multicolumn{2}{c} {Mo$_{38}$V$_2$} & \multicolumn{2}{c} {Mo$_{36}$V$_4$} & \multicolumn{2}{c} {Mo$_{34}$V$_6$} & \multicolumn{2}{c} {Mo$_{32}$V$_8$} & \multicolumn{2}{c} {Mo$_{30}$V$_{10}$} & \multicolumn{2}{c} {Mo$_{28}$V$_{12}$} & \multicolumn{2}{c} {Mo$_{26}$V$_{14}$} & \multicolumn{2}{c} {Mo$_{24}$V$_{16}$} \\
Atom & Charge & Atom & Charge & Atom & Charge & Atom & Charge & Atom & Charge & Atom & Charge & Atom & Charge & Atom & Charge & Atom & Charge \\ \hline
Mo1 & 2.3995 & Mo1 & 2.4092 & Mo1 & 2.4324 & Mo1 & 2.4175 & Mo1 & 2.4177 & Mo1 & 2.4370 & Mo1 & 2.4434 & Mo1 & 2.4568 & Mo1 & 2.5008 \\
Mo2 & 2.4026 & Mo2 & 2.4402 & Mo2 & 2.4070 & Mo2 & 2.4495 & Mo2 & 2.4131 & Mo2 & 2.4367 & Mo2 & 2.4580 & Mo2 & 2.4491 & Mo2 & 2.3436 \\
Mo3 & 2.4143 & Mo3 & 2.4455 & Mo3 & 2.4370 & Mo3 & 2.4265 & Mo3 & 2.4312 & Mo3 & 2.4543 & Mo3 & 2.3998 & Mo3 & 2.4556 & Mo3 & 2.4115 \\
Mo4 & 2.4076 & Mo4 & 2.4232 & Mo4 & 2.4481 & Mo4 & 2.4130 & Mo4 & 2.4479 & Mo4 & 2.4686 & Mo4 & 2.4691 & Mo4 & 2.4740 & Mo4 & 2.4927 \\
Mo5 & 2.4518 & Mo5 & 2.4018 & Mo5 & 2.4261 & Mo5 & 2.4533 & Mo5 & 2.4697 & Mo5 & 2.4427 & Mo5 & 2.4579 & Mo5 & 2.4814 & Mo5 & 2.4600 \\
Mo6 & 2.4759 & Mo6 & 2.3961 & Mo6 & 2.4679 & Mo6 & 2.4813 & Mo6 & 2.4973 & Mo6 & 2.4991 & Mo6 & 2.4838 & Mo6 & 2.5096 & Mo6 & 2.4719 \\
Mo7 & 2.2926 & Mo7 & 2.2987 & Mo7 & 2.3321 & Mo7 & 2.4627 & Mo7 & 2.4541 & Mo7 & 2.4772 & Mo7 & 2.4432 & Mo7 & 2.4691 & Mo7 & 2.4605 \\
Mo8 & 2.4161 & Mo8 & 2.4108 & Mo8 & 2.4325 & Mo8 & 2.4396 & Mo8 & 2.4559 & Mo8 & 2.4495 & Mo8 & 2.4360 & Mo8 & 2.4764 & Mo8 & 2.2722 \\
Mo9 & 2.4322 & Mo9 & 2.4422 & Mo9 & 2.4440 & Mo9 & 2.4503 & Mo9 & 2.4307 & Mo9 & 2.4388 & Mo9 & 2.4279 & Mo9 & 2.4589 & Mo9 & 2.5343 \\
Mo10 & 2.4065 & Mo10 & 2.4334 & Mo10 & 2.4348 & Mo10 & 2.4332 & Mo10 & 2.4276 & Mo10 & 2.4473 & Mo10 & 2.4514 & Mo10 & 2.4530 & Mo10 & 2.4416 \\
Mo11 & 2.4457 & Mo11 & 2.4438 & Mo11 & 2.4523 & Mo11 & 2.4622 & Mo11 & 2.4686 & Mo11 & 2.4680 & Mo11 & 2.4487 & Mo11 & 2.4580 & Mo11 & 2.4996 \\
Mo12 & 2.4701 & Mo12 & 2.4679 & Mo12 & 2.4721 & Mo12 & 2.4616 & Mo12 & 2.4742 & Mo12 & 2.4780 & Mo12 & 2.4619 & Mo12 & 2.4614 & Mo12 & 2.4379 \\
Mo13 & 2.4017 & Mo13 & 2.4302 & Mo13 & 2.4868 & Mo13 & 2.4842 & Mo13 & 2.5009 & Mo13 & 2.4955 & Mo13 & 2.3748 & Mo13 & 2.4694 & Mo13 & 2.4947 \\
Mo14 & 2.4579 & Mo14 & 2.4525 & Mo14 & 2.4632 & Mo14 & 2.4671 & Mo14 & 2.4765 & Mo14 & 2.4697 & Mo14 & 2.4464 & Mo14 & 2.4550 & Mo14 & 2.2675 \\
Mo15 & 2.4162 & Mo15 & 2.4105 & Mo15 & 2.4256 & Mo15 & 2.4502 & Mo15 & 2.4560 & Mo15 & 2.4494 & Mo15 & 2.4489 & Mo15 & 2.4645 & Mo15 & 2.5537 \\
Mo16 & 2.4321 & Mo16 & 2.4422 & Mo16 & 2.4240 & Mo16 & 2.4231 & Mo16 & 2.4310 & Mo16 & 2.4392 & Mo16 & 2.4314 & Mo16 & 2.4523 & Mo16 & 2.3885 \\
Mo17 & 2.4065 & Mo17 & 2.4332 & Mo17 & 2.4487 & Mo17 & 2.4252 & Mo17 & 2.4285 & Mo17 & 2.4437 & Mo17 & 2.4645 & Mo17 & 2.4462 & Mo17 & 2.4856 \\
Mo18 & 2.4457 & Mo18 & 2.4439 & Mo18 & 2.4559 & Mo18 & 2.4587 & Mo18 & 2.4684 & Mo18 & 2.4678 & Mo18 & 2.4067 & Mo18 & 2.4638 & Mo18 & 2.4241 \\
Mo19 & 2.4702 & Mo19 & 2.4684 & Mo19 & 2.4810 & Mo19 & 2.4242 & Mo19 & 2.4733 & Mo19 & 2.4782 & Mo19 & 2.4564 & Mo19 & 2.4593 & Mo19 & 2.4525 \\
Mo20 & 2.4018 & Mo20 & 2.4302 & Mo20 & 2.4841 & Mo20 & 2.4633 & Mo20 & 2.5009 & Mo20 & 2.4961 & Mo20 & 2.3690 & Mo20 & 2.5110 & Mo20 & 2.3456 \\
Mo21 & 2.4578 & Mo21 & 2.4527 & Mo21 & 2.4521 & Mo21 & 2.4582 & Mo21 & 2.4766 & Mo21 & 2.4700 & Mo21 & 2.4539 & Mo21 & 2.4664 & Mo21 & 2.3909 \\
Mo22 & 2.3996 & Mo22 & 2.4091 & Mo22 & 2.4044 & Mo22 & 2.4096 & Mo22 & 2.4174 & Mo22 & 2.4379 & Mo22 & 2.4078 & Mo22 & 2.4417 & Mo22 & 2.5317 \\
Mo23 & 2.4026 & Mo23 & 2.4402 & Mo23 & 2.4062 & Mo23 & 2.4027 & Mo23 & 2.4129 & Mo23 & 2.4356 & Mo23 & 2.4586 & Mo23 & 2.4471 & Mo23 & 2.4673 \\
Mo24 & 2.4142 & Mo24 & 2.4454 & Mo24 & 2.4375 & Mo24 & 2.4257 & Mo24 & 2.4333 & Mo24 & 2.4550 & Mo24 & 2.3783 & Mo24 & 2.4748 & Mo24 & 2.5180 \\
Mo25 & 2.4076 & Mo25 & 2.4236 & Mo25 & 2.4469 & Mo25 & 2.4411 & Mo25 & 2.4478 & Mo25 & 2.4682 & Mo25 & 2.5002 & Mo25 & 2.4565 & V1 & 2.0051 \\
Mo26 & 2.4517 & Mo26 & 2.4017 & Mo26 & 2.4136 & Mo26 & 2.3983 & Mo26 & 2.4691 & Mo26 & 2.4429 & Mo26 & 2.4535 & Mo26 & 2.5163 & V2 & 1.9890 \\
Mo27 & 2.4758 & Mo27 & 2.3963 & Mo27 & 2.4514 & Mo27 & 2.4840 & Mo27 & 2.4973 & Mo27 & 2.4982 & Mo27 & 2.5034 & V1 & 2.0121 & V3 & 2.0022 \\
Mo28 & 2.2927 & Mo28 & 2.2988 & Mo28 & 2.3309 & Mo28 & 2.3349 & Mo28 & 2.4542 & Mo28 & 2.4768 & Mo28 & 2.4428 & V2 & 2.0232 & V4 & 1.9903 \\
Mo29 & 2.4629 & Mo29 & 2.4493 & Mo29 & 2.3022 & Mo29 & 2.4585 & Mo29 & 2.4692 & Mo29 & 2.4861 & V1 & 2.0076 & V3 & 2.0011 & V5 & 2.0294 \\
Mo30 & 2.4628 & Mo30 & 2.4494 & Mo30 & 2.3957 & Mo30 & 2.4422 & Mo30 & 2.4695 & Mo30 & 2.4870 & V2 & 1.9865 & V4 & 1.9860 & V6 & 1.9990 \\
Mo31 & 2.3824 & Mo31 & 2.2982 & Mo31 & 2.3842 & Mo31 & 2.3247 & Mo31 & 2.4927 & V1 & 2.9086 & V3 & 1.9487 & V5 & 2.0503 & V7 & 2.0516 \\
Mo32 & 2.3824 & Mo32 & 2.2976 & Mo32 & 2.3719 & Mo32 & 2.3896 & Mo32 & 2.4926 & V2 & 2.9086 & V4 & 1.9379 & V6 & 2.0472 & V8 & 2.0155 \\
Mo33 & 2.2860 & Mo33 & 2.4394 & Mo33 & 2.4060 & Mo33 & 2.3132 & V1 & 1.9493 & V3 & 2.8973 & V5 & 1.8979 & V7 & 1.9274 & V9 & 1.9447 \\
Mo34 & 2.2860 & Mo34 & 2.4396 & Mo34 & 2.3389 & Mo34 & 2.3938 & V2 & 1.9491 & V4 & 1.8981 & V6 & 1.8937 & V8 & 1.9550 & V10 & 1.9452 \\
Mo35 & 2.3610 & Mo35 & 2.3747 & Mo35 & 2.4355 & V1 & 1.9598 & V3 & 1.9660 & V5 & 1.9122 & V7 & 2.0179 & V9 & 2.0055 & V11 & 2.0404 \\
Mo36 & 2.3610 & Mo36 & 2.3747 & Mo36 & 2.4696 & V2 & 1.9446 & V4 & 1.9774 & V6 & 1.9120 & V8 & 2.0075 & V10 & 2.0062 & V12 & 1.9964 \\
Mo37 & 2.3848 & Mo37 & 2.3625 & V1 & 1.9090 & V3 & 1.9413 & V5 & 1.9774 & V7 & 2.0443 & V9 & 1.9442 & V11 & 2.0543 & V13 & 2.0414 \\
Mo38 & 2.3849 & Mo38 & 2.3623 & V2 & 1.9204 & V4 & 1.9645 & V6 & 1.9663 & V8 & 2.0433 & V10 & 1.9594 & V12 & 2.0299 & V14 & 2.0437 \\
Mo39 & 2.3779 & V1 & 1.8934 & V3 & 1.8958 & V5 & 1.9819 & V7 & 1.9633 & V9 & 2.0272 & V11 & 2.0257 & V13 & 2.0207 & V15 & 2.0462 \\
Mo40 & 2.3781 & V2 & 1.8937 & V4 & 1.8939 & V6 & 2.0264 & V8 & 1.9633 & V10 & 2.0288 & V12 & 2.0413 & V14 & 2.0091 & V16 & 2.0500 \\
O1 & -0.8406 & O1 & -0.7337 & O1 & -0.7796 & O1 & -0.7943 & O1 & -0.8042 & O1 & -1.7976 & O1 & -0.7615 & O1 & -0.8130 & O1 & -0.7901 \\
O2 & -0.8238 & O2 & -0.8223 & O2 & -0.7797 & O2 & -0.7254 & O2 & -0.6975 & O2 & -1.6428 & O2 & -0.6364 & O2 & -0.6697 & O2 & -0.6673 \\
O3 & -0.7560 & O3 & -0.7696 & O3 & -0.7806 & O3 & -0.7766 & O3 & -0.7013 & O3 & -1.6777 & O3 & -0.6525 & O3 & -0.7163 & O3 & -0.6677 \\
O4 & -0.9727 & O4 & -0.9708 & O4 & -0.9738 & O4 & -0.9687 & O4 & -0.9747 & O4 & -0.9737 & O4 & -1.0029 & O4 & -0.9662 & O4 & -0.9795 \\
O5 & -0.9690 & O5 & -0.9656 & O5 & -0.9677 & O5 & -0.9653 & O5 & -0.9596 & O5 & -0.9668 & O5 & -0.9599 & O5 & -0.9644 & O5 & -0.9444 \\
O6 & -0.8700 & O6 & -0.8865 & O6 & -0.8864 & O6 & -0.8914 & O6 & -0.8945 & O6 & -0.8947 & O6 & -0.8854 & O6 & -0.8616 & O6 & -0.6907 \\
O7 & -0.8659 & O7 & -0.8727 & O7 & -0.8745 & O7 & -0.8735 & O7 & -0.8794 & O7 & -0.8821 & O7 & -0.8870 & O7 & -0.8574 & O7 & -0.8130 \\
O8 & -0.8571 & O8 & -0.8602 & O8 & -0.8553 & O8 & -0.8639 & O8 & -0.8393 & O8 & -0.8329 & O8 & -0.8352 & O8 & -0.8103 & O8 & -0.8092 \\
O9 & -0.9703 & O9 & -0.9658 & O9 & -0.9665 & O9 & -0.9696 & O9 & -0.9647 & O9 & -0.9652 & O9 & -1.0052 & O9 & -0.9591 & O9 & -1.0199 \\
O10 & -0.8634 & O10 & -0.8301 & O10 & -0.8324 & O10 & -0.8240 & O10 & -0.8116 & O10 & -0.8169 & O10 & -0.8052 & O10 & -0.8312 & O10 & -0.7211 \\
O11 & -0.8684 & O11 & -0.8620 & O11 & -0.8224 & O11 & -0.8207 & O11 & -0.8237 & O11 & -0.8156 & O11 & -0.8370 & O11 & -0.8371 & O11 & -0.8374 \\
O12 & -0.9704 & O12 & -0.9667 & O12 & -0.9619 & O12 & -0.9657 & O12 & -0.9619 & O12 & -0.9616 & O12 & -0.9608 & O12 & -0.9609 & O12 & -0.9489 \\
O13 & -0.8653 & O13 & -0.8782 & O13 & -0.8801 & O13 & -0.8701 & O13 & -0.8410 & O13 & -0.8520 & O13 & -0.8419 & O13 & -0.8528 & O13 & -0.8270 \\
O14 & -0.8483 & O14 & -0.8534 & O14 & -0.8579 & O14 & -0.8745 & O14 & -0.8488 & O14 & -0.8725 & O14 & -0.8399 & O14 & -0.7946 & O14 & -0.7991 \\
O15 & -0.7331 & O15 & -0.7460 & O15 & -0.7422 & O15 & -0.7717 & O15 & -0.7714 & O15 & -0.7657 & O15 & -0.7473 & O15 & -0.8029 & O15 & -0.7216 \\
O16 & -0.9711 & O16 & -0.9772 & O16 & -0.9743 & O16 & -0.9710 & O16 & -0.9655 & O16 & -0.9648 & O16 & -0.9585 & O16 & -0.9682 & O16 & -0.9185 \\
O17 & -0.8360 & O17 & -0.8189 & O17 & -0.7735 & O17 & -0.8088 & O17 & -0.7729 & O17 & -0.7726 & O17 & -0.7462 & O17 & -0.8039 & O17 & -0.7571 \\
O18 & -0.8140 & O18 & -0.8075 & O18 & -0.7974 & O18 & -0.8096 & O18 & -0.8109 & O18 & -0.7945 & O18 & -0.7653 & O18 & -0.7890 & O18 & -0.8413 \\
O19 & -0.8160 & O19 & -0.7761 & O19 & -0.7854 & O19 & -0.8113 & O19 & -0.8184 & O19 & -0.7929 & O19 & -0.8178 & O19 & -0.8244 & O19 & -0.8651 \\
O20 & -0.7780 & O20 & -0.7821 & O20 & -0.8200 & O20 & -0.7714 & O20 & -0.8177 & O20 & -0.8118 & O20 & -0.7753 & O20 & -0.8208 & O20 & -0.8198 \\
O21 & -0.8161 & O21 & -0.8123 & O21 & -0.7987 & O21 & -0.8004 & O21 & -0.8042 & O21 & -0.8087 & O21 & -0.7638 & O21 & -0.8135 & O21 & -0.8057 \\
O22 & -0.8610 & O22 & -0.8283 & O22 & -0.8260 & O22 & -0.8237 & O22 & -0.8326 & O22 & -0.8201 & O22 & -0.8555 & O22 & -0.8218 & O22 & -0.8504 \\
O23 & -0.8861 & O23 & -0.8801 & O23 & -0.8902 & O23 & -0.8866 & O23 & -0.8108 & O23 & -0.8253 & O23 & -0.8196 & O23 & -0.7495 & O23 & -0.8062 \\
O24 & -0.8591 & O24 & -0.8748 & O24 & -0.8716 & O24 & -0.8690 & O24 & -0.8399 & O24 & -0.8494 & O24 & -0.8302 & O24 & -0.7749 & O24 & -0.8348 \\
O25 & -0.8713 & O25 & -0.8566 & O25 & -0.8345 & O25 & -0.8322 & O25 & -0.8355 & O25 & -0.8406 & O25 & -0.8532 & O25 & -0.8505 & O25 & -0.8267 \\
O26 & -0.8614 & O26 & -0.8581 & O26 & -0.8662 & O26 & -0.8637 & O26 & -0.8348 & O26 & -0.8366 & O26 & -0.8392 & O26 & -0.8289 & O26 & -0.8406 \\
O27 & -0.8759 & O27 & -0.8675 & O27 & -0.8800 & O27 & -0.8093 & O27 & -0.8153 & O27 & -0.7948 & O27 & -0.8191 & O27 & -0.8118 & O27 & -0.7700 \\
O28 & -0.7576 & O28 & -0.7580 & O28 & -0.7904 & O28 & -0.7867 & O28 & -0.8009 & O28 & -0.7958 & O28 & -0.6455 & O28 & -0.7742 & O28 & -0.7300 \\
O29 & -0.8065 & O29 & -0.7519 & O29 & -0.7179 & O29 & -0.7308 & O29 & -0.6955 & O29 & -0.6820 & O29 & -0.6308 & O29 & -0.6596 & O29 & -0.6508 \\
O30 & -0.7782 & O30 & -0.7870 & O30 & -0.7301 & O30 & -0.7798 & O30 & -0.7949 & O30 & -0.6768 & O30 & -0.6467 & O30 & -0.6831 & O30 & -0.6943 \\
O31 & -0.9725 & O31 & -0.9711 & O31 & -0.9718 & O31 & -0.9714 & O31 & -0.9706 & O31 & -0.9670 & O31 & -0.9929 & O31 & -0.9647 & O31 & -0.9569 \\
O32 & -0.9709 & O32 & -0.9651 & O32 & -0.9658 & O32 & -0.9639 & O32 & -0.9628 & O32 & -0.9678 & O32 & -0.9635 & O32 & -0.9584 & O32 & -0.9353 \\
O33 & -0.8522 & O33 & -0.8683 & O33 & -0.8696 & O33 & -0.8759 & O33 & -0.8818 & O33 & -0.8855 & O33 & -0.8931 & O33 & -0.8526 & O33 & -0.6250 \\
O34 & -0.8713 & O34 & -0.8688 & O34 & -0.8854 & O34 & -0.8764 & O34 & -0.8796 & O34 & -0.8540 & O34 & -0.8651 & O34 & -0.8544 & O34 & -0.8129 \\
O35 & -0.8644 & O35 & -0.8738 & O35 & -0.8643 & O35 & -0.8682 & O35 & -0.8807 & O35 & -0.8490 & O35 & -0.8562 & O35 & -0.8127 & O35 & -0.8236 \\
O36 & -0.9730 & O36 & -0.9657 & O36 & -0.9691 & O36 & -0.9727 & O36 & -0.9706 & O36 & -0.9674 & O36 & -0.9508 & O36 & -0.9662 & O36 & -0.9011 \\ \hline
\end{tabular}}
\caption{Calculated Bader charges for all systems.}
\label{bader}
\end{table}

\begin{table}
%\addcontentsline{toc}{table}{Table \ref{bader}. Calculated bader charges for all systems}
\centering
\resizebox{0.98\linewidth}{!}{
\begin{tabular}{cccccccccccccccccc} \hline
\multicolumn{2}{c} {Mo$_{40}$} & \multicolumn{2}{c} {Mo$_{38}$V$_2$} & \multicolumn{2}{c} {Mo$_{36}$V$_4$} & \multicolumn{2}{c} {Mo$_{34}$V$_6$} & \multicolumn{2}{c} {Mo$_{32}$V$_8$} & \multicolumn{2}{c} {Mo$_{30}$V$_{10}$} & \multicolumn{2}{c} {Mo$_{28}$V$_{12}$} & \multicolumn{2}{c} {Mo$_{26}$V$_{14}$} & \multicolumn{2}{c} {Mo$_{24}$V$_{16}$} \\
Atom & Charge & Atom & Charge & Atom & Charge & Atom & Charge & Atom & Charge & Atom & Charge & Atom & Charge & Atom & Charge & Atom & Charge \\ \hline
O37 & -0.8720 & O37 & -0.8372 & O37 & -0.8423 & O37 & -0.8412 & O37 & -0.8336 & O37 & -0.8333 & O37 & -0.8510 & O37 & -0.8391 & O37 & -0.7673 \\
O38 & -0.8589 & O38 & -0.8590 & O38 & -0.8094 & O38 & -0.8262 & O38 & -0.8152 & O38 & -0.8397 & O38 & -0.8181 & O38 & -0.8381 & O38 & -0.8296 \\
O39 & -0.9718 & O39 & -0.9730 & O39 & -0.9671 & O39 & -0.9684 & O39 & -0.9665 & O39 & -0.9656 & O39 & -0.9698 & O39 & -0.9608 & O39 & -0.9831 \\
O40 & -0.8709 & O40 & -0.8803 & O40 & -0.8822 & O40 & -0.8759 & O40 & -0.8381 & O40 & -0.8406 & O40 & -0.8760 & O40 & -0.8528 & O40 & -0.8938 \\
O41 & -0.8027 & O41 & -0.8236 & O41 & -0.8109 & O41 & -0.8332 & O41 & -0.8185 & O41 & -0.8415 & O41 & -0.7953 & O41 & -0.7908 & O41 & -0.7290 \\
O42 & -0.7815 & O42 & -0.7686 & O42 & -0.7654 & O42 & -0.7711 & O42 & -0.7665 & O42 & -0.7696 & O42 & -0.7301 & O42 & -0.7955 & O42 & -0.7261 \\
O43 & -0.9697 & O43 & -0.9705 & O43 & -0.9746 & O43 & -0.9720 & O43 & -0.9667 & O43 & -0.9681 & O43 & -0.9998 & O43 & -0.9590 & O43 & -0.9976 \\
O44 & -0.8062 & O44 & -0.8077 & O44 & -0.8025 & O44 & -0.8073 & O44 & -0.7893 & O44 & -0.7813 & O44 & -0.8017 & O44 & -0.8254 & O44 & -0.8428 \\
O45 & -0.8105 & O45 & -0.8043 & O45 & -0.7992 & O45 & -0.8096 & O45 & -0.8136 & O45 & -0.8115 & O45 & -0.7399 & O45 & -0.7905 & O45 & -0.7912 \\
O46 & -0.8222 & O46 & -0.8142 & O46 & -0.8104 & O46 & -0.8117 & O46 & -0.8179 & O46 & -0.8101 & O46 & -0.7495 & O46 & -0.8044 & O46 & -0.7646 \\
O47 & -0.8193 & O47 & -0.8194 & O47 & -0.8157 & O47 & -0.8237 & O47 & -0.8204 & O47 & -0.8125 & O47 & -0.7818 & O47 & -0.8106 & O47 & -0.8157 \\
O48 & -0.8145 & O48 & -0.7945 & O48 & -0.8037 & O48 & -0.8083 & O48 & -0.8092 & O48 & -0.7839 & O48 & -0.7683 & O48 & -0.8196 & O48 & -0.8293 \\
O49 & -0.8579 & O49 & -0.8347 & O49 & -0.8287 & O49 & -0.8277 & O49 & -0.8319 & O49 & -0.8509 & O49 & -0.8511 & O49 & -0.8186 & O49 & -0.8167 \\
O50 & -0.8805 & O50 & -0.8716 & O50 & -0.8835 & O50 & -0.8811 & O50 & -0.8305 & O50 & -0.8247 & O50 & -0.8181 & O50 & -0.7668 & O50 & -0.7928 \\
O51 & -0.8582 & O51 & -0.8624 & O51 & -0.8750 & O51 & -0.8692 & O51 & -0.8763 & O51 & -0.8000 & O51 & -0.8190 & O51 & -0.8118 & O51 & -0.7869 \\
O52 & -0.8662 & O52 & -0.8640 & O52 & -0.8209 & O52 & -0.8267 & O52 & -0.8261 & O52 & -0.8547 & O52 & -0.8675 & O52 & -0.8435 & O52 & -0.8609 \\
O53 & -0.8595 & O53 & -0.8603 & O53 & -0.8624 & O53 & -0.8650 & O53 & -0.8132 & O53 & -0.8176 & O53 & -0.8168 & O53 & -0.8292 & O53 & -0.8300 \\
O54 & -0.8843 & O54 & -0.8739 & O54 & -0.8823 & O54 & -0.8309 & O54 & -0.8767 & O54 & -0.8237 & O54 & -0.8336 & O54 & -0.8115 & O54 & -0.7781 \\
O55 & -0.7576 & O55 & -0.7580 & O55 & -0.7866 & O55 & -0.7571 & O55 & -0.8010 & O55 & -0.7957 & O55 & -0.6698 & O55 & -0.8143 & O55 & -0.7460 \\
O56 & -0.8065 & O56 & -0.7514 & O56 & -0.6904 & O56 & -0.6821 & O56 & -0.6953 & O56 & -0.6814 & O56 & -0.6241 & O56 & -0.6663 & O56 & -0.6454 \\
O57 & -0.7782 & O57 & -0.7872 & O57 & -0.7871 & O57 & -0.6741 & O57 & -0.7949 & O57 & -0.6761 & O57 & -0.6389 & O57 & -0.7077 & O57 & -0.7122 \\
O58 & -0.9725 & O58 & -0.9709 & O58 & -0.9710 & O58 & -0.9707 & O58 & -0.9706 & O58 & -0.9673 & O58 & -0.9557 & O58 & -0.9605 & O58 & -0.9178 \\
O59 & -0.9710 & O59 & -0.9651 & O59 & -0.9650 & O59 & -0.9659 & O59 & -0.9627 & O59 & -0.9683 & O59 & -0.9757 & O59 & -0.9576 & O59 & -0.9395 \\
O60 & -0.8522 & O60 & -0.8682 & O60 & -0.8820 & O60 & -0.8794 & O60 & -0.8818 & O60 & -0.8858 & O60 & -0.8919 & O60 & -0.8284 & O60 & -0.6735 \\
O61 & -0.8713 & O61 & -0.8688 & O61 & -0.8650 & O61 & -0.8706 & O61 & -0.8790 & O61 & -0.8540 & O61 & -0.8865 & O61 & -0.8180 & O61 & -0.8271 \\
O62 & -0.8644 & O62 & -0.8739 & O62 & -0.8813 & O62 & -0.8431 & O62 & -0.8804 & O62 & -0.8487 & O62 & -0.8230 & O62 & -0.8288 & O62 & -0.7636 \\
O63 & -0.9730 & O63 & -0.9658 & O63 & -0.9714 & O63 & -0.9704 & O63 & -0.9704 & O63 & -0.9669 & O63 & -0.9667 & O63 & -0.9636 & O63 & -0.9278 \\
O64 & -0.8720 & O64 & -0.8374 & O64 & -0.8266 & O64 & -0.8288 & O64 & -0.8336 & O64 & -0.8337 & O64 & -0.8455 & O64 & -0.8335 & O64 & -0.7639 \\
O65 & -0.8589 & O65 & -0.8590 & O65 & -0.8672 & O65 & -0.8637 & O65 & -0.8148 & O65 & -0.8397 & O65 & -0.8050 & O65 & -0.8335 & O65 & -0.8128 \\
O66 & -0.9718 & O66 & -0.9727 & O66 & -0.9727 & O66 & -0.9702 & O66 & -0.9664 & O66 & -0.9663 & O66 & -0.9559 & O66 & -0.9604 & O66 & -0.9474 \\
O67 & -0.8709 & O67 & -0.8802 & O67 & -0.8485 & O67 & -0.8447 & O67 & -0.8391 & O67 & -0.8404 & O67 & -0.8693 & O67 & -0.8465 & O67 & -0.8605 \\
O68 & -0.8027 & O68 & -0.8235 & O68 & -0.8089 & O68 & -0.8110 & O68 & -0.8182 & O68 & -0.8417 & O68 & -0.8173 & O68 & -0.7598 & O68 & -0.7489 \\
O69 & -0.7815 & O69 & -0.7685 & O69 & -0.7651 & O69 & -0.7670 & O69 & -0.7666 & O69 & -0.7694 & O69 & -0.7441 & O69 & -0.6740 & O69 & -0.7273 \\
O70 & -0.9697 & O70 & -0.9706 & O70 & -0.9651 & O70 & -0.9647 & O70 & -0.9669 & O70 & -0.9678 & O70 & -0.9658 & O70 & -0.9618 & O70 & -0.9633 \\
O71 & -0.8062 & O71 & -0.8078 & O71 & -0.7816 & O71 & -0.7867 & O71 & -0.7895 & O71 & -0.7813 & O71 & -0.7928 & O71 & -0.8274 & O71 & -0.8098 \\
O72 & -0.8105 & O72 & -0.8041 & O72 & -0.8073 & O72 & -0.8037 & O72 & -0.8134 & O72 & -0.8085 & O72 & -0.7254 & O72 & -0.7908 & O72 & -0.7750 \\
O73 & -0.8222 & O73 & -0.8152 & O73 & -0.8129 & O73 & -0.7585 & O73 & -0.8174 & O73 & -0.8101 & O73 & -0.7764 & O73 & -0.8119 & O73 & -0.8190 \\
O74 & -0.8193 & O74 & -0.8194 & O74 & -0.8057 & O74 & -0.8093 & O74 & -0.8208 & O74 & -0.8140 & O74 & -0.8042 & O74 & -0.8149 & O74 & -0.8660 \\
O75 & -0.8146 & O75 & -0.7944 & O75 & -0.8095 & O75 & -0.8091 & O75 & -0.8090 & O75 & -0.7839 & O75 & -0.7462 & O75 & -0.8035 & O75 & -0.7977 \\
O76 & -0.8578 & O76 & -0.8349 & O76 & -0.8510 & O76 & -0.8382 & O76 & -0.8328 & O76 & -0.8507 & O76 & -0.8434 & O76 & -0.8130 & O76 & -0.8463 \\
O77 & -0.8804 & O77 & -0.8716 & O77 & -0.8780 & O77 & -0.8321 & O77 & -0.8305 & O77 & -0.8242 & O77 & -0.7933 & O77 & -0.7808 & O77 & -0.7575 \\
O78 & -0.8582 & O78 & -0.8622 & O78 & -0.8586 & O78 & -0.8092 & O78 & -0.8762 & O78 & -0.8000 & O78 & -0.8448 & O78 & -0.8184 & O78 & -0.8132 \\
O79 & -0.8662 & O79 & -0.8640 & O79 & -0.8682 & O79 & -0.8743 & O79 & -0.8260 & O79 & -0.8548 & O79 & -0.8560 & O79 & -0.8424 & O79 & -0.8755 \\
O80 & -0.8595 & O80 & -0.8603 & O80 & -0.8252 & O80 & -0.8274 & O80 & -0.8135 & O80 & -0.8164 & O80 & -0.8074 & O80 & -0.8230 & O80 & -0.8020 \\
O81 & -0.8844 & O81 & -0.8740 & O81 & -0.8735 & O81 & -0.8811 & O81 & -0.8770 & O81 & -0.8238 & O81 & -0.8083 & O81 & -0.7730 & O81 & -0.7176 \\
O82 & -0.8406 & O82 & -0.7337 & O82 & -0.7583 & O82 & -0.7898 & O82 & -0.8042 & O82 & -0.7976 & O82 & -0.7853 & O82 & -0.8180 & O82 & -0.8322 \\
O83 & -0.8238 & O83 & -0.8223 & O83 & -0.6959 & O83 & -0.7004 & O83 & -0.6975 & O83 & -0.6427 & O83 & -0.6203 & O83 & -0.6574 & O83 & -0.6418 \\
O84 & -0.7560 & O84 & -0.7695 & O84 & -0.7606 & O84 & -0.6951 & O84 & -0.7013 & O84 & -0.6777 & O84 & -0.6445 & O84 & -0.6916 & O84 & -0.6965 \\
O85 & -0.9724 & O85 & -0.9708 & O85 & -0.9741 & O85 & -0.9728 & O85 & -0.9745 & O85 & -0.9734 & O85 & -0.9732 & O85 & -0.9585 & O85 & -0.9734 \\
O86 & -0.9689 & O86 & -0.9655 & O86 & -0.9653 & O86 & -0.9659 & O86 & -0.9598 & O86 & -0.9658 & O86 & -0.9983 & O86 & -0.9571 & O86 & -0.9726 \\
O87 & -0.8701 & O87 & -0.8866 & O87 & -0.8888 & O87 & -0.8899 & O87 & -0.8947 & O87 & -0.8942 & O87 & -0.8948 & O87 & -0.7935 & O87 & -0.6659 \\
O88 & -0.8659 & O88 & -0.8728 & O88 & -0.8634 & O88 & -0.8636 & O88 & -0.8792 & O88 & -0.8820 & O88 & -0.8722 & O88 & -0.8220 & O88 & -0.7931 \\
O89 & -0.8571 & O89 & -0.8602 & O89 & -0.8704 & O89 & -0.8403 & O89 & -0.8395 & O89 & -0.8325 & O89 & -0.8676 & O89 & -0.8300 & O89 & -0.8259 \\
O90 & -0.9703 & O90 & -0.9659 & O90 & -0.9718 & O90 & -0.9665 & O90 & -0.9642 & O90 & -0.9651 & O90 & -0.9945 & O90 & -0.9672 & O90 & -1.0039 \\
O91 & -0.8633 & O91 & -0.8301 & O91 & -0.8223 & O91 & -0.8266 & O91 & -0.8115 & O91 & -0.8172 & O91 & -0.8263 & O91 & -0.8400 & O91 & -0.7550 \\
O92 & -0.8682 & O92 & -0.8619 & O92 & -0.8713 & O92 & -0.8733 & O92 & -0.8240 & O92 & -0.8153 & O92 & -0.8594 & O92 & -0.8348 & O92 & -0.8523 \\
O93 & -0.9705 & O93 & -0.9666 & O93 & -0.9730 & O93 & -0.9673 & O93 & -0.9621 & O93 & -0.9608 & O93 & -0.9979 & O93 & -0.9645 & O93 & -1.0016 \\
O94 & -0.8652 & O94 & -0.8780 & O94 & -0.8471 & O94 & -0.8416 & O94 & -0.8407 & O94 & -0.8529 & O94 & -0.8724 & O94 & -0.8520 & O94 & -0.8721 \\
O95 & -0.8483 & O95 & -0.8534 & O95 & -0.8581 & O95 & -0.8592 & O95 & -0.8491 & O95 & -0.8730 & O95 & -0.8396 & O95 & -0.6843 & O95 & -0.7670 \\
O96 & -0.7331 & O96 & -0.7459 & O96 & -0.7501 & O96 & -0.7479 & O96 & -0.7712 & O96 & -0.7667 & O96 & -0.7381 & O96 & -0.7017 & O96 & -0.7434 \\
O97 & -0.9709 & O97 & -0.9773 & O97 & -0.9632 & O97 & -0.9663 & O97 & -0.9649 & O97 & -0.9649 & O97 & -0.9565 & O97 & -0.9627 & O97 & -0.9404 \\
O98 & -0.8359 & O98 & -0.8190 & O98 & -0.7724 & O98 & -0.7782 & O98 & -0.7730 & O98 & -0.7727 & O98 & -0.7592 & O98 & -0.8272 & O98 & -0.7657 \\
O99 & -0.8140 & O99 & -0.8073 & O99 & -0.8075 & O99 & -0.8047 & O99 & -0.8132 & O99 & -0.7934 & O99 & -0.7718 & O99 & -0.7933 & O99 & -0.8495 \\
O100 & -0.8160 & O100 & -0.7761 & O100 & -0.7636 & O100 & -0.7537 & O100 & -0.8181 & O100 & -0.7925 & O100 & -0.7906 & O100 & -0.7975 & O100 & -0.8426 \\
O101 & -0.7781 & O101 & -0.7825 & O101 & -0.8092 & O101 & -0.8136 & O101 & -0.8175 & O101 & -0.8115 & O101 & -0.7526 & O101 & -0.8247 & O101 & -0.7679 \\
O102 & -0.8161 & O102 & -0.8123 & O102 & -0.8158 & O102 & -0.7955 & O102 & -0.8043 & O102 & -0.8082 & O102 & -0.7717 & O102 & -0.8074 & O102 & -0.8194 \\
O103 & -0.8611 & O103 & -0.8285 & O103 & -0.8424 & O103 & -0.8320 & O103 & -0.8324 & O103 & -0.8216 & O103 & -0.7982 & O103 & -0.8181 & O103 & -0.7747 \\
O104 & -0.8861 & O104 & -0.8800 & O104 & -0.8833 & O104 & -0.8261 & O104 & -0.8106 & O104 & -0.8245 & O104 & -0.8196 & O104 & -0.7618 & O104 & -0.8198 \\
O105 & -0.8591 & O105 & -0.8745 & O105 & -0.8629 & O105 & -0.8313 & O105 & -0.8400 & O105 & -0.8507 & O105 & -0.8314 & O105 & -0.8131 & O105 & -0.8175 \\
O106 & -0.8714 & O106 & -0.8567 & O106 & -0.8743 & O106 & -0.8743 & O106 & -0.8352 & O106 & -0.8409 & O106 & -0.8085 & O106 & -0.8439 & O106 & -0.7927 \\
O107 & -0.8615 & O107 & -0.8582 & O107 & -0.8232 & O107 & -0.8311 & O107 & -0.8348 & O107 & -0.8363 & O107 & -0.8440 & O107 & -0.8240 & O107 & -0.8625 \\
O108 & -0.8758 & O108 & -0.8675 & O108 & -0.8790 & O108 & -0.8680 & O108 & -0.8151 & O108 & -0.7951 & O108 & -0.8181 & O108 & -0.7465 & O108 & -0.7823 \\
O109 & -0.8021 & O109 & -0.7966 & O109 & -0.7458 & O109 & -0.7483 & O109 & -0.8061 & O109 & -0.7741 & O109 & -0.6035 & O109 & -0.6814 & O109 & -0.6416 \\
O110 & -0.8021 & O110 & -0.7966 & O110 & -0.7883 & O110 & -0.7589 & O110 & -0.8059 & O110 & -0.7735 & O110 & -0.5813 & O110 & -0.6317 & O110 & -0.6231 \\
O111 & -0.7541 & O111 & -0.6646 & O111 & -0.6586 & O111 & -0.6937 & O111 & -0.6978 & O111 & -0.6612 & O111 & -0.6421 & O111 & -0.6961 & O111 & -0.6793 \\
O112 & -0.7541 & O112 & -0.6647 & O112 & -0.6575 & O112 & -0.6852 & O112 & -0.6979 & O112 & -0.6611 & O112 & -0.6454 & O112 & -0.6911 & O112 & -0.6664 \\ \hline
\end{tabular}}
%\caption{Calculated bader charges for all systems.}
%\label{bader}
\end{table}

\clearpage
\begin{figure}[!h]
\addcontentsline{toc}{figure}{Figure \ref{hubbardUBenchmarking}. Benchmark of Hubbard U values for Mo, V, and O atoms in Mo$_{30}$V$_{10}$}
    \centering
    \includegraphics[width=0.99\linewidth]{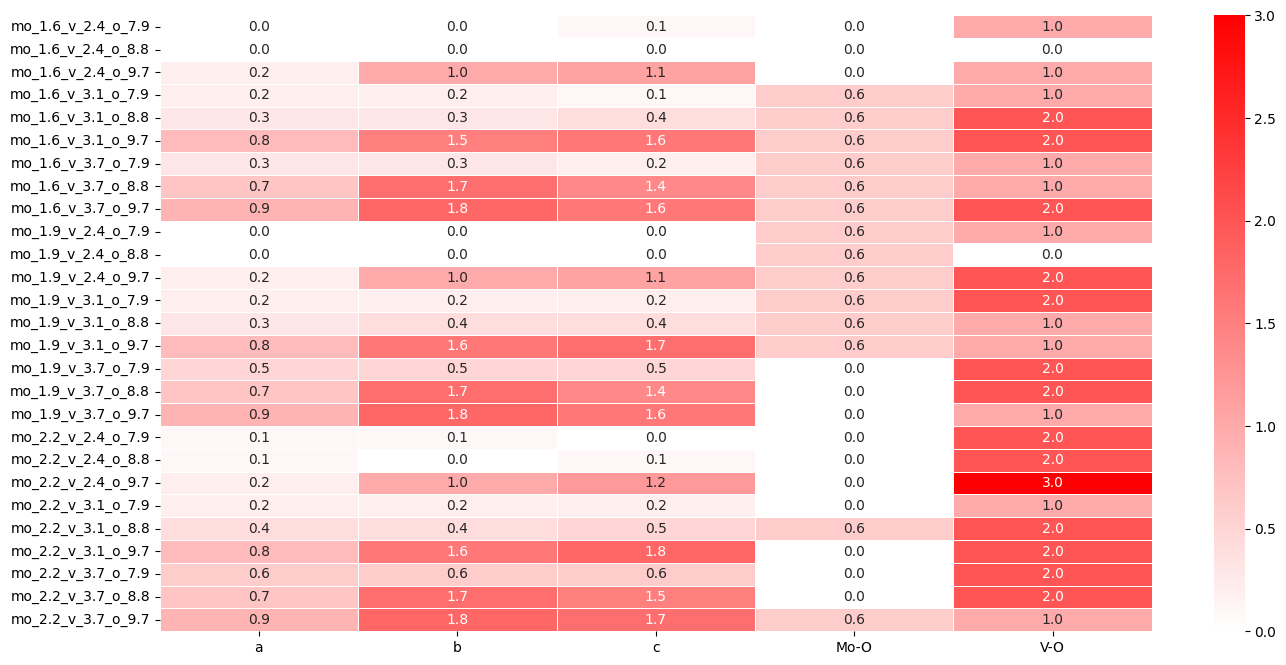}
    \caption{Calculated \%error in unit cell vectors and Mo-O and V-O distances of the Mo$_{30}$V$_{10}$ system using different Hubbard U values for Mo, V, and O atoms compared to Li \textit{et al.}\cite{li2018distribution}.}
    \label{hubbardUBenchmarking}
\end{figure}

\clearpage
\begin{figure}[!h]
\addcontentsline{toc}{figure}{Figure \ref{fitting}. Calculated Kr (left column) and Xe (right column) uptakes at 298 K for Mo$_{30}$V$_{10}$O$_{112}$ using different $\epsilon_{Kr-O}$ and $\epsilon_{Xe-O}$ values compared to experiment with using a different number of equilibration and production cycles}
    \centering
    \includegraphics[width=0.8\linewidth]{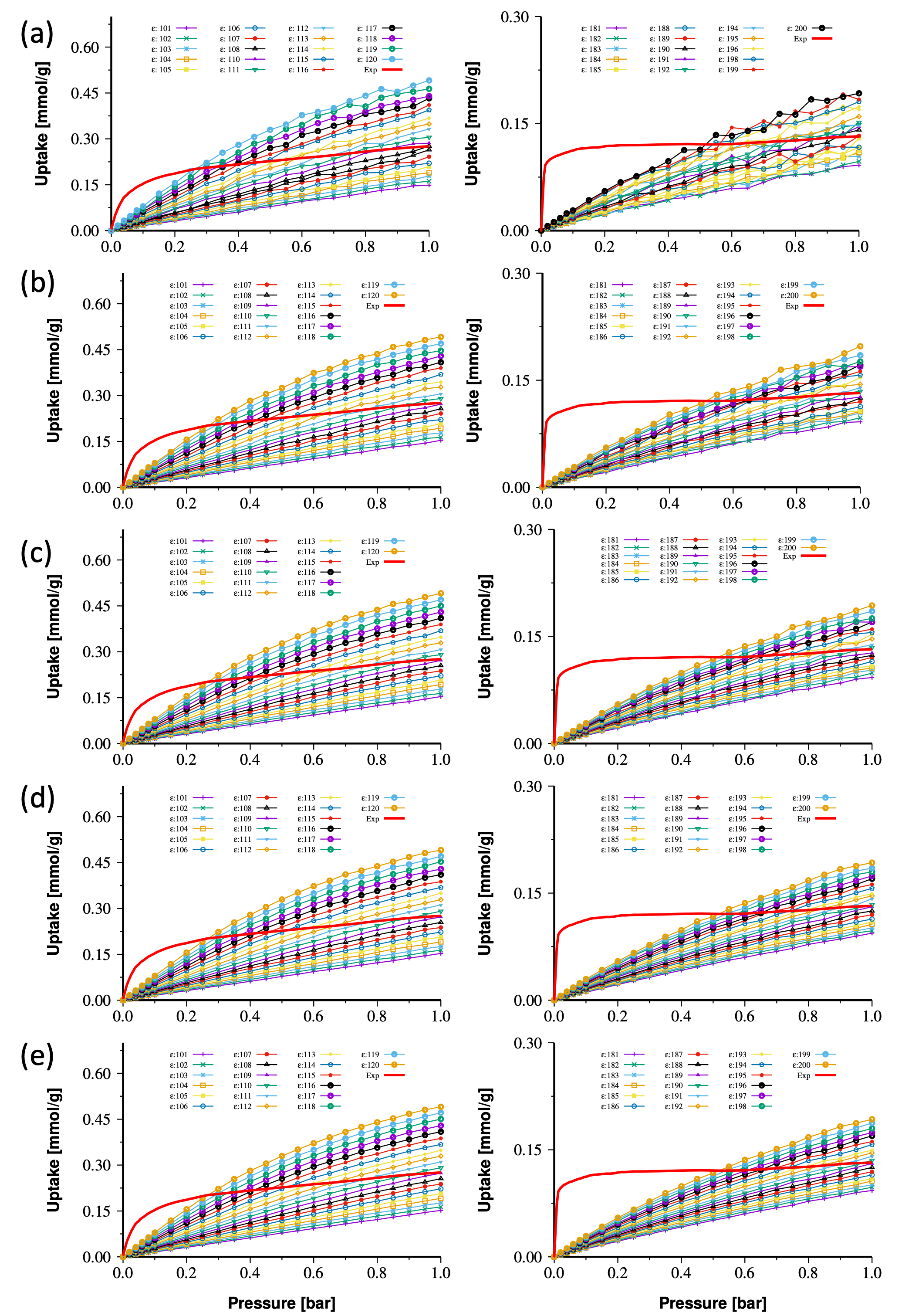}
    \caption{Calculated Kr (left column) and Xe (right column) uptakes at 298 K for Mo$_{30}$V$_{10}$O$_{112}$ using different $\epsilon_{Kr-O}$ and $\epsilon_{Xe-O}$ values compared to experiment (red curves) with using (a) 1$\times$10$^5$, (b) 5$\times$10$^5$, (c) 1$\times$10$^6$, (d) 2$\times$10$^6$, and (e) 5$\times$10$^6$ equilibration and production cycles.}
    \label{fitting}
\end{figure}

\clearpage
\begin{table}
\addcontentsline{toc}{table}{Table \ref{method_benchmark}. Benchmark of different exchange correlation functionals and dispersion and Hubbard U correction methods}
\centering
\resizebox{0.99\linewidth}{!}{
\begin{tabular}{cccccccccccc}  \hline
Method & $a$  & $b$  & $c$ & $\alpha$  & $\beta$  & $\gamma$  & V-V & V-Mo & V-O & Mo-Mo & Mo-O \\ \hline
PBE & 0.5 & 0.1 & 1.5 & 0.0 & 0.0 & 0.0 & -1.5 & -0.5 & -1.0 & 0 & -1.1 \\
RPBE & 0.5 & 0.2 & 7.7 & -0.1 & 0.0 & 0.0 & -1.6 & 0.5 & 1.0 & 0.6 & 4.5 \\
PBEsol & 0.0 & -0.4 & -2.1 & -0.1 & 0.0 & 0.0 & -1.5 & -2.6 & -4.9 & -0.5 & -1.0 \\
PBE-D2 & 0.1 & -0.2 & -0.3 & -0.1 & 0.0 & 0.0 & -1.5 & 0.2 & -5.9 & -0.5 & -1.0 \\
PBE-D3(Zero) & 0.2 & -0.1 & 0.8 & -0.1 & 0.0 & 0.0 & -1.5 & 0.1 & -5.9 & -0.5 & -1.0 \\
PBE-D3(BJ) & 0.2 & -0.2 & 0.7 & -0.1 & 0.0 & 0.0 & -1.5 & -0.4 & -4.9 & 0.1 & -2.1 \\
PBE-U-D3-BJ & -0.3 & -0.2 & 0.6 & 0.1 & 0.0 & 0.0 & 0.0 & 0.3 & 0.0 & 0.0 & -1.7 \\ \hline
\end{tabular}}
\caption{Calculated \%error in unit cell lengths and angles as well as V-V, V-Mo, V-O, Mo-Mo, and Mo-O distances in Mo$_{30}$V$_{10}$ using different exchange correlation functionals with and without dispersion and Hubbard U corrections compared to Li \textit{et al.}\cite{li2018distribution}.}
\label{method_benchmark}
\end{table}

\clearpage
\begin{figure}[!h]
\addcontentsline{toc}{figure}{Figure \ref{mo40kr}. Benchmark of GCMC simulated Kr uptakes at 298 K for Mo$_{40}$ }
    \centering
    \includegraphics[width=0.99\linewidth]{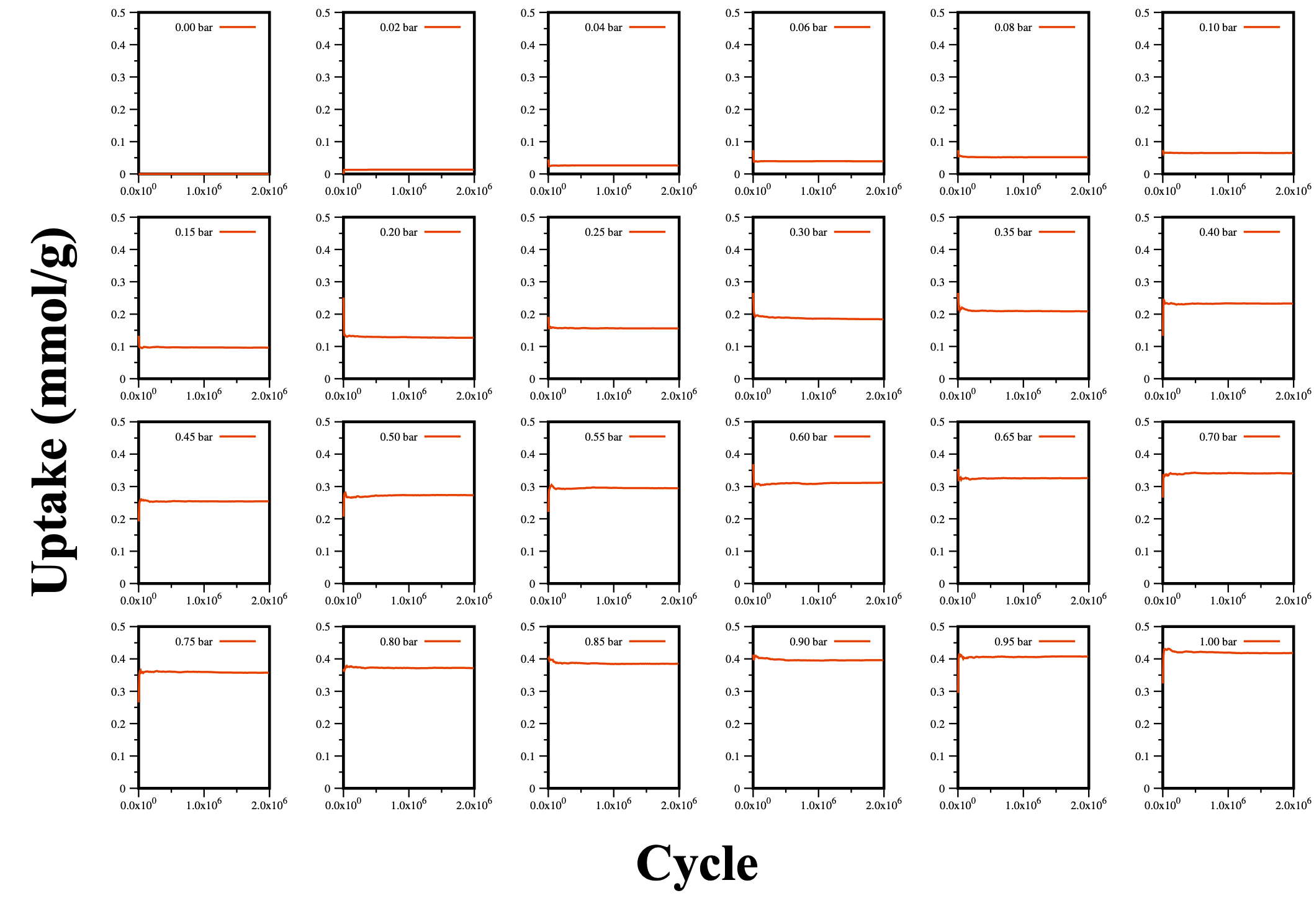}
    \caption{Benchmark of GCMC simulated Kr uptakes at 298 K for Mo$_{40}$ using different numbers of equilibration and production cycles.}
    \label{mo40kr}
\end{figure}

\clearpage
\begin{figure}[!h]
\addcontentsline{toc}{figure}{Figure \ref{mo40xe}. Benchmark of GCMC simulated Xe uptakes at 298 K for Mo$_{40}$ }
    \centering
    \includegraphics[width=0.99\linewidth]{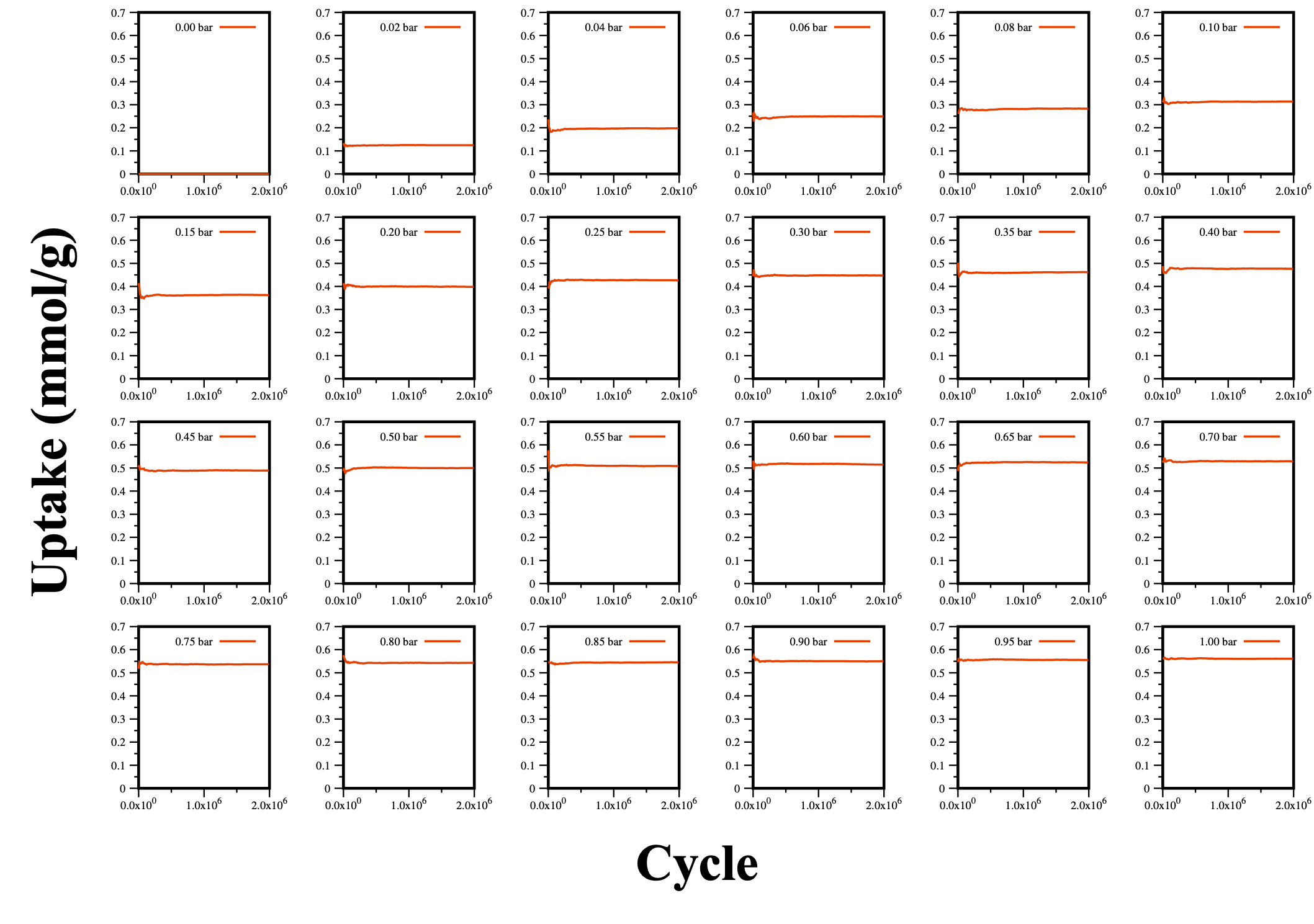}
    \caption{ Benchmark of GCMC simulated Xe uptakes at 298 K for Mo$_{40}$ using different numbers of equilibration and production cycles.}
    \label{mo40kr}
    \label{mo40xe}
\end{figure}

\clearpage
\begin{figure}[!h]
\addcontentsline{toc}{figure}{Figure \ref{mo38kr}. Benchmark of GCMC simulated Kr uptakes at 298 K for Mo$_{38}$V$_2$ }
    \centering
    \includegraphics[width=0.99\linewidth]{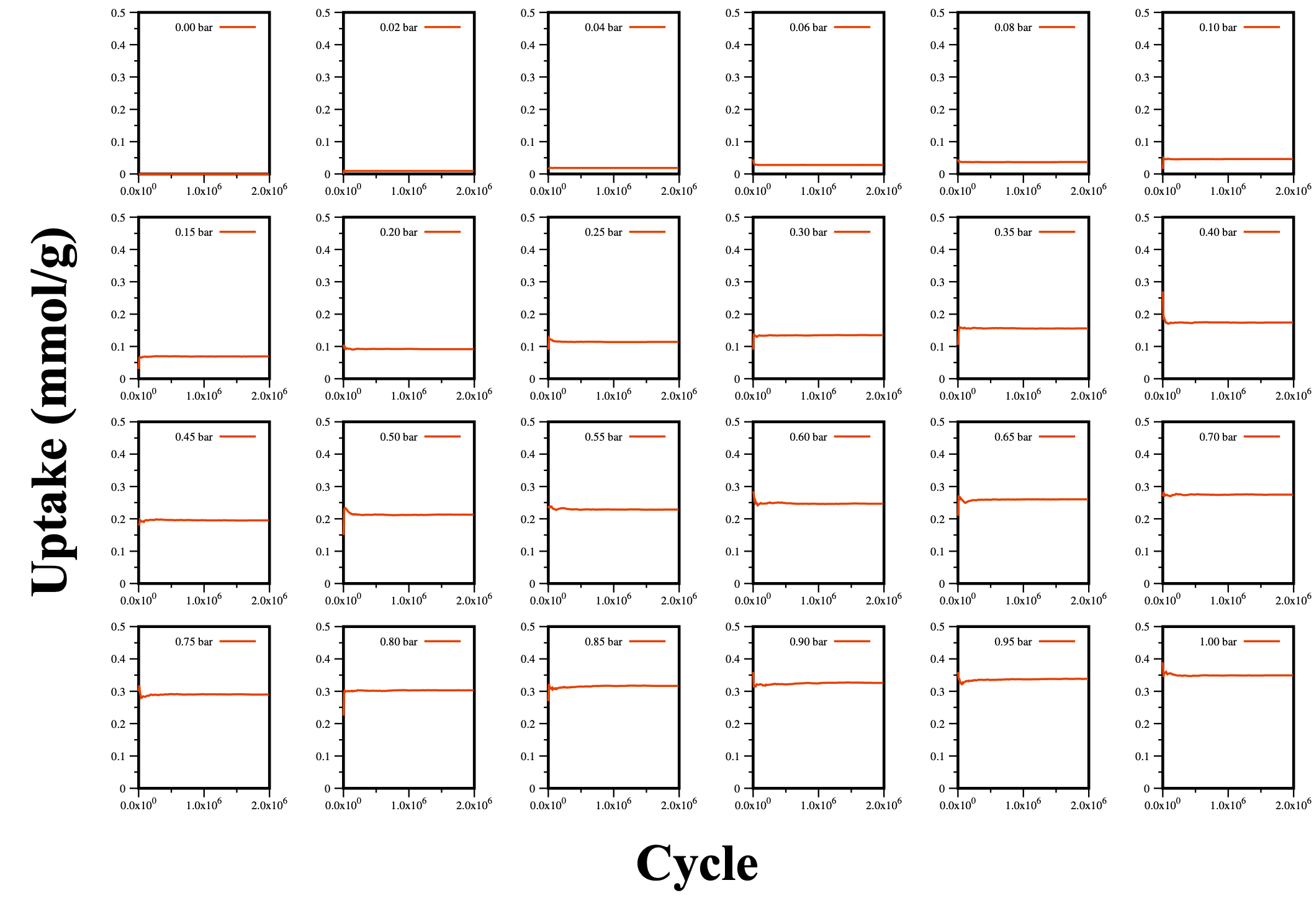}
    \caption{ Benchmark of GCMC simulated Kr uptakes at 298 K for Mo$_{38}$V$_2$ using different number of equilibration and production cycles.}
    \label{mo38kr}
\end{figure}

\clearpage
\begin{figure}[!h]
\addcontentsline{toc}{figure}{Figure \ref{mo38xe}. Benchmark of GCMC simulated Xe uptakes at 298 K for Mo$_{38}$V$_2$ }
    \centering
    \includegraphics[width=0.99\linewidth]{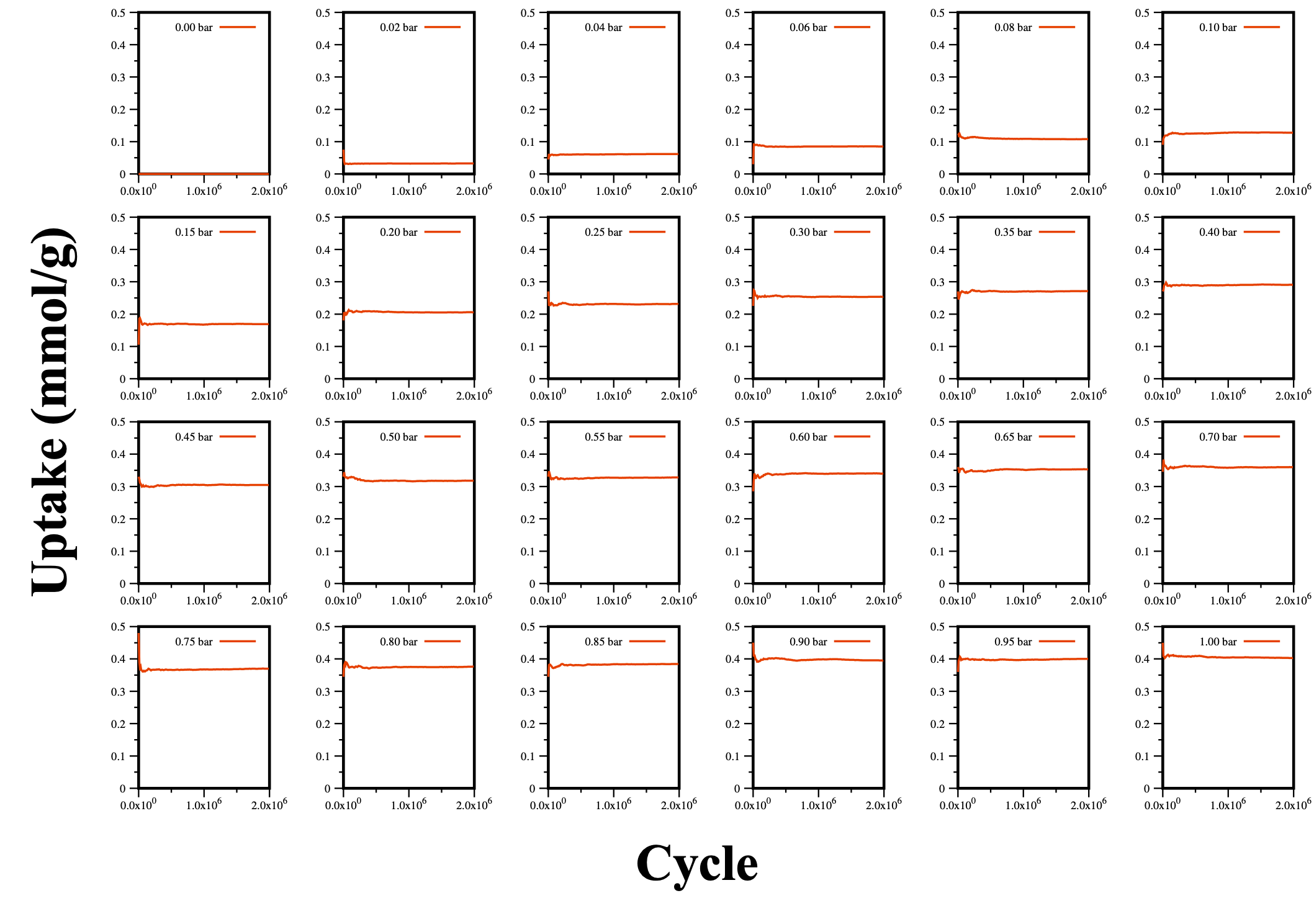}
    \caption{Benchmark of GCMC simulated Xe uptakes at 298 K for Mo$_{38}$V$_2$ using different number of equilibration and production cycles.}
    \label{mo38xe}
\end{figure}

\clearpage
\begin{figure}[!h]
\addcontentsline{toc}{figure}{Figure \ref{mo36kr}. Benchmark of GCMC simulated Kr uptakes at 298 K for Mo$_{36}$V$_4$  }
    \centering
    \includegraphics[width=0.99\linewidth]{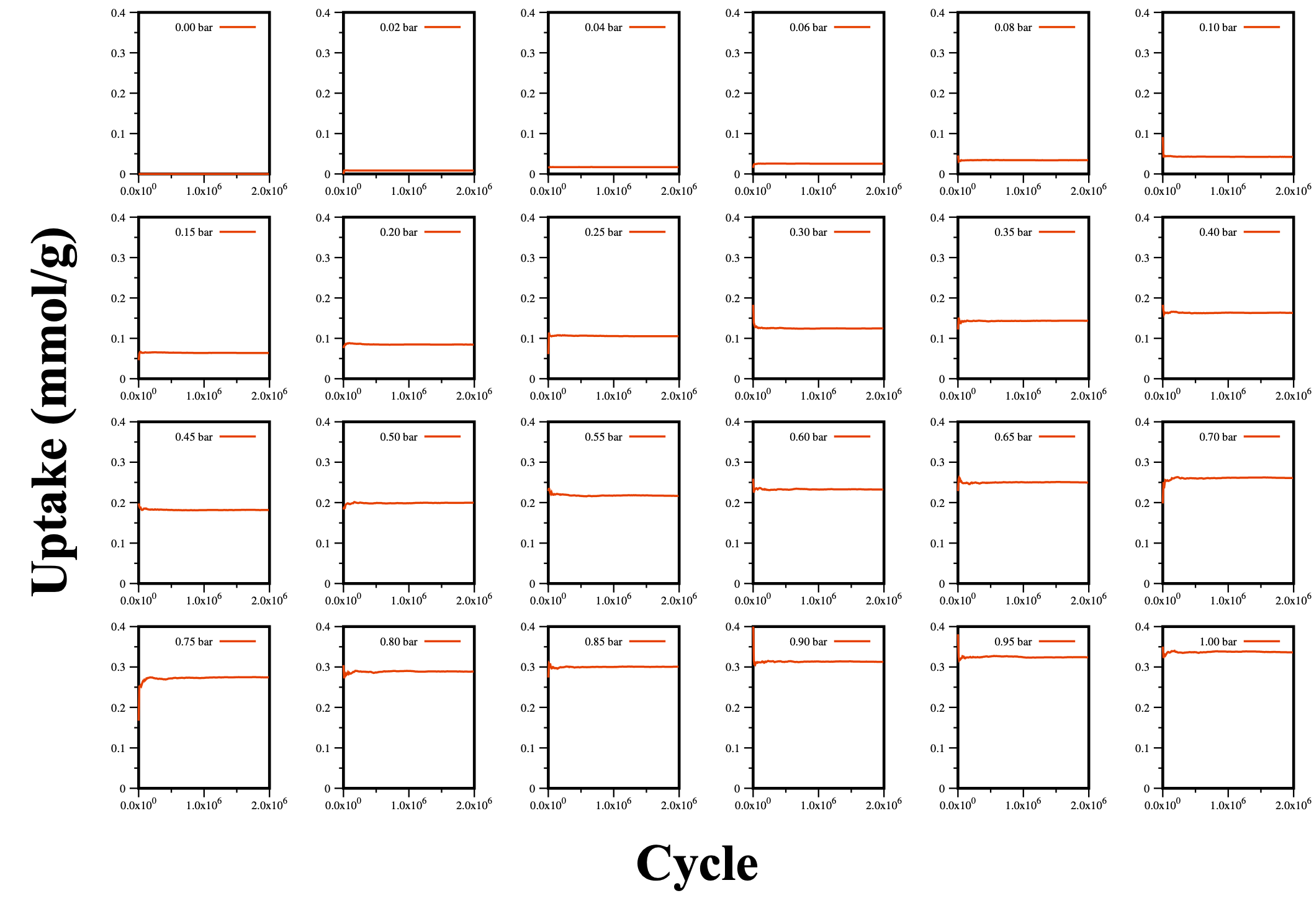}
    \caption{Benchmark of GCMC simulated Kr uptakes at 298 K for Mo$_{36}$V$_4$ using different number of equilibration and production cycles.}
    \label{mo36kr}
\end{figure}

\clearpage
\begin{figure}[!h]
\addcontentsline{toc}{figure}{Figure \ref{mo36xe}. Benchmark of GCMC simulated Xe uptakes at 298 K for Mo$_{36}$V$_4$ }
    \centering
    \includegraphics[width=0.99\linewidth]{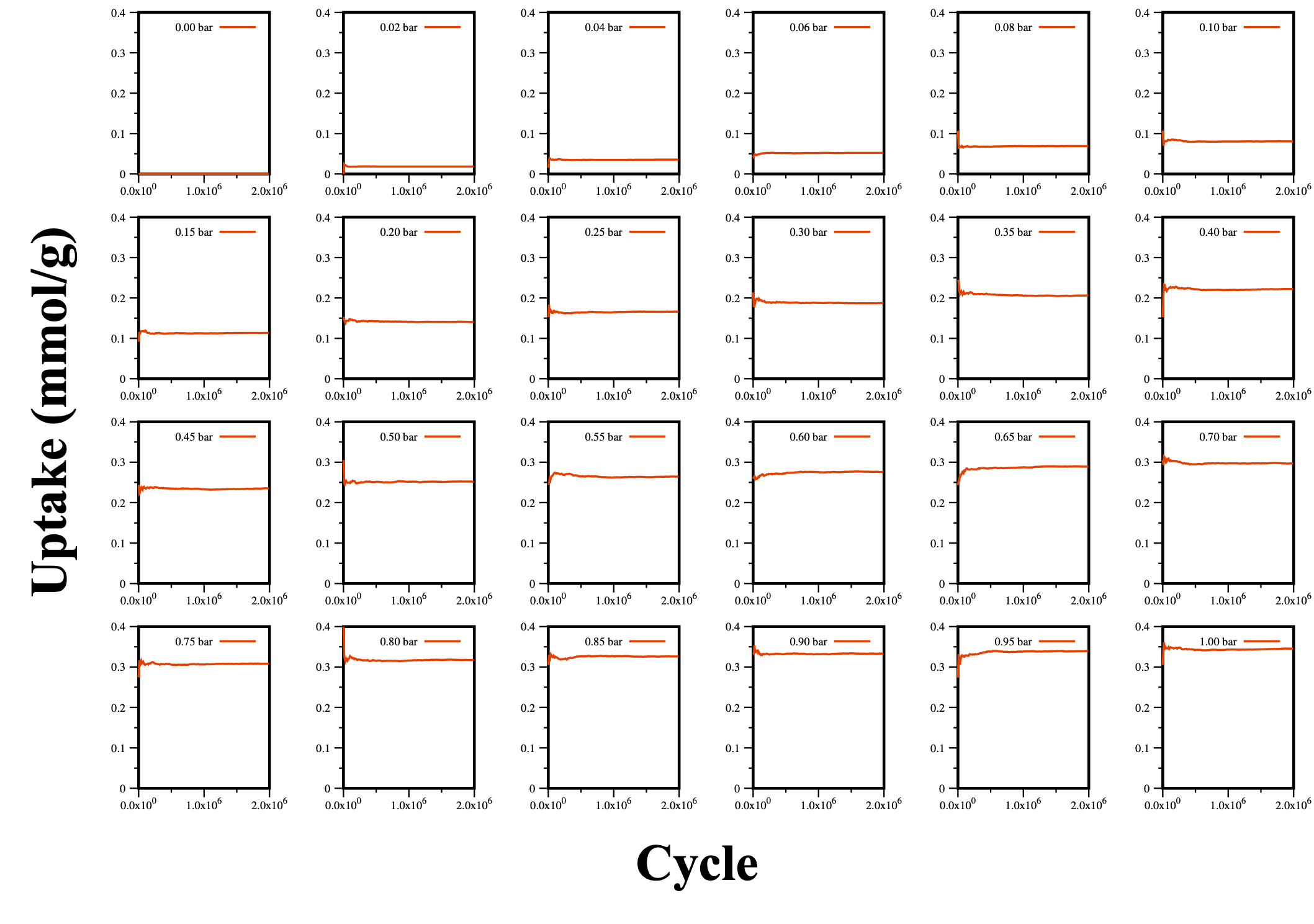}
    \caption{Benchmark of GCMC simulated Xe uptakes at 298 K for Mo$_{36}$V$_4$ using different number of equilibration and production cycles.}
    \label{mo36xe}
\end{figure}

\clearpage
\begin{figure}[!h]
\addcontentsline{toc}{figure}{Figure \ref{mo34kr}. Benchmark of GCMC simulated Kr uptakes at 298 K for Mo$_{34}$V$_6$ }
    \centering
    \includegraphics[width=0.99\linewidth]{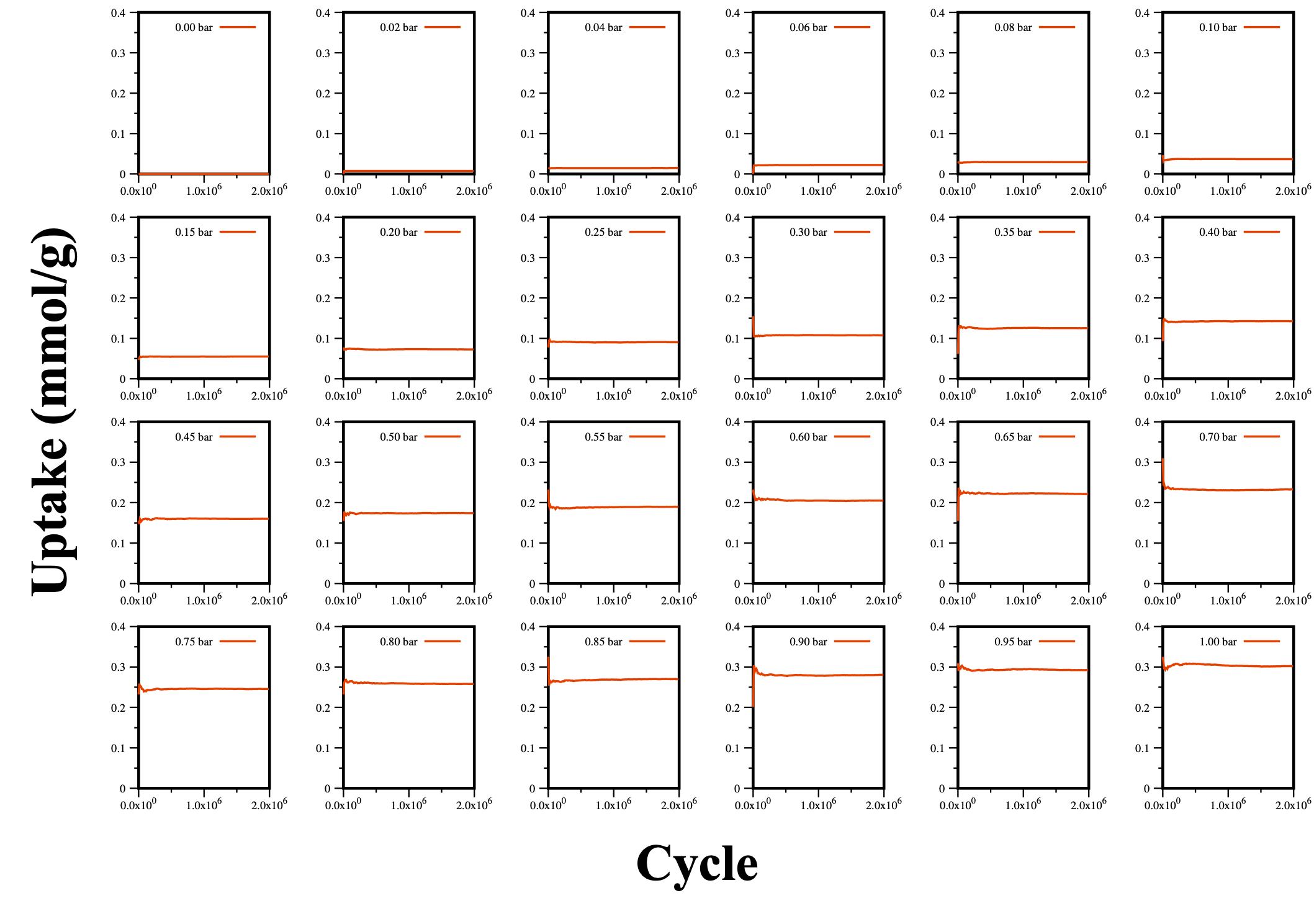}
    \caption{Benchmark of GCMC simulated Kr uptakes at 298 K for Mo$_{34}$V$_6$ using different number of equilibration and production cycles.}
    \label{mo34kr}
\end{figure}

\clearpage 
\begin{figure}[!h]
\addcontentsline{toc}{figure}{Figure \ref{mo34xe}. Benchmark of GCMC simulated Xe uptakes at 298 K for Mo$_{34}$V$_6$ }
    \centering
    \includegraphics[width=0.99\linewidth]{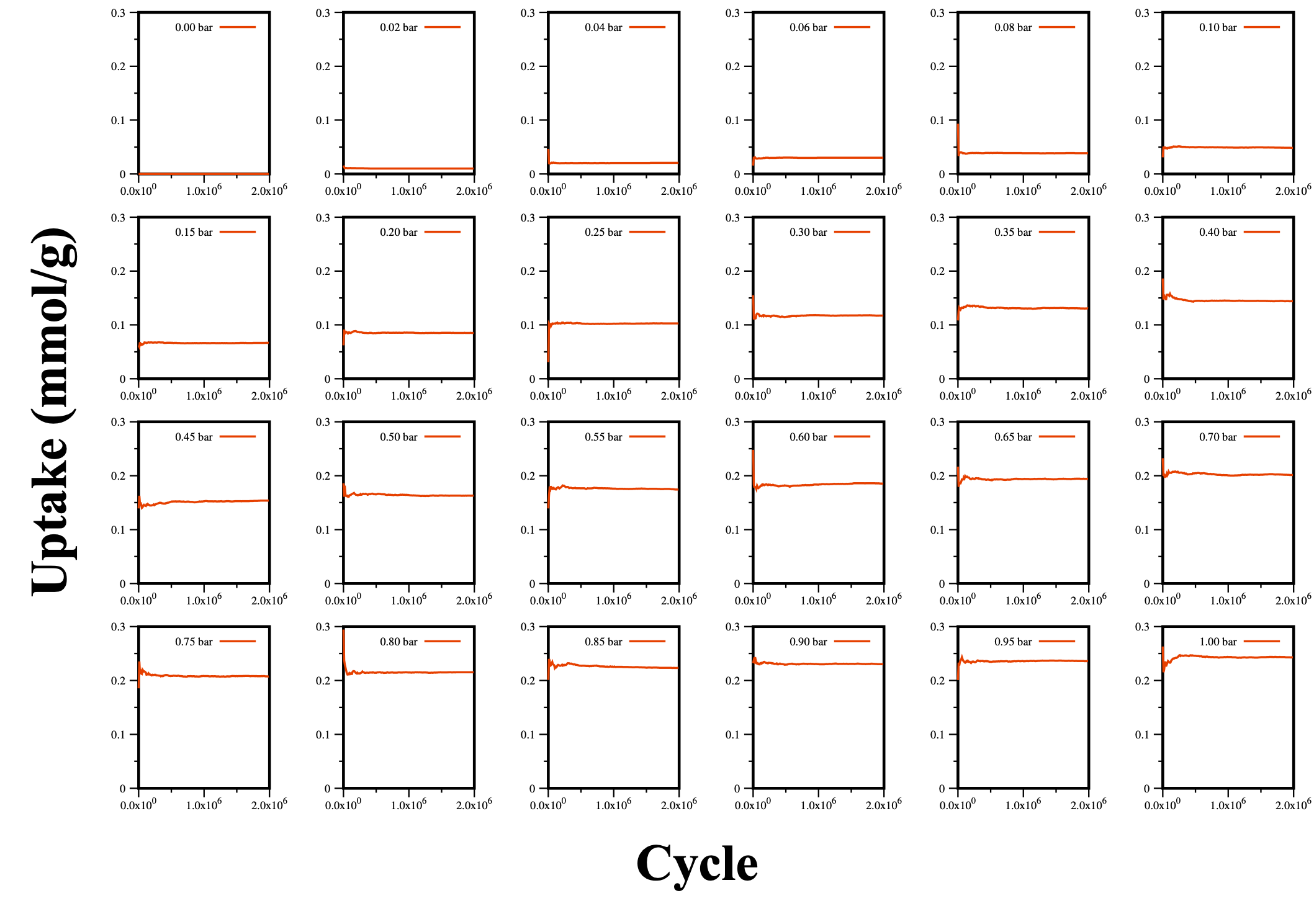}
    \caption{Benchmark of GCMC simulated Xe uptakes at 298 K for Mo$_{34}$V$_6$ using different numbers of equilibration and production cycles.}
    \label{mo34xe}
\end{figure}

\clearpage
\begin{figure}[!h]
\addcontentsline{toc}{figure}{Figure \ref{mo32kr}. Benchmark of GCMC simulated Kr uptakes at 298 K for Mo$_{32}$V$_8$ }
    \centering
    \includegraphics[width=0.99\linewidth]{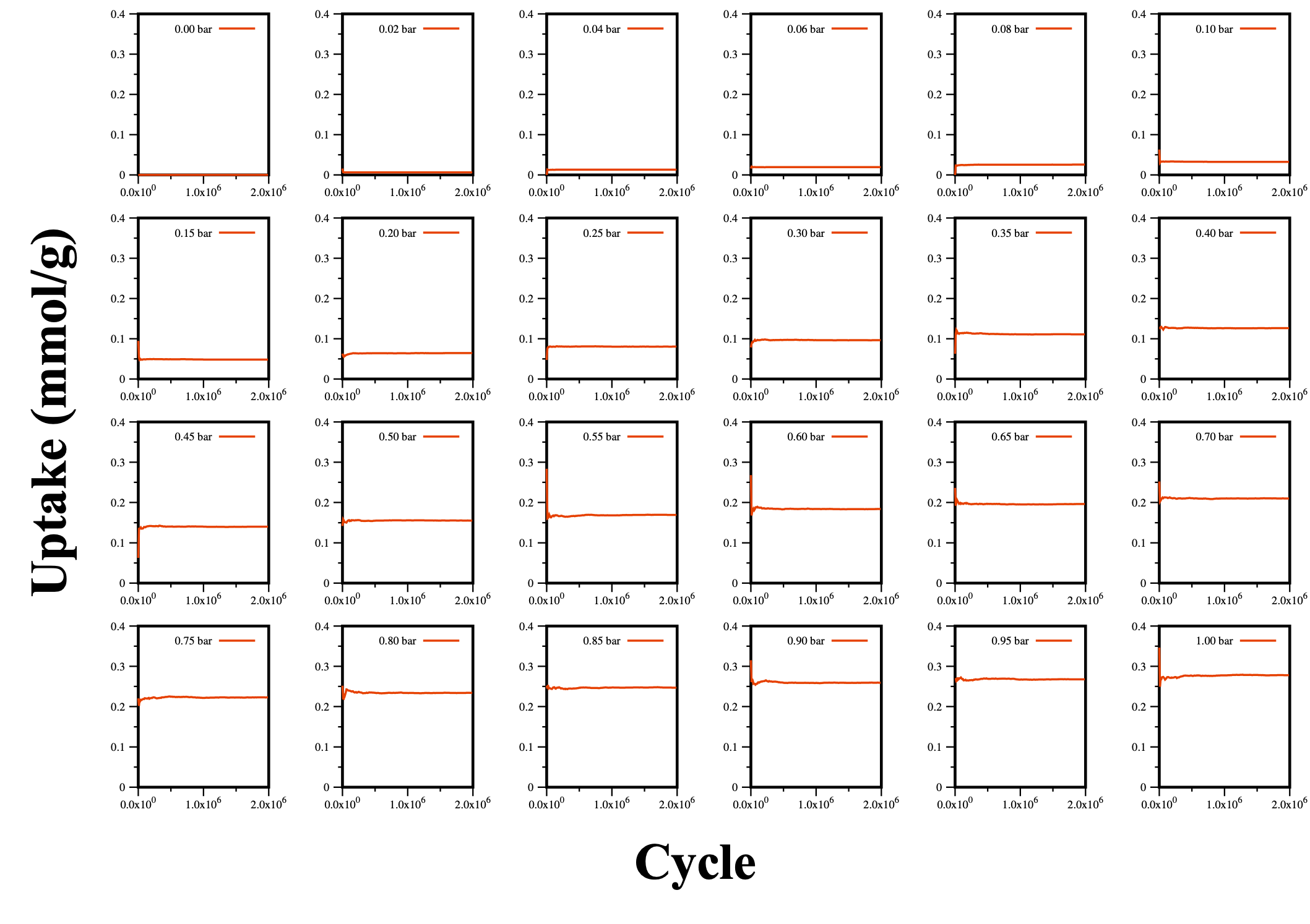}
    \caption{Benchmark of GCMC simulated Kr uptakes at 298 K for Mo$_{32}$V$_8$ using different number of equilibration and production cycles.}
    \label{mo32kr}
\end{figure}

\clearpage
\begin{figure}[!h]
\addcontentsline{toc}{figure}{Figure \ref{mo32xe}. Benchmark of GCMC simulated Xe uptakes at 298 K for Mo$_{32}$V$_8$ }
    \centering
    \includegraphics[width=0.99\linewidth]{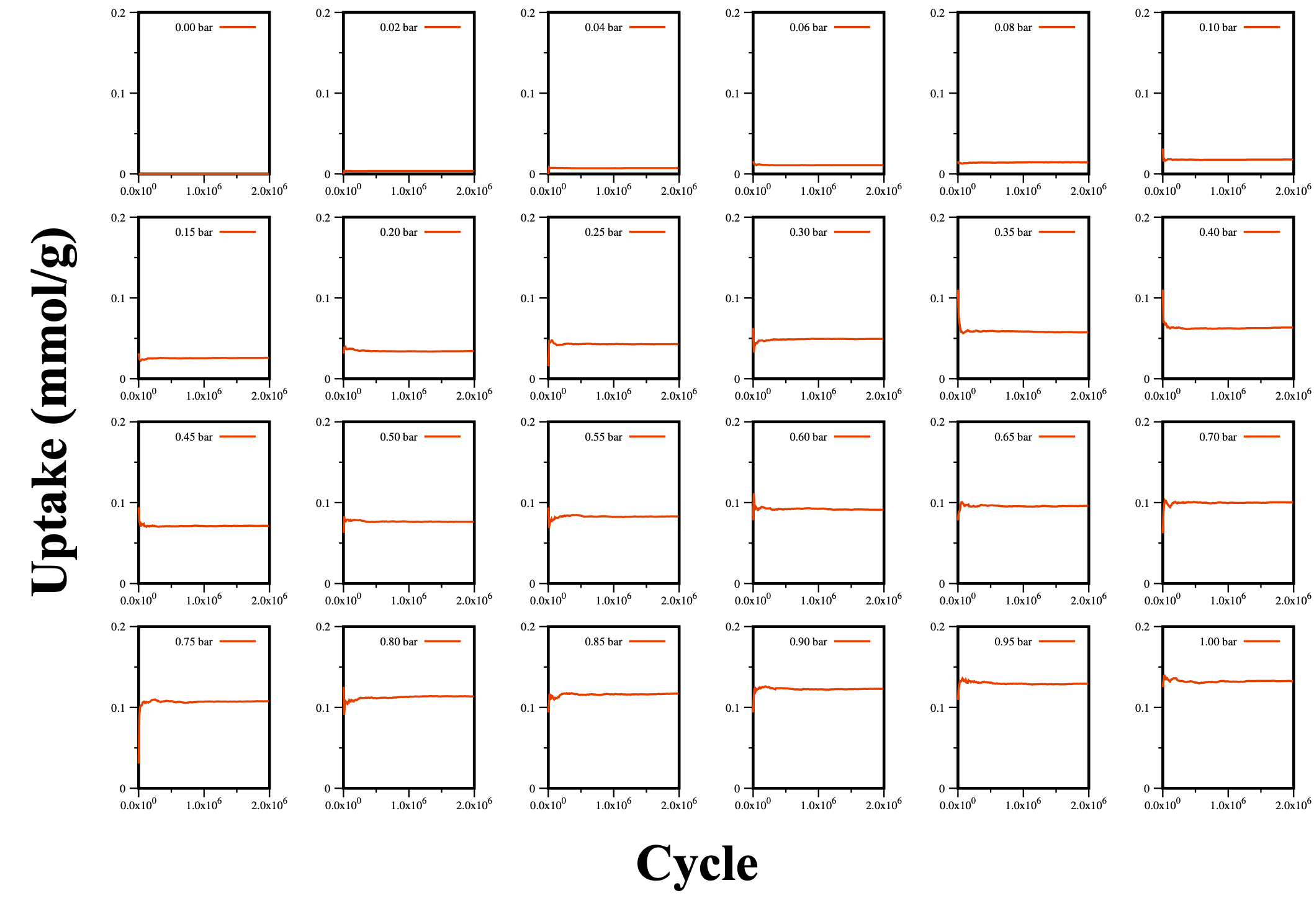}
    \caption{Benchmark of GCMC simulated Xe uptakes at 298 K for Mo$_{32}$V$_8$ using different number of equilibration and production cycles.}
    \label{mo32xe}
\end{figure}

\clearpage
\begin{figure}[!h]
\addcontentsline{toc}{figure}{Figure \ref{mo30kr}. Benchmark of GCMC simulated Kr uptakes at 298 K for Mo$_{30}$V$_{10}$ }
    \centering
    \includegraphics[width=0.99\linewidth]{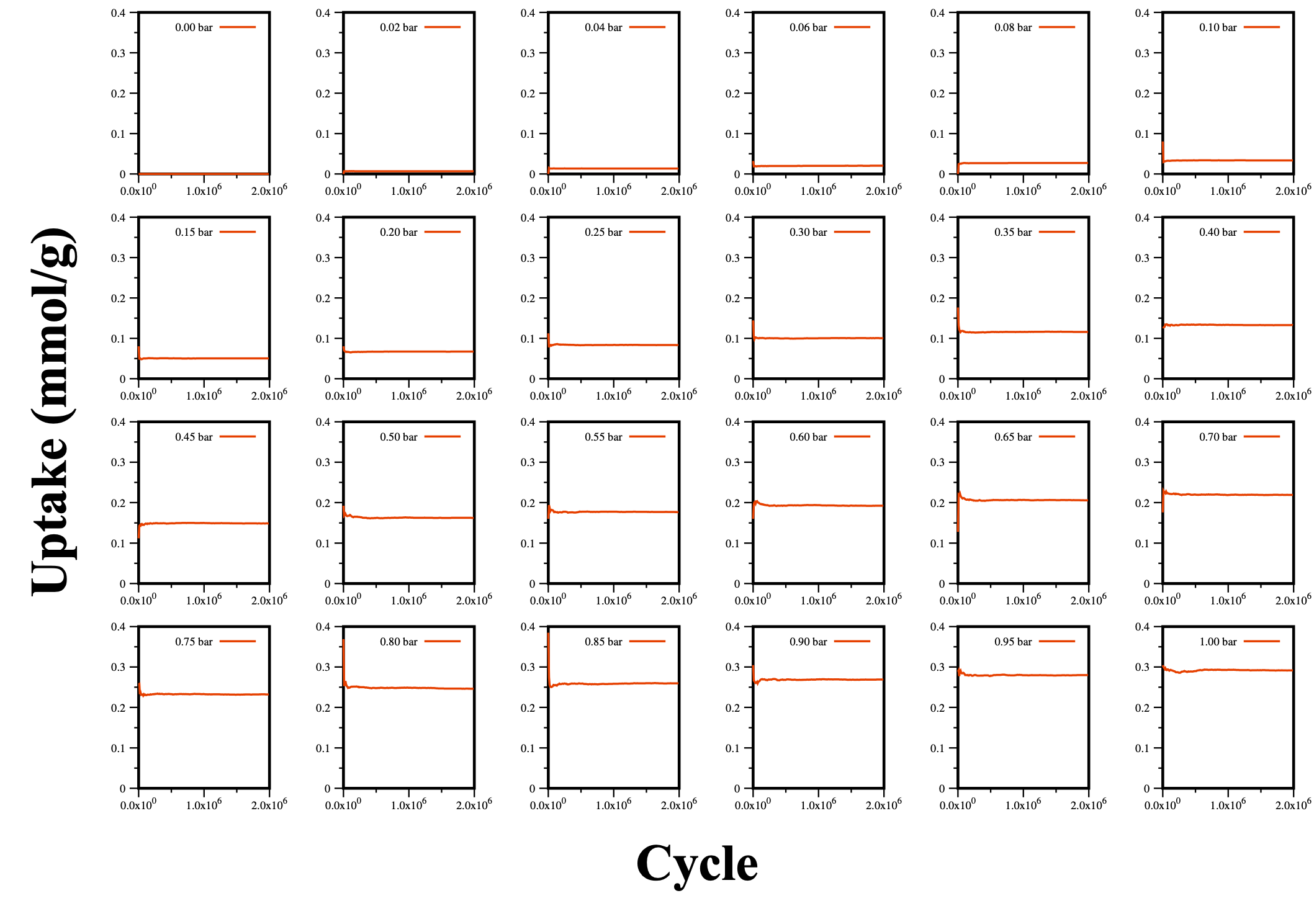}
    \caption{Benchmark of GCMC simulated Kr uptakes at 298 K for Mo$_{30}$V$_{10}$ using different number of equilibration and production cycles.}
    \label{mo30kr}
\end{figure}

\clearpage
\begin{figure}[!h]
\addcontentsline{toc}{figure}{Figure \ref{mo30xe}. Benchmark of GCMC simulated Xe uptakes at 298 K for Mo$_{30}$V$_{10}$ }
    \centering
    \includegraphics[width=0.99\linewidth]{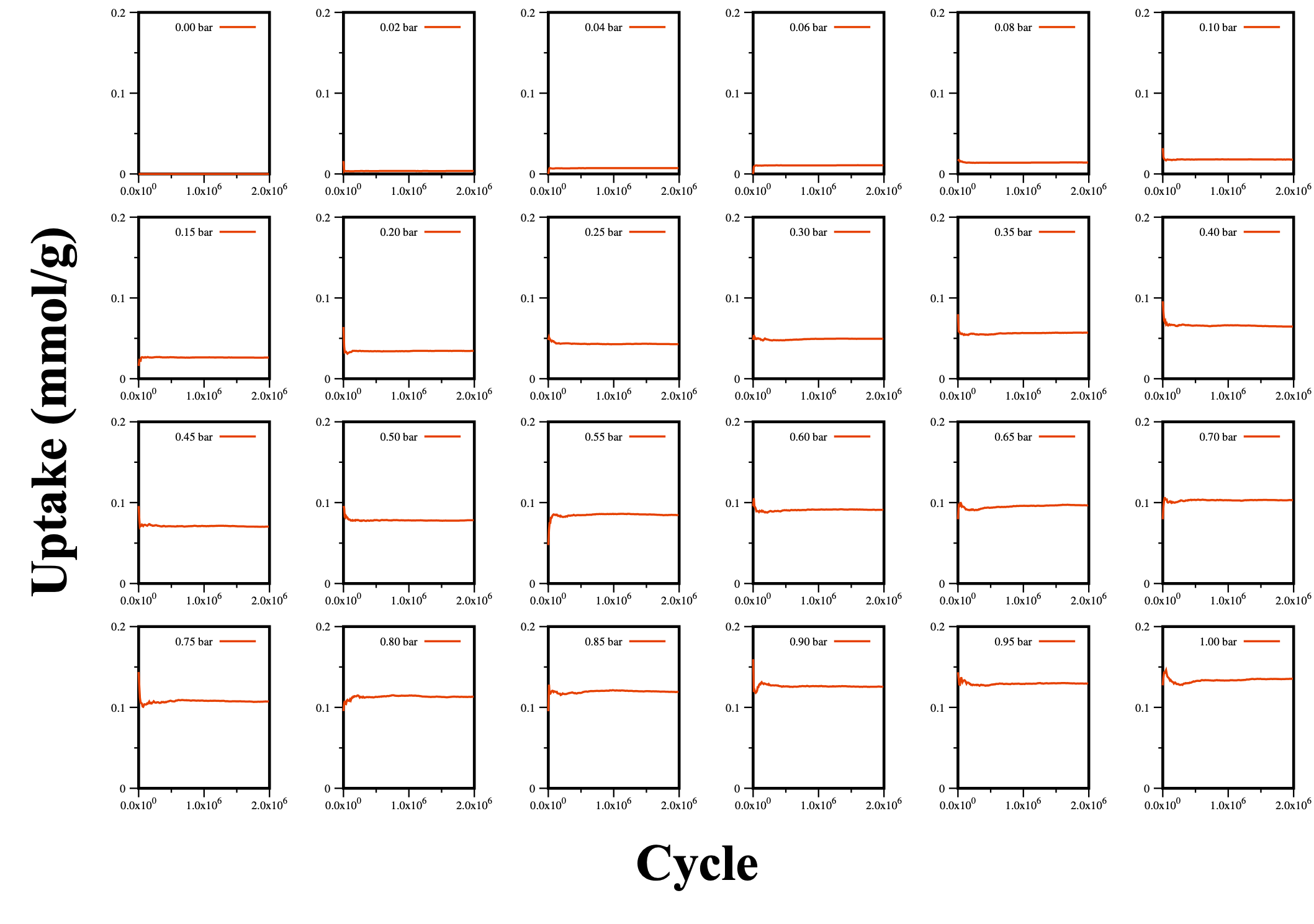}
    \caption{Benchmark of GCMC simulated Xe uptakes at 298 K for Mo$_{30}$V$_{10}$ using different number of equilibration and production cycles.}
    \label{mo30xe}
\end{figure}

\clearpage
\begin{figure}[!h]
\addcontentsline{toc}{figure}{Figure \ref{mo28kr}. Benchmark of GCMC simulated Kr uptakes at 298 K for Mo$_{28}$V$_{12}$ }
    \centering
    \includegraphics[width=0.99\linewidth]{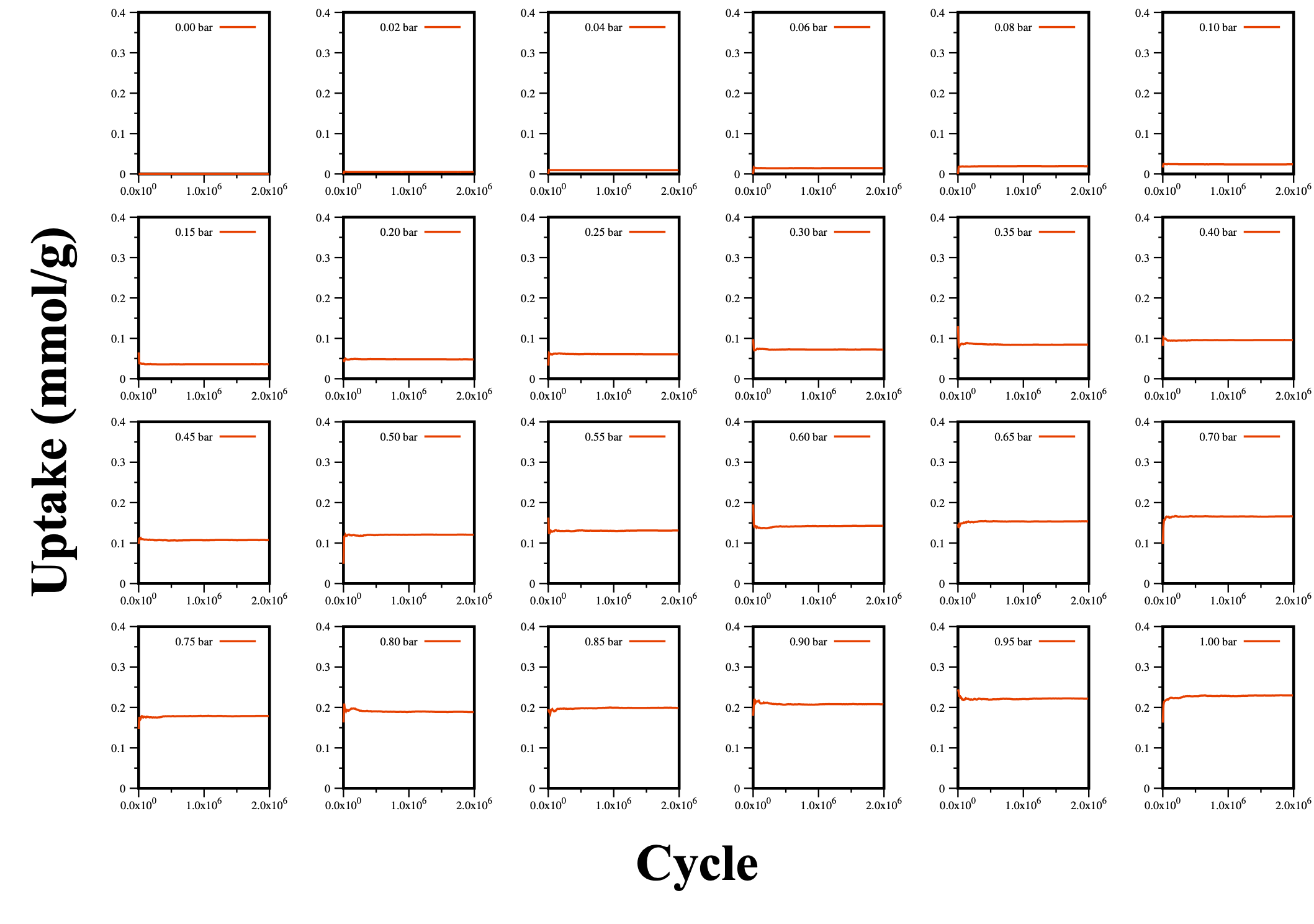}
    \caption{Benchmark of GCMC simulated Kr uptakes at 298 K for Mo$_{28}$V$_{12}$ using different number of equilibration and production cycles.}
    \label{mo28kr}
\end{figure}

\clearpage
\begin{figure}[!h]
\addcontentsline{toc}{figure}{Figure \ref{mo28xe}. Benchmark of GCMC simulated Xe uptakes at 298 K for Mo$_{28}$V$_{12}$  }
    \centering
    \includegraphics[width=0.99\linewidth]{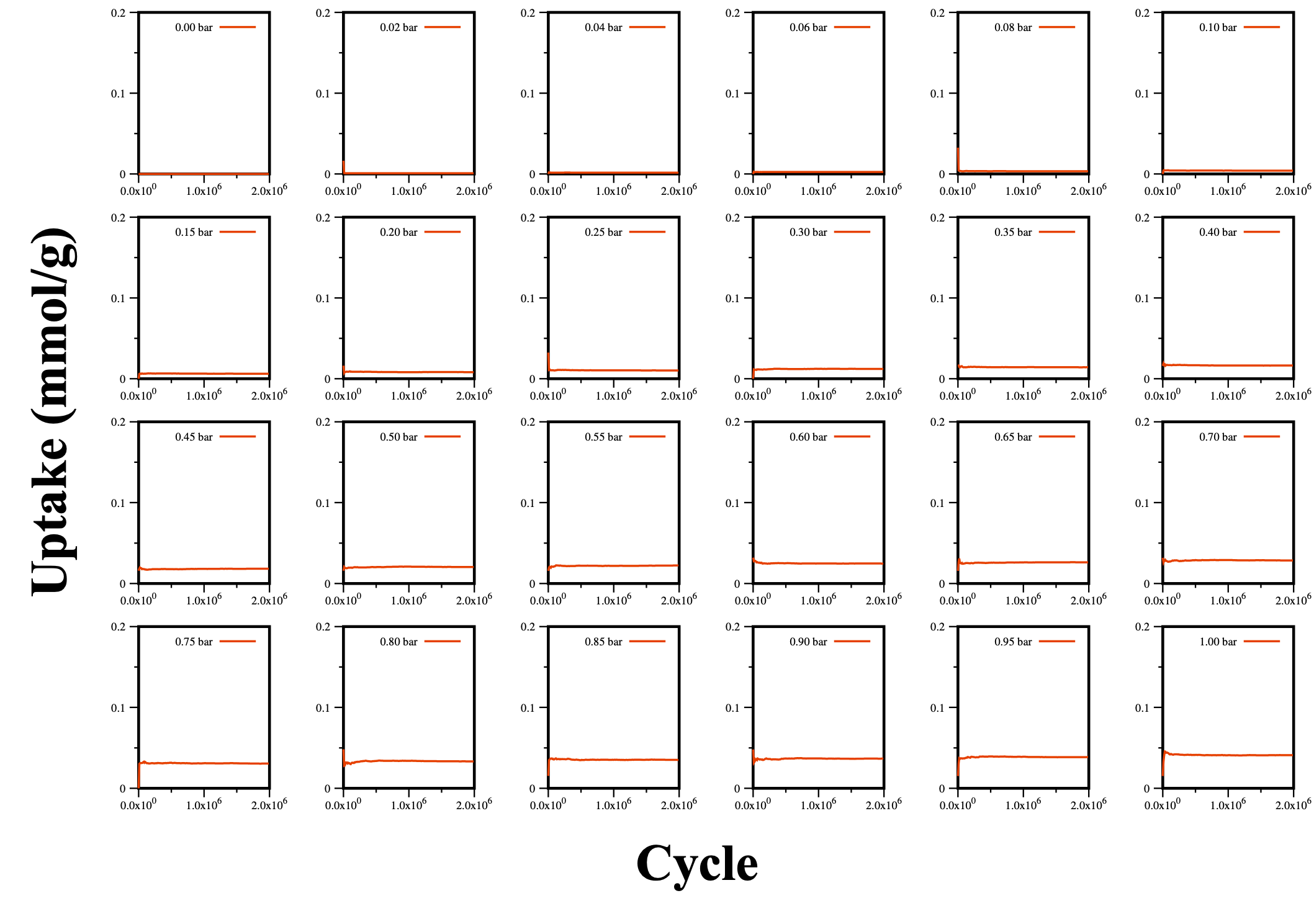}
    \caption{Benchmark of GCMC simulated Xe uptakes at 298 K for Mo$_{28}$V$_{12}$ using different number of equilibration and production cycles.}
    \label{mo28xe}
\end{figure}

\clearpage
\begin{figure}[!h]
\addcontentsline{toc}{figure}{Figure \ref{mo26kr}. Benchmark of GCMC simulated Kr uptakes at 298 K for Mo$_{26}$V$_{14}$ }
    \centering
    \includegraphics[width=0.99\linewidth]{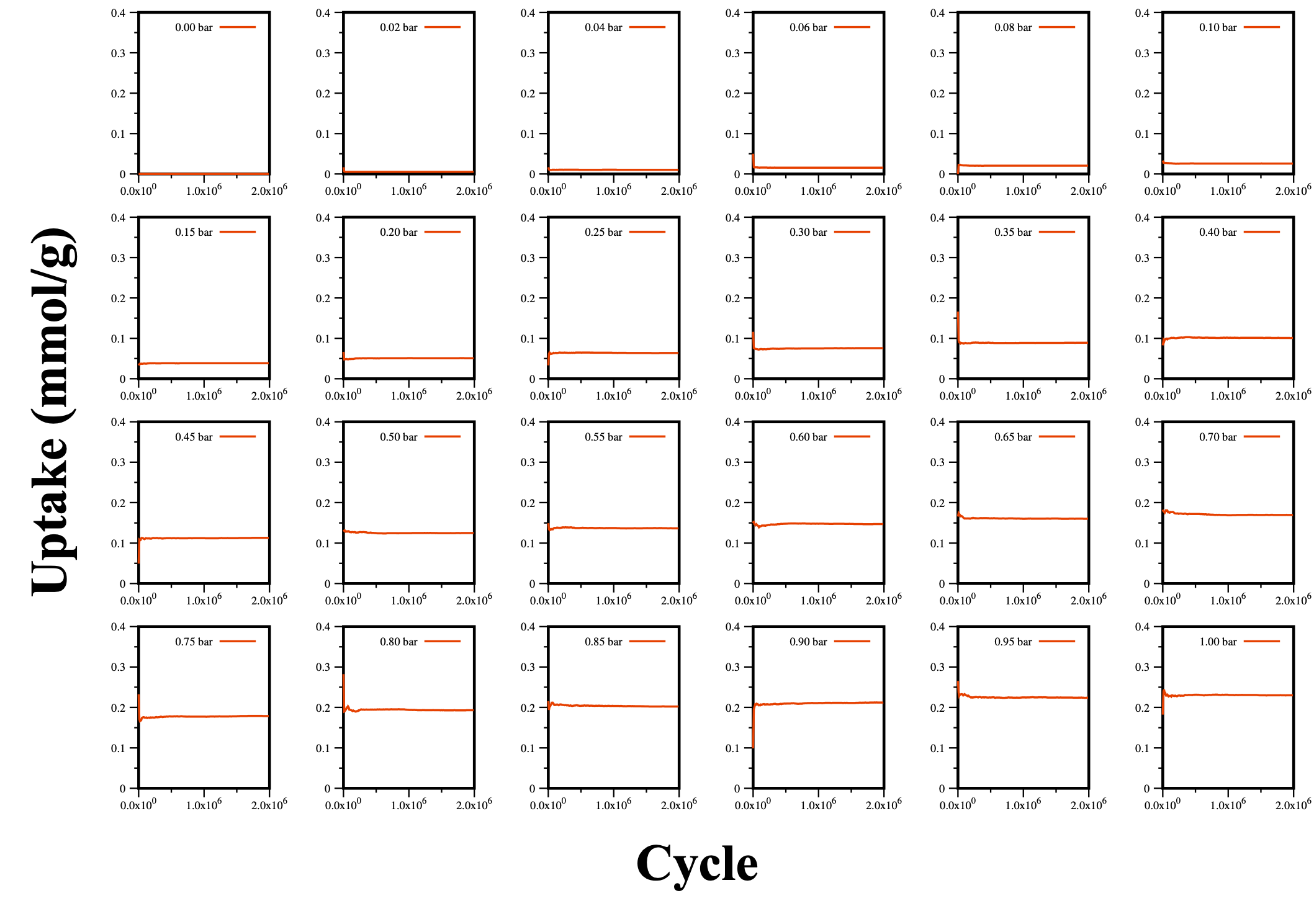}
    \caption{Benchmark of GCMC simulated Kr uptakes at 298 K for Mo$_{26}$V$_{14}$ using different number of equilibration and production cycles.}
    \label{mo26kr}
\end{figure}

\clearpage
\begin{figure}[!h]
\addcontentsline{toc}{figure}{Figure \ref{mo26xe}. Benchmark of GCMC simulated Xe uptakes at 298 K for Mo$_{26}$V$_{14}$ }
    \centering
    \includegraphics[width=0.99\linewidth]{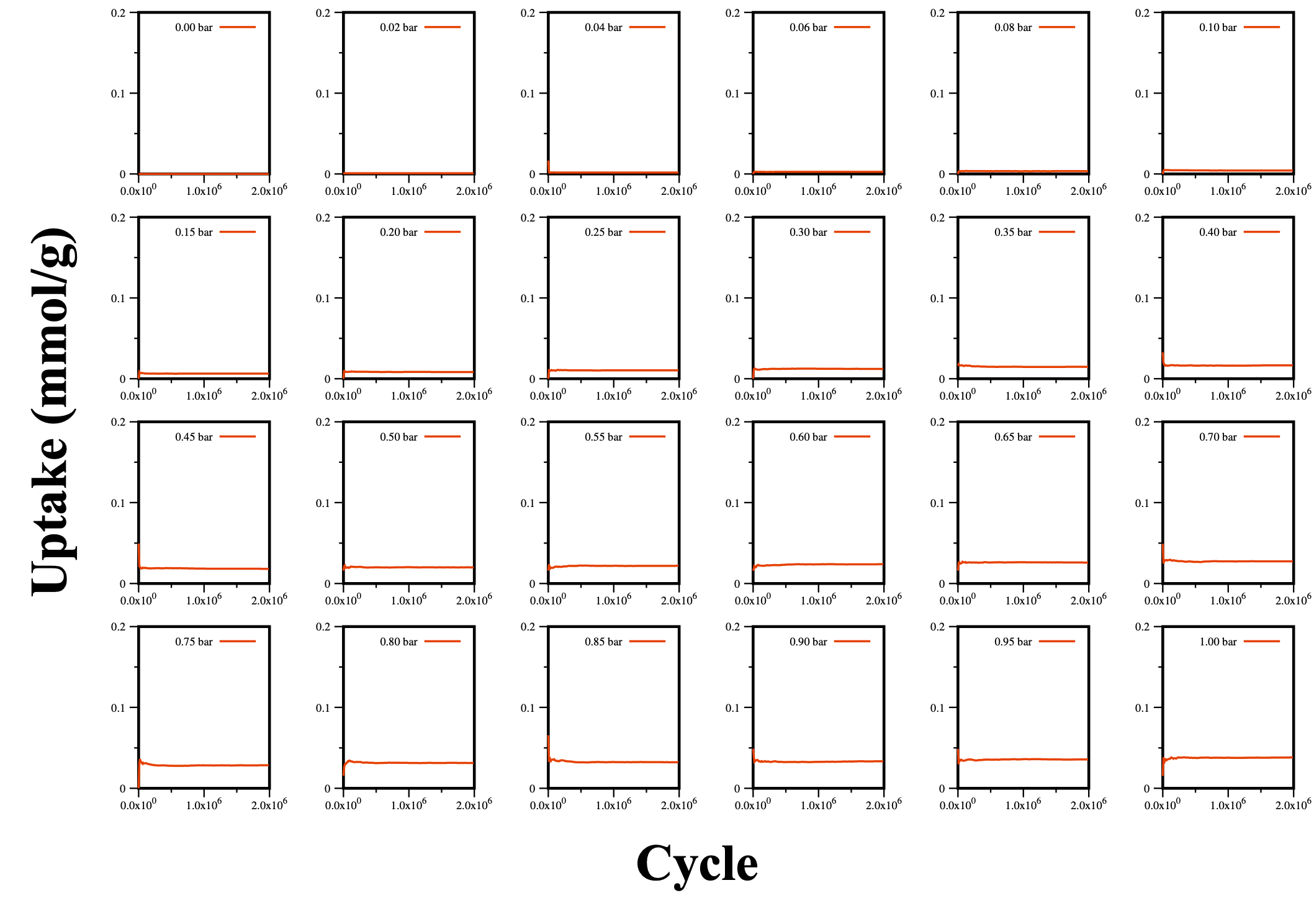}
    \caption{Benchmark of GCMC simulated Xe uptakes at 298 K for Mo$_{26}$V$_{14}$ using different number of equilibration and production cycles.}
    \label{mo26xe}
\end{figure}

\clearpage
\begin{figure}[!h]
\addcontentsline{toc}{figure}{Figure \ref{mo24kr}. Benchmark of GCMC simulated Kr uptakes at 298 K for Mo$_{24}$V$_{16}$ }
    \centering
    \includegraphics[width=0.99\linewidth]{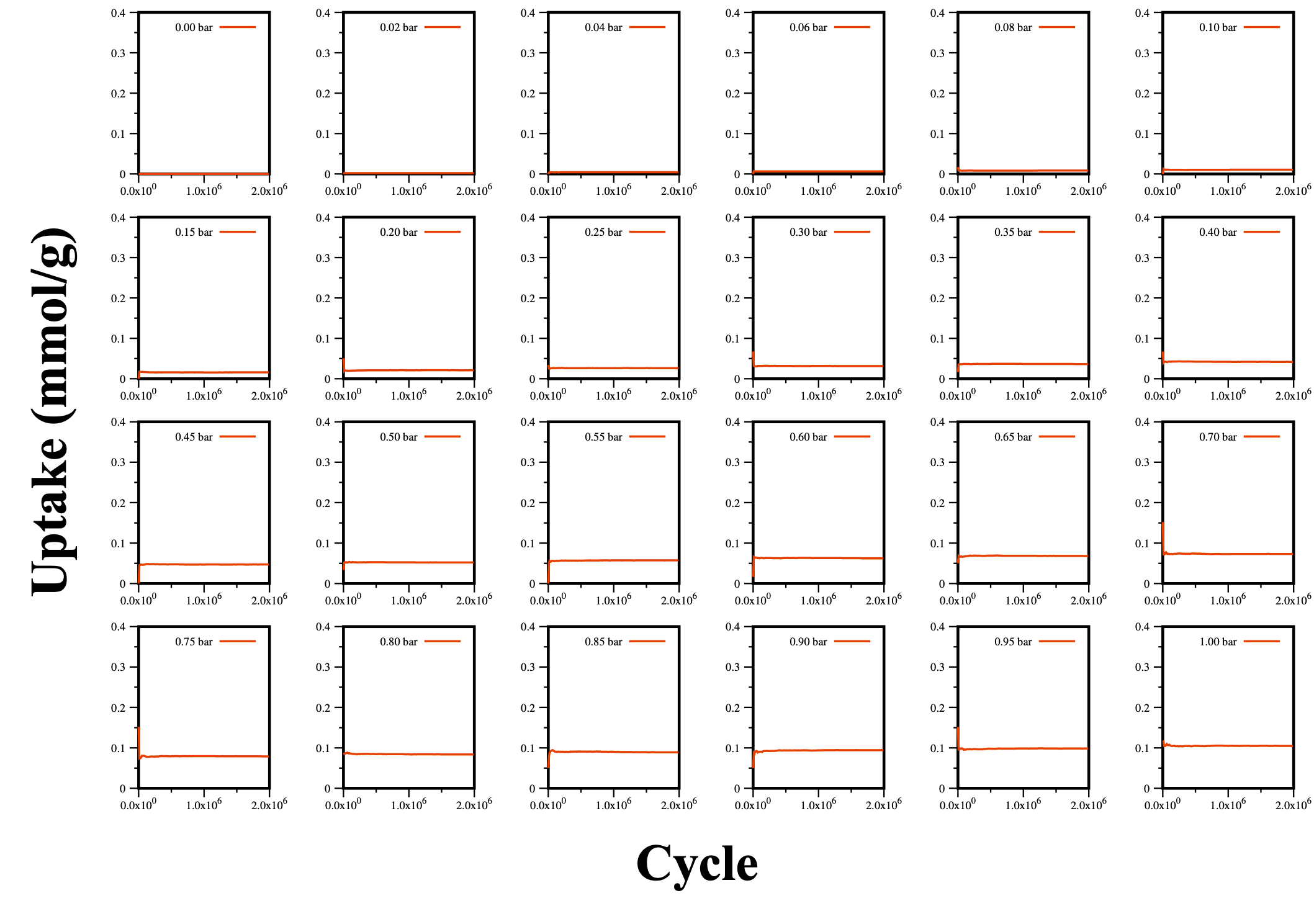}
    \caption{Benchmark of GCMC simulated Kr uptakes at 298 K for Mo$_{24}$V$_{16}$ using different number of equilibration and production cycles}
    \label{mo24kr}
\end{figure}

\clearpage
\begin{figure}[!h]
\addcontentsline{toc}{figure}{Figure \ref{selectivity}. GCMC calculated Kr/Xe selectivity for the 80:20 mixture at 298 K. }
    \centering
    \includegraphics[width=0.7\linewidth]{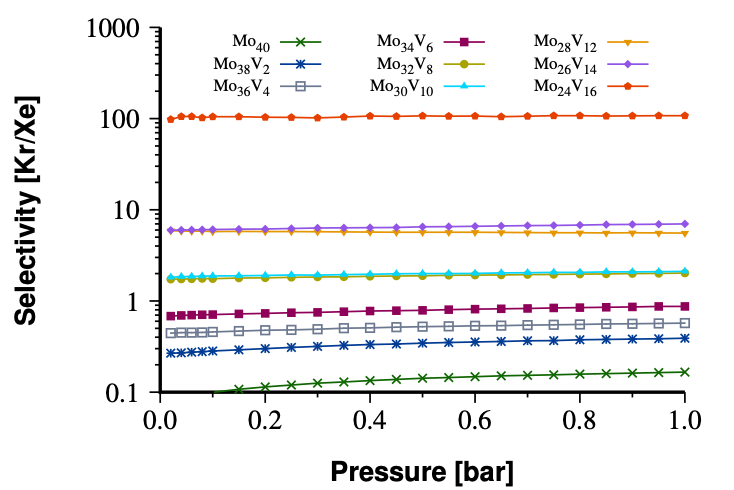}
    \caption{GCMC calculated Kr/Xe selectivity for the 80:20 mixture at 298 K.}
    \label{selectivity}
\end{figure}

\clearpage
\begin{figure*}[!h]
\addcontentsline{toc}{figure}{Figure \ref{trend}. GCMC calculated relationships between Kr/Xe uptakes (top) and selectivity (bottom) against porosity and LCD. }
    \centering
    \includegraphics[width=0.99\linewidth]{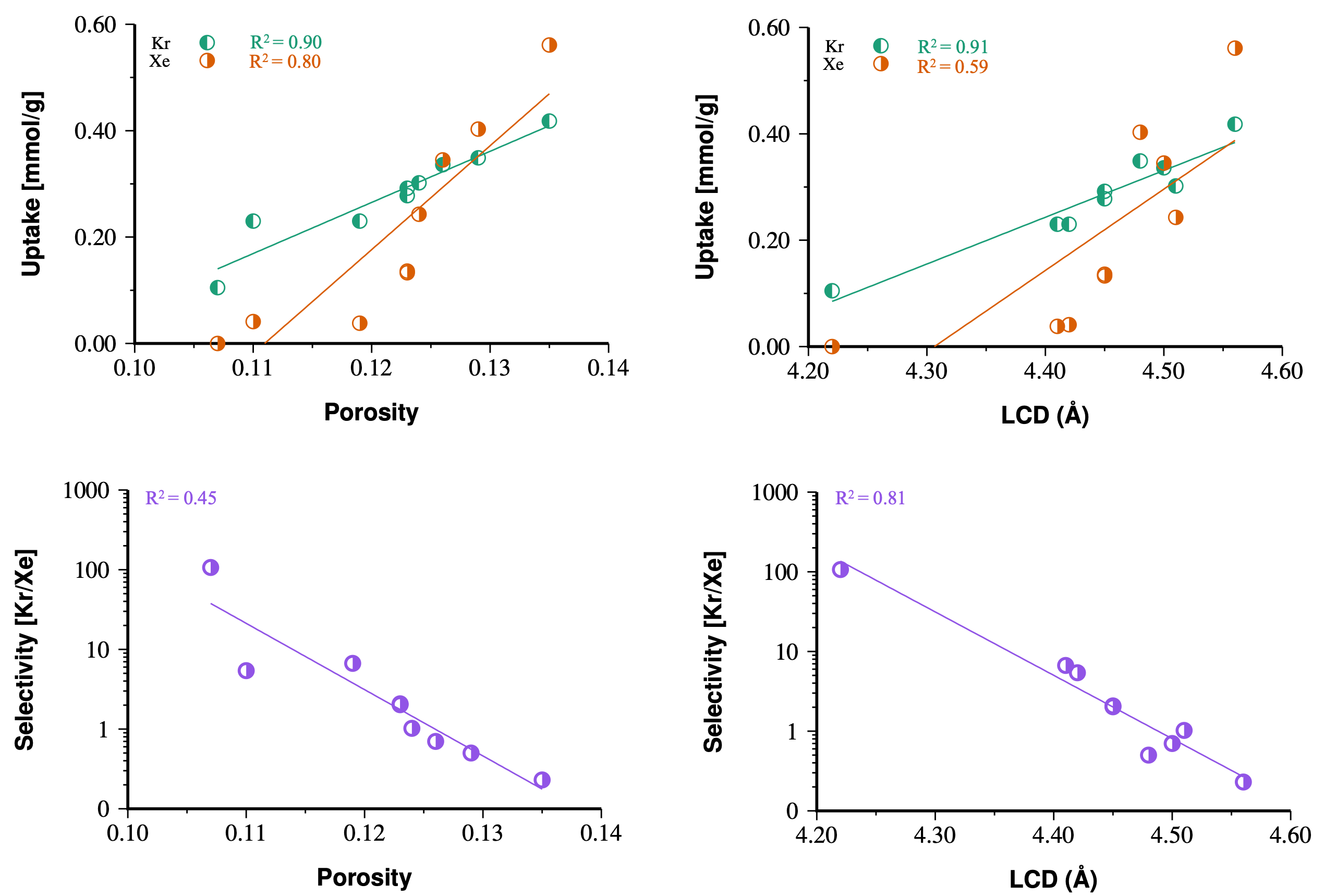}
    \caption{GCMC calculated relationships between Kr/Xe uptakes (top) and selectivity (bottom) against porosity and LCD.}
    \label{trend}
\end{figure*}

\clearpage
\begin{figure}[!h]
\addcontentsline{toc}{figure}{Figure \ref{conv_co2_n2}. Benchmark of GCMC simulated CO$_2$ and N$_2$ uptakes for Mo$_{30}$V$_{10}$ at 298 K }
    \centering
    \includegraphics[width=0.99\linewidth]{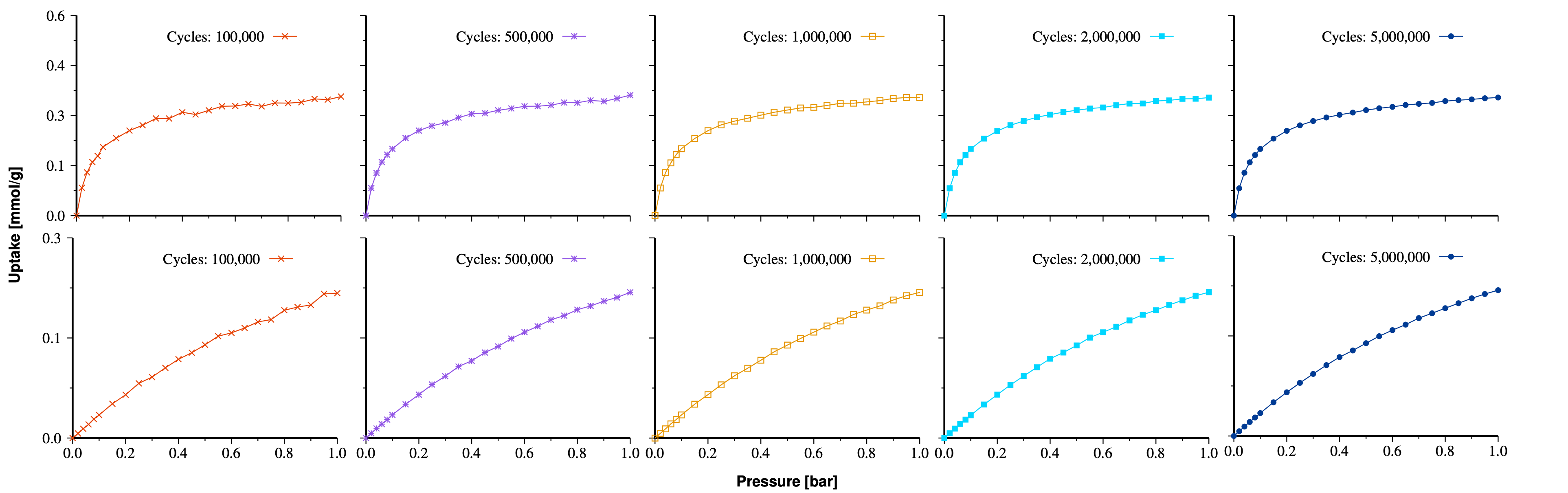}
    \caption{Benchmark of GCMC calculated CO$_2$ (top row) and N$_2$ (bottom row) uptakes using different number of equilibration and production cycles for the Mo$_{30}$V$_{10}$ system at 298 K.}
    \label{conv_co2_n2}
\end{figure}

\clearpage
\begin{figure*}[!h]
\addcontentsline{toc}{figure}{Figure \ref{rdf}. GCMC calculated Kr-O and Xe-O RDFs}
    \centering
    \includegraphics[width=0.99\linewidth]{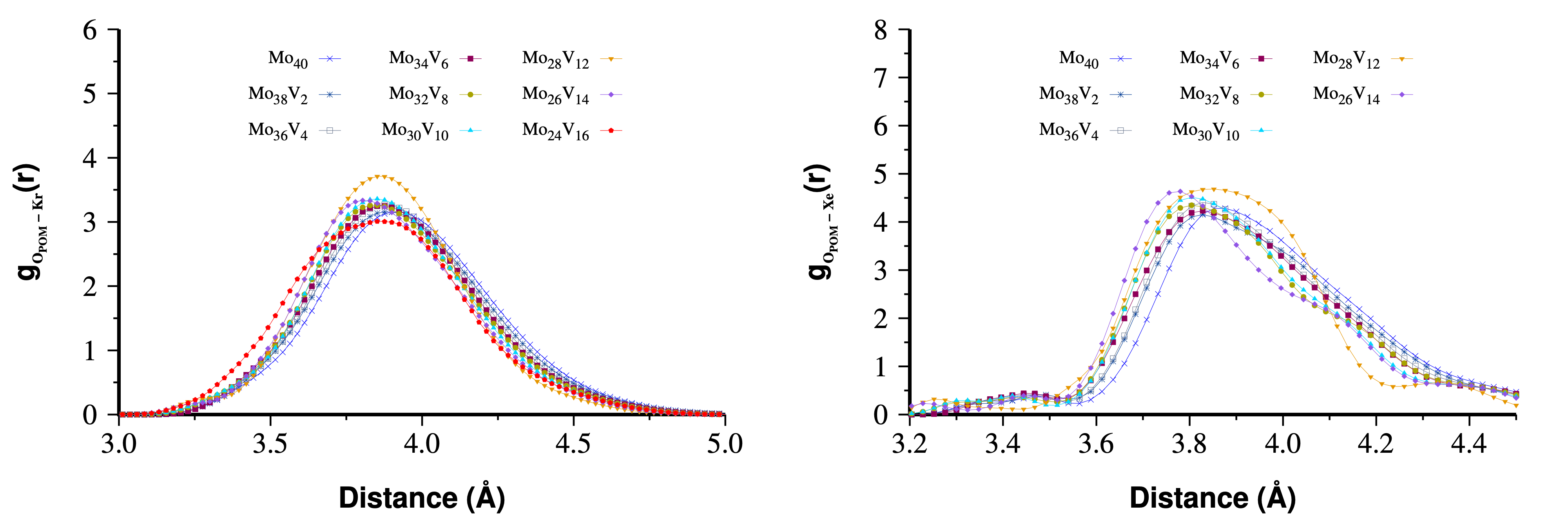}
    \caption{GCMC calculated Kr-O and Xe-O RDFs of all considered systems.}
    \label{rdf}
\end{figure*}

\clearpage
\begin{figure*}[!h]
\addcontentsline{toc}{figure}{Figure \ref{co2_n2_isotherm}. GCMC calculated CO$_{2}$ and N$_{2}$ adsorption isotherms (a and b) and Kr/CO$_2$ and Kr/N$_2$ selectivities for 50:50 mixtures (c and d) at 298 K}
    \centering
    \includegraphics[width=0.99\linewidth]{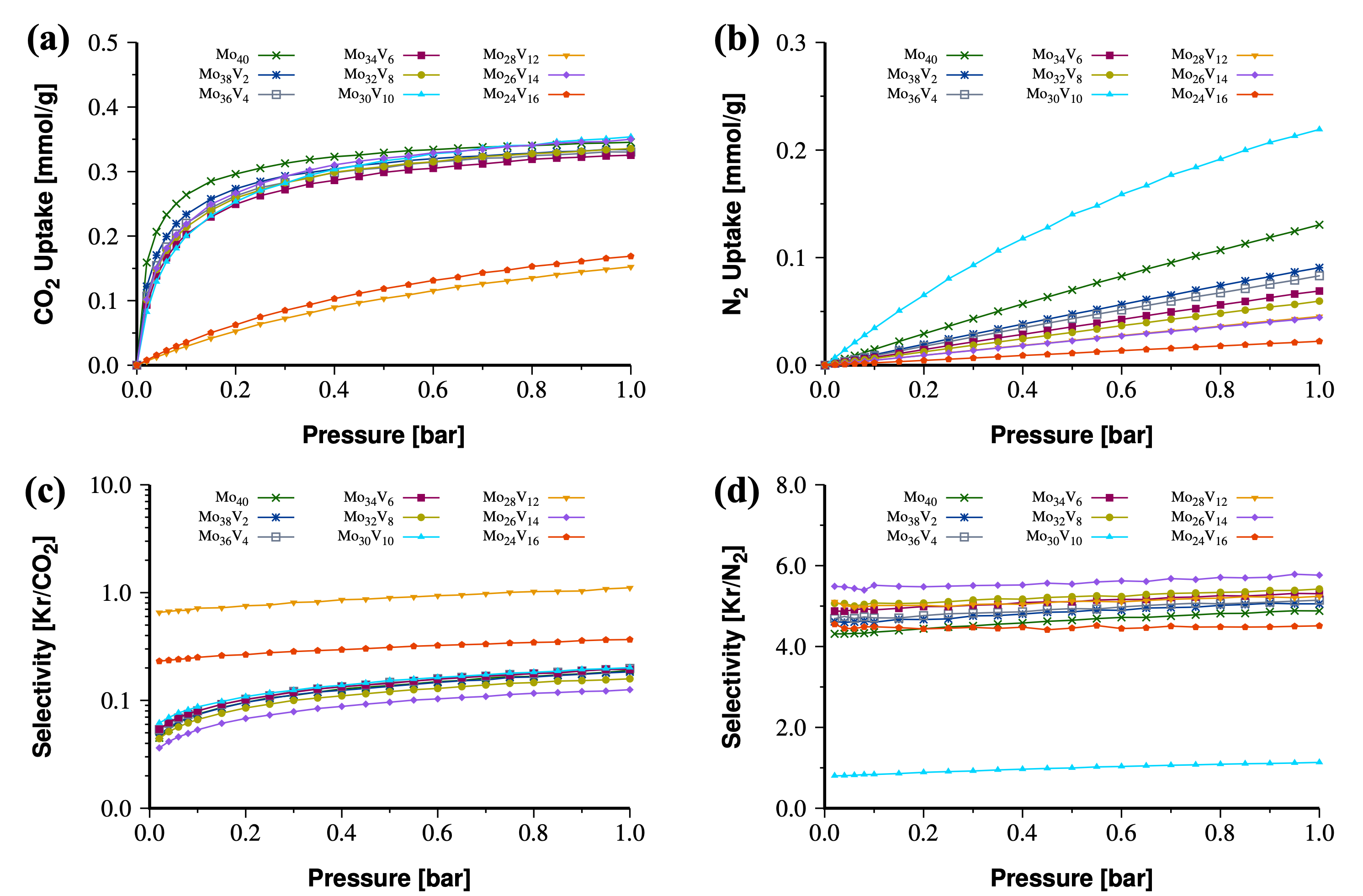}
    \caption{GCMC calculated CO$_{2}$ and N$_{2}$ adsorption isotherms (a and b) and Kr/CO$_2$ and Kr/N$_2$ selectivities for 50:50 mixtures (c and d) at 298 K for all systems considered in this work.}
    \label{co2_n2_isotherm}
\end{figure*}

\clearpage
\begin{figure*}[!h]
\addcontentsline{toc}{figure}{Figure \ref{pore}. Calculated (a) Kr and (b)Xe density differences in Mo$_{30}$V$_{10}$}
    \centering
    \includegraphics[width=0.99\linewidth]{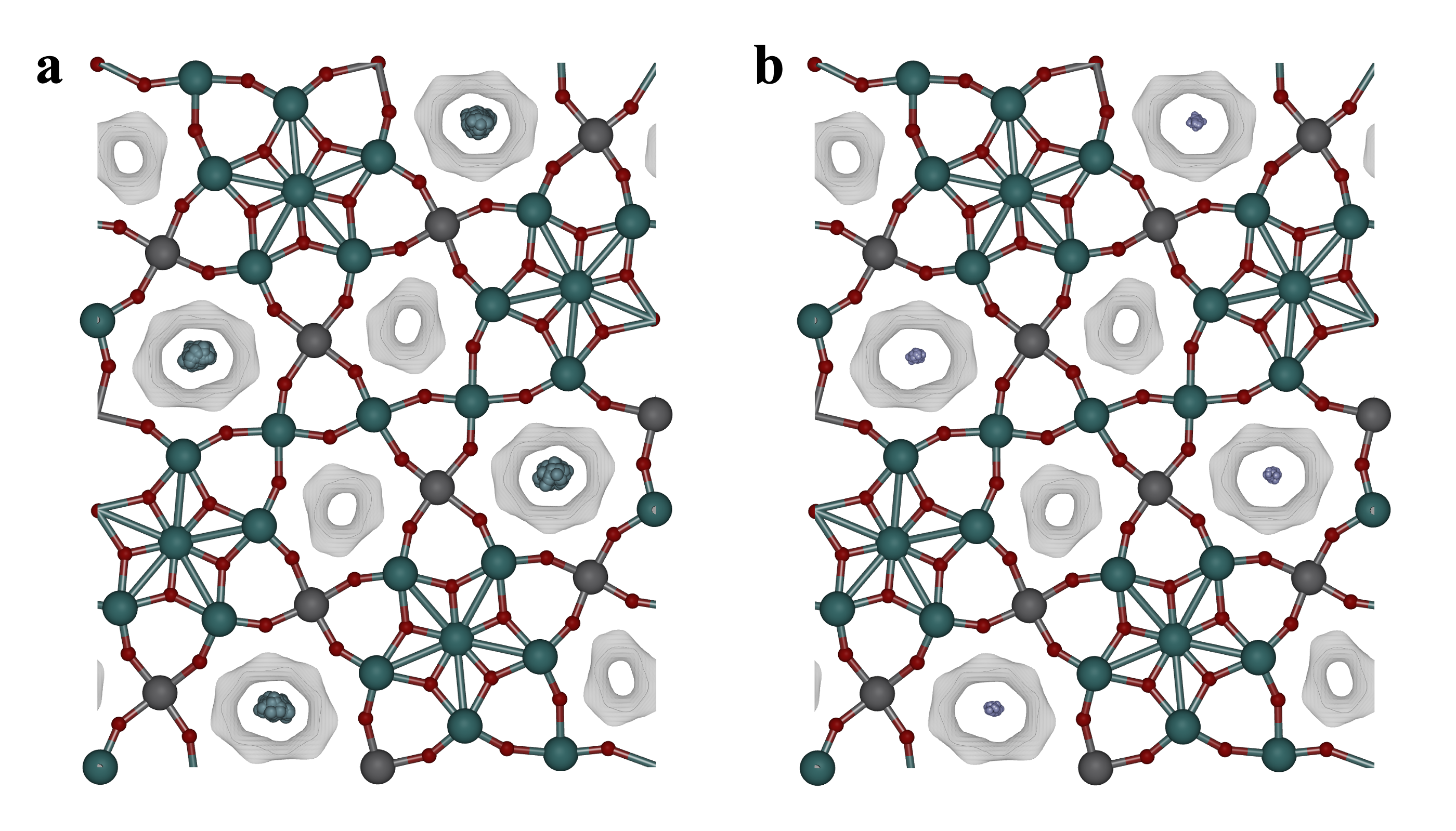}
    \caption{Visualization of post adsorption (a) Kr and (b)Xe distribution in the heptagonal pores in Mo$_{30}$V$_{10}$. Only the heptagonal pores contribute to the adsorption.}
    \label{pore}
\end{figure*}

\clearpage
\begin{landscape}
\begin{table*}[!h]
\addcontentsline{toc}{table}{Table \ref{zeo}. Calculated density, porosity, GSA, VSA, LCD, PLD, Kr and Xe uptakes, and selectivity for all systems considered in this work}
\centering
\resizebox{0.99\linewidth}{!}{
\begin{tabular}{lcccccccccccccc} \hline
\textbf{System} & \textbf{Density} & \textbf{Porosity} & \textbf{GSA} & \textbf{VSA} & \textbf{LCD} & \textbf{PLD} & \textbf{Kr Uptake} & \textbf{Xe Uptake} & \textbf{Kr/Xe 50:50} & \textbf{Kr/Xe 80:20} & \textbf{CO$_{2}$ Uptake} & \textbf{N$_{2}$ Uptake} & \textbf{Kr/CO$_{2}$ 50:50} & \textbf{Kr/N$_{2}$ 50:50}\\
& \textbf{(g/cm$^3$)} & & \textbf{(m$^2$/g)} & \textbf{(m$^2$/cm$^3$)} & \textbf{(\AA)} & \textbf{(\AA)} & \textbf{(mmol/g)} & \textbf{(mmol/g)} & \textbf{Selectivity} & \textbf{Selectivity} & \textbf{(mmol/g)} & \textbf{(mmol/g)} & \textbf{Selectivity} & \textbf{Selectivity} \\ \hline
Mo$_{40}$ & 4.02 & 0.14 & 90.10 & 361.75 & 4.56 & 3.43 & 0.41 & 0.54 & 0.23 & 0.17 & 0.35 & 0.13 & 0.19 & 4.88 \\
Mo$_{38}$V$_2$ & 3.97 & 0.13 & 87.99 & 349.07 & 4.48 & 3.34 & 0.31 & 0.41 & 0.50 & 0.39 & 0.33 & 0.09 & 0.18 & 5.06 \\
Mo$_{36}$V$_4$ & 3.92 & 0.13 & 91.60 & 358.90 & 4.50 & 3.34 & 0.35 & 0.35 & 0.70 & 0.57 & 0.33 & 0.08 & 0.20 & 5.15 \\
Mo$_{34}$V$_6$ & 3.88 & 0.12 & 84.28 & 327.17 & 4.51 & 3.35 & 0.32 & 0.25 & 1.02 & 0.87 & 0.33 & 0.07 & 0.19 & 5.31 \\
Mo$_{32}$V$_8$ & 3.85 & 0.12 & 86.90 & 334.88 & 4.45 & 3.31 & 0.29 & 0.13 & 2.02 & 2.02 & 0.34 & 0.06 & 0.16 & 5.42 \\
Mo$_{30}$V$_{10}$ & 3.80 & 0.12 & 88.55 & 336.74 & 4.45 & 3.28 & 0.31 & 0.14 & 2.08 & 2.11 & 0.35 & 0.22 & 0.20 & 1.13 \\
Mo$_{28}$V$_{12}$ & 3.62 & 0.11 & 89.56 & 324.46 & 4.42 & 3.09 & 0.22 & 0.05 & 5.41 & 5.56 & 0.15 & 0.05 & 1.11 & 5.24 \\
Mo$_{26}$V$_{14}$ & 3.73 & 0.12 & 85.38 & 318.66 & 4.41 & 3.28 & 0.25 & 0.04 & 6.69 & 7.01 & 0.35 & 0.04 & 0.13 & 5.76 \\
Mo$_{24}$V$_{16}$ & 3.66 & 0.11 & 87.83 & 321.41 & 4.22 & 3.05 & 0.10 & 0.00 & 106.53 & 107.36 & 0.17 & 0.02 & 0.37 & 4.51 \\ \hline
\end{tabular}}
\caption{Calculated density, porosity, GSA, VSA, LCD, PLD, Kr and Xe uptakes, and selectivity for all systems considered in this work.}
\label{zeo}
\end{table*}
\end{landscape}

\begin{table*}[!h]
\addcontentsline{toc}{table}{Table \ref{isosteric}. Isosteric heat of adsorption for all systems}
\centering
\resizebox{0.55\linewidth}{!}{
\begin{tabular}{lcccc} \hline    
\textbf{System}  & \textbf{Q$_{st}$ Kr} & \textbf{Q$_{st}$ Xe} & \textbf{Q$_{st}$ N$_2$} & \textbf{Q$_{st}$ CO$_2$} \\ 
 & \textbf{(kJ/mol)} & \textbf{(kJ/mol)} & \textbf{(kJ/mol)} & \textbf{(kJ/mol)} \\ \hline
Mo$_{40}$ & 31.7 & 39.2 & 27.4 & 40.5 \\
Mo$_{38}$V$_2$ & 30.8 & 36.5 & 26.6 & 42.7 \\
Mo$_{36}$V$_4$ & 30.7 & 35.7 & 26.5 & 43.3 \\
Mo$_{34}$V$_6$ & 30.3 & 34.6 & 26.1 & 43.3 \\
Mo$_{32}$V$_8$ & 30.1 & 32.2 & 25.8 & 44.2 \\
Mo$_{30}$V$_{10}$ & 30.2 & 32.2 & 32.9 & 43.7 \\
Mo$_{28}$V$_{12}$ & 29.2 & 29.0 & 25.5 & 39.4 \\
Mo$_{26}$V$_{14}$ & 29.5 & 29.2 & 25.3 & 44.5\\
Mo$_{24}$V$_{16}$ & 27.2 & 20.0 & 24.6 & 39.0 \\ \hline
\end{tabular}}
\caption{Isosteric heat of adsorption for all systems.}
\label{isosteric}
\end{table*}

\clearpage
\bibliography{bib}